\RequirePackage{fix-cm}
\documentclass[smallextended]{svjour3}       

\smartqed

\usepackage[T1]{fontenc}
\usepackage{tgbonum}

\usepackage{amssymb}

\usepackage{mathptmx}

\usepackage{titlesec}

\titleformat{\title}{\normalfont\LARGE\bfseries}{\thesection}{1em}{}

\usepackage[a4paper, top=1in, bottom=1in, left=1in, right=1in]{geometry}

\usepackage[square,numbers]{natbib}
\usepackage{url}
\usepackage{listings}
\usepackage{multirow, tabularx, balance, stfloats, booktabs}
\usepackage{enumitem}
\usepackage{hyperref}
\usepackage{lscape}
\usepackage{makecell}
\usepackage[table]{xcolor}
\usepackage{soul}
\usepackage{subcaption}
\usepackage{graphicx}
\usepackage[normalem]{ulem}

\usepackage[most]{tcolorbox}

\definecolor{boxheader}{RGB}{99, 140, 109}
\definecolor{boxcontent}{RGB}{231, 251, 180}

\definecolor{redul}{RGB}{255,0,0}
\newcommand{\reduline}[1]{%
  {\setulcolor{redul}\ul{#1}}%
}

\usepackage{titlesec} 

\titleformat{\section}
  {\bfseries\Large}  
  {\thesection}{1em}{}

\titleformat{\subsection}
  {\bfseries\large}  
  {\thesubsection}{1em}{}

\smartqed  
\usepackage{graphicx}
\newcommand{\starcoder}{StarCoder2 }
\newcommand{\granite}{Granite }
\newcommand{\passk}{pass@k }
\newcommand{\robustpass}{RobustPass@k }
\newcommand{\passone}{pass@1 }
\newcommand{\passten}{pass@10 }

\usepackage[most]{tcolorbox}
\tcbset{textmarker/.style={%
        enhanced, breakable,
        parbox=false,boxrule=0mm,boxsep=0mm,arc=0mm,
        outer arc=0mm,left=1.8mm,right=1.8mm,top=7pt,bottom=7pt,
        toptitle=1mm,bottomtitle=1mm,oversize}}

\newtcolorbox{noteBox}{
    breakable,
    enhanced,
    textmarker,
    borderline west={3pt}{0pt}{gray},
    colback=gray!10!white
}

\def\BibTeX{{\rm B\kern-.05em{\sc i\kern-.025em b}\kern-.08em
    T\kern-.1667em\lower.7ex\hbox{E}\kern-.125emX}}

\definecolor{dkgreen}{rgb}{0,0.6,0}
\definecolor{gray}{rgb}{0.5,0.5,0.5}
\definecolor{mauve}{rgb}{0.58,0,0.82}

\journalname{Empirical Software Engineering}
\begin{document}
\fontsize{12pt}{14pt}\selectfont

\title{\textbf{MergeRepair: An Exploratory Study on Merging Task-Specific Adapters in Code LLMs for Automated Program Repair}}

\titlerunning{MergeRepair}        

\author{Meghdad Dehghan \and Jie JW Wu \and Fatemeh H. Fard \and Ali Ouni 
}


\institute{Meghdad Dehghan \at
              University of British Columbia, Department of Computer Science \\
              \email{meghdad.dehghan@ubc.ca}           
           \and
           Jie JW Wu \at
           University of British Columbia, Department of Computer Science \\ 
           \email{jie.jw.wu@acm.org}
           \and
           Fatemeh Fard \at
           University of British Columbia, Department of Computer Science \\
           \email{fatemeh.fard@ubc.ca}
           \and
           Ali Ouni \at
           École de technologie supérieure, Department of Software and IT Engineering \\
           \email{ali.ouni@etsmtl.ca}
}

\date{Received: date / Accepted: date}

\maketitle

\begin{abstract}
Large Language Models (LLMs) have shown high capabilities in several software development-related tasks such as program repair, documentation, code refactoring, debugging, and testing. However, training these models requires massive amount of data and significant computational resources. Adapters are specialized, small modules designed for parameter efficient fine-tuning of LLMs for specific tasks, domains, or applications without requiring extensive retraining of the entire model. These adapters offer a more efficient way to customize LLMs for particular needs, leveraging the pre-existing capabilities of the large model. 
Model (and adapter) merging have emerged as a technique to develop one model capable of multiple tasks, with minimal or no training required. 
Although model and adapter merging has shown promising performance in domains such as natural language processing and computer vision, its applicability to software engineering tasks remains underexplored. In this paper, we investigate the effectiveness of merged adapters within the context of software engineering, with a particular focus on the Automated Program Repair (APR) task, through our approach, \textbf{MergeRepair}. In particular, we merge multiple task-specific adapters using three different merging methods, including weight-averaging, ties, and dare-ties, and evaluate the performance of the merged adapter on the APR task. 
We introduce a \textbf{continual merging} approach, a novel method in which we sequentially merge the task-specific adapters where the order and weight of the merged adapters play a significant role. 
We further compare the performance of our approach with a baseline method consisting of equal-weight merging applied on parameters of different adapters, where all adapters are of equal importance. 
Through continual merging, we explore the capability of merged adapters and the effect of task order, as it occurs in real-world software projects.

To evaluate our approach, we consider two LLMs, \starcoder and \granite models on an APR benchmark, HumanEvalFix, with various combinations of the merged adapters trained on a subset of the CommitPackFT dataset that contains four tasks and APR in Python. The results show that merging task-specific adapters can enhance the performance of models on the APR task, \textit{without additional training}. In particular, our approach achieves up to 2.38\% 
performance improvement over APR task in terms of \passone score when merging APR with Improvement and Misc task-adapters on top of the \starcoder model. Performance rises to 4.01\% for \passten score with the same model and merged adapters. This improvement could be significantly influenced by the base model and the specific task adapters merged, rather than only the number of adapters. The proposed continual merging approach can be effective if the order of merging is efficiently determined. Results show that performance improves when the most effective task-adapter is added last in a continual merging process. Moreover, we can achieve on-par or even better performance when APR adapter is not used in the merged model, emphasizing the generalizeability of the approach. 
These findings suggest that leveraging knowledge from other code-related tasks through adapter merging can improve performance on new tasks like APR, even with limited data. 


\end{abstract}

\section{Introduction}
\label{intro}


Significant Software Engineering (SE) efforts have been devoted to automate code-related tasks and facilitate developers' tasks using deep learning models. 
Bug prediction \cite{bugpred}, bug fixing \cite{sequencer, coconut}, code generation \cite{codex, codegen}, comment generation \cite{commentgen}, and commit message generation \cite{atom} are among these tasks. 
In practice, developers often need to train, test and validate different models for each of these tasks separately along with data collection and pre-processing, feature engineering, hyper-parameter tuning, and so on, to ensure efficient models are developed. However, repeating the process for each new task can be time-consuming and cumbersome.
To train a new model that can perform well on a downstream task, there are several requirements, including an (often domain-specific) dataset, and sufficient computational resources. 
Prior research has shown that these requirements are not trivial to fulfill \cite{se-adapter1}. 
First, for each new task, a new model should be trained and deployed, which is costly. 
Additionally, by following this process, practitioners will end up with several task-specific models, which are unable to leverage the strengths of other models, while they could benefit from related tasks to improve the performance within or outside their domain \cite{mergekit, zipit}.

The main question that we intend to investigate in this study is \textit{``Can we re-use previously trained code-related adapters for a new SE task by merging them?''}
Adapters are small modules used in Parameter-Efficient Fine-Tuning (PEFT)~\cite{houlsby} of Large Language Models (LLM) and have results on par with or better than fully fine tuning language models~\cite{ding2023parameter, weyssow2023exploring}. 
The idea behind our study is rooted in the foundations of transfer learning~\cite{transferrrr}, where the learned knowledge of a language model can be adapted to new tasks, domains, or languages \cite{t5}. 
The advantages of transfer learning in software engineering and code intelligence have been shown by several recent works adopting LLMs \cite{repairllama, weyssow2023exploring}. 
Although fine-tuning is one of the main approaches for using language models, there are concerns about using several task-specific adapters or models. The issue of catastrophic forgetting~\cite{l3} is another concern when fine-tuning a language model.
Additionally, fully fine-tuning LLMs \cite{fine-tuning} becomes inefficient as multiple instances of {the model} for different tasks should be trained and deployed. 

To address the above-mentioned issues, recent research
showed the efficiency of merging models. Merging models enhances their performance compared to task-specific models~\cite{mergekit, zipit}. 
~\citet{mergekit} showed that merging models combines the parameters of individual task-specific models into a single model in order to leverage the knowledge of other models. Note that merging models is intrinsically different from Multi-Task Learning (MTL)~\cite{mtl, mtl-nl}. In MTL, one model is jointly trained on two or more tasks; while in merging models, there exist multiple models, each of them trained separately on distinct tasks. Then, all of these models are merged together. Though there are multiple approaches for model merging, we choose the techniques that merge models \textit{without additional training}~\cite{yang2024model}. 

Similar techniques are developed for merging adapters. 
There are studies that merge multiple models~\cite{fisher, rebasin, dare} or adapters in natural language processing (NLP) domain~\cite{adaptersoup, generaliz}. 
Adapters were also used for Multi-Task Learning~\cite{fusion}. Others adopted the idea of mixture-of-experts to inject a set of expert adapter layers in the transformer-based models in order to train multi-task models~\cite{adamix}. 
However, 
the research on merging adapters is scarce, and
there is a research gap in investigating the merging ability of adapters for code-related tasks. 
This is specially important with the advent of Code LLMs--LLMs that are pre-trained  specifically to understand and generate code in various programming languages. 
Though these models have shown promising results for many SE tasks, their computational cost is not negligible for companies and researchers. 
Thus, it is beneficial to re-use the trained models/adapters for new tasks by merging them, \textit{without additional training}. There is no prior research that have investigated whether current approaches affect code-related tasks in the same way as in NLP and whether they would improve the performance of each task in the merged model.

The main goal of this study is to \textit{explore the performance of merged task-specific adapters and continual merging, in the context of Automated Program Repair (APR) task.}
We investigated the effect of merging adapters for different SE tasks, using our framework, \textbf{MergeRepair}. 
First, we examined three merging techniques from the existing literature, being weight-space averaging\cite{modelsoup}, TIES-Merging~\cite{ties} and DARE~\cite{dare}, in which all adapters are of \textit{equal} importance in the merged model. 
We studied the effect of merging on APR, and the generalizability of merging adapters to APR, i.e., when the merged model is tested on APR without being trained on it. 
Second, we explored our proposed merging paradigm, \textbf{continual merging} using all three merging techniques (i.e., weight-space averaging, TIES-Merging, and DARE). In continual merging, the \textit{order and weights} of the adapters play an important factor, and the merged adapters do not have equal importance.
Our proposed approach is motivated by the practical considerations of merged models, when new models are often introduced sequentially. Additionally, not all tasks could benefit the new target task equally. The continual merging addresses these two points by considering different weights based on the order of the tasks in the merged model. 

To evaluate our approach, we utilize the Low-Rank Adaptation (LoRA)~\cite{lora} as adapter modules and train one instance of LoRA per task. LoRA optimizes low-rank decomposition of the weight matrices and, in this way, can achieve comparable results to fully fine-tuning~\cite{ding2023parameter, weyssow2023exploring}.  
In our framework, \textbf{MergeRepair}, we explore the merging capability of adapters injected in Code LLMs to improve the performance of a selected task, automatic program repair. APR is an active research area aiming to reduce the manual effort for software developers~\cite{critical-apr}, and facilitates software development and maintenance~\cite{lower-cost}. Specifically, APR automates the process of generating the fixed code or patch, for a given code that contains bugs. 
We chose the broader categories of the tasks from CommitPackFT dataset~\cite{octopack}, being Development, Test \& QA, Improvement, and Misc (includes tasks like configuration, formatting, dependencies, and documentation). 
CommitPackFT dataset~\cite{octopack} is chosen as it is a widely used dataset for code related tasks and code repair~\cite{starcoder2, octopack, zhuo2024astraios} for our study, as merging tasks to APR (i.e., Bug category of the CommitPackFT dataset).
The dataset contains 59K samples, with $\sim 19\%$ belonging to APR. 

\textbf{Findings.} Our results show that the base models and the constituent tasks of the merged adapter play an important role in improving the performance of APR in a merged model. Additionally, the performance of the merged adapter depends more on the task-adapter being used in the merging than the number of task-adapters being merged. 
Furthermore, we observe that merged adapters originating from non-APR tasks can still perform well on the APR benchmark, provided their constituent adapters achieve strong APR scores.
In some cases, the merged adapter even outperforms the APR-specific adapter, demonstrating the generalizability of the merging approach. 
Finally, in the continual merging approach, the order of merging task-specific adapters affects the performance of the merged adapters significantly. 
For most of the merged task-specific adapters, we observed an improved performance when using continual merging, considering the best order of the adapter, compared to their equal-weight
averaging. 

The contributions of this study are as follows: 
\begin{itemize}
    \item We are the first to study merging adapters for automatic program repair task. 
    \item We empirically study the effect of different equal-weight averaging merging techniques on automatic program repair task, using our framework, \textbf{MergeRepair}. 
    \item We propose continual merging, a new technique that considers different weights for the adapters in the merged model. 
    \item  Our code and replication package are made public\footnote{Replication package: \url{https://github.com/mqddd/mergerepair}} to
support open data and open science principles. 
\end{itemize}

Our study offers actionable insights for both practitioners and researchers interested in model reuse through merging techniques across various software engineering tasks.
Since the merging methods explored in this work do not require training on datasets, they can support data confidentiality by enabling model sharing without exposing sensitive data. 
Additionally, our findings highlight the role of different tasks in influencing the effectiveness of model merging for automated program repair, potentially opening new avenues for future research.

The rest of the paper is structured as follows. Section \ref{sec:rl} describes the background and related work of our research.
Section \ref{sec:design} explains the design of our empirical study. Section \ref{sec:res} summarizes the results for RQs. Section \ref{sec:disc} includes
more analysis and discussions on the results. Implications of study and threats to validity are explained in sections \ref{sec:implications} and \ref{sec:threats}, respectively. Finally, Section \ref{sec:con} concludes this work.

\section{Related Works}
\label{sec:rl}
 

\subsection{Automated Program Repair}
Early research in APR includes searched-based methods which are studied in several works~\cite{search-based1, search-based2, simfix, varfix, game}. These studies search for the most probable places in the code that are prone to be buggy and then generate candidate patches to fix the buggy codes. To localize the buggy part, various techniques are utilized. In \cite{game}, the authors treat the program repair task as a game and propose a heuristic method to search in the search space. More recent techniques such as SimFix~\cite{simfix} and VarFix~\cite{varfix} use code similarity metrics to find the correct patch among a set of candidate patches. More specifically, SimFix~\cite{simfix} uses the Abstract Syntax Tree (AST) of the similar code snippets and existing candidate patches and searches in the intersection of these two sets to find the correct patch. VarFix~\cite{varfix} uses the variational execution technique in which a program is executed multiple times that allows systematically exploration in large search spaces.

Constraint-based and template-based methods are other approaches for APR that are introduced following search-based methods. In constraint-based methods, the focus is on generating patch candidates based on a set of rules or constraints. Such techniques make the search space more limited compared to other methods~\cite{autofix, semfix, nopol}. Template-based methods are introduced to alleviate the issues with the previous methods. Such tools utilize a predefined program template for generating fixed patches. For example, in FixMiner~\cite{fixminer}, authors proposed an approach to mine fix patterns leveraging the changes within patches that could be used in patch generation systems.  
Later works on automatic program repair train deep neural net~\cite{coconut, deepdelta} and use pre-trained language models~\cite{vulrepair, practicalrepair}. By the emergence of LLMs, the research towards adapting these models to downstream SE tasks has shifted to instruction-tuning them to be aligned with the target domain of interest~\cite{vulrepair, practicalrepair}.

In this research, to develop the task-specific models, we train adapters using instruction-tuning datasets and evaluate the models, containing the merged adapter, on the APR task.
Although language models and PEFT methods are studied for APR in previous works \cite{repairllama}, there is no current research that leverages the combination of similar task-specific adapters using merging methods in order to study the performance of the APR task.

\subsection{Adapters}
Adapters are light-weight modules that are inserted between the Transformer layers and used as a technique for parameter-efficient fine-tuning \cite{houlsby2019parameter}. Adapters have emerged as a technique to optimize memory usage by reducing the number of trainable parameters during fine-tuning \cite{houlsby}. 
They have shown promising results compared to fully fine-tuned models while efficiently adapting large language models to downstream tasks. 
Subsequently, several adapter architectures have been proposed, including LoRA \cite{lora}
and IA3 \cite{tfew}, each targeting specific layers and parameters of the transformer-based models.
There are empirical studies on the utilization of adapters for SE tasks \cite{se-adapter1, se-adapter3}. Notably, recent studies have adopted adapters as a primary alternative method for efficiently fine-tuning large language models in specialized domains such as program repair \cite{repairllama}. Apart from APR and SE tasks, adapters and PEFT methods have been utilized primarily in other domains such as natural language processing~\cite{houlsby, adapters-nlp}, computer vision~\cite{peft-cv1, peft-cv2, peft-cv3}, and speech recognition~\cite{speech1, speech2}.
Though adapters and LORA have been used in several software engineering studies~\cite{zou2025experimental, haque2025systematic, mannisto-etal-2025-comparative}, merging them for APR is not studied. 

In this study, we opt for LoRA as the adapter for fine-tuning the models across all target tasks. LoRA stands out as one of the most prevalent and widely used methods in the research community~\cite{ding2023parameter, weyssow2023exploring}. We then apply the merging techniques on the trained LORA adapters. 

\subsection{Model Merging}
Merging models involves combining the parameters of two or more specialized models to create a unified multi-task model~\cite{fisher}. 
Though multi-task model is one of the main motivations of model merging, this technique has been used for model unlearning, knowledge transfer among different languages, detoxification of LLMs, adversarial learning, federated learning, and multi-modal and cross-modal fusions as well~\cite{yang2024model}. 
\citet{zhang2023composing} proposed to compose different PEFT modules with arithmetic operations without retraining and demonstrated the benefits of such new PEFT in several experiments.   
Several approaches are proposed to enhance model merging, including EMR-merging~\cite{huang2024emr}, LoraHub~\cite{huang2023lorahub}, mixture of LORA~\cite{wu2024mixture}, 
AdaMerging~\cite{yang2023adamerging}, AdaMergX~\cite{zhao2024adamergex}, and
LoraSoups~\cite{prabhakar2024lorasoups}. 
These works merge models or PEFT with or without training, each trying to address some limitations of the existing works.

Recent work has demonstrated that merging Llama2-7b-chat, a general-purpose chat model, with Meditron-7b, specialized for the medical domain, resulted in a merged model that outperformed its constituent models across both general and medical benchmarks \cite{mergekit}. 
The increasing number of merged models on the Open LLM leaderboard \cite{leaderboard} further proves the success of applying this approach to various benchmarks. 
Several studies have proposed different merging techniques, such as weight-space averaging \cite{modelsoup}, Fisher-Weighted averaging \cite{fisher}, Git Re-Basin \cite{rebasin}, TIES-Merging \cite{ties}, and DARE \cite{dare}, to efficiently merge models trained on distinct domains. 
Considering the trend of scaling models to billions of parameters and the cost of fully fine-tuning these models, there is an increasing need to explore the feasibility of merging multiple adapters within a single language model when adapters are used to fine-tune the models for the SE domain. 
Works such as Adaptersoup \cite{adaptersoup} have investigated this approach for both in-domain and out-of-domain evaluation of mixed adapters. 
Another study has conducted a comprehensive analysis of the effectiveness of different adapter architectures for the in-domain generalizability of merged adapters \cite{generaliz}.

In this work, we employ three merging methods, i.e., weight-space merging \cite{modelsoup}, TIES-Merging \cite{ties}, and DARE \cite{dare}, to merge multiple task-specific adapters. 
Specifically, we train multiple LoRA instances on various SE tasks and assess the performance of the merged LoRA in  equal-weight merging and continual merging scenarios for all three merging methods. 
Though there are several model merging studies that merge models or PEFT approaches with or without training, there is no work that investigates the applicability of this technique in software engineering and specifically for automatic program repair. 

\section{Study Design}
\label{sec:design}

\subsection{Research Questions}
\label{sec:rqs}

\begin{figure}[t]
\centering
\includegraphics[width=\textwidth]{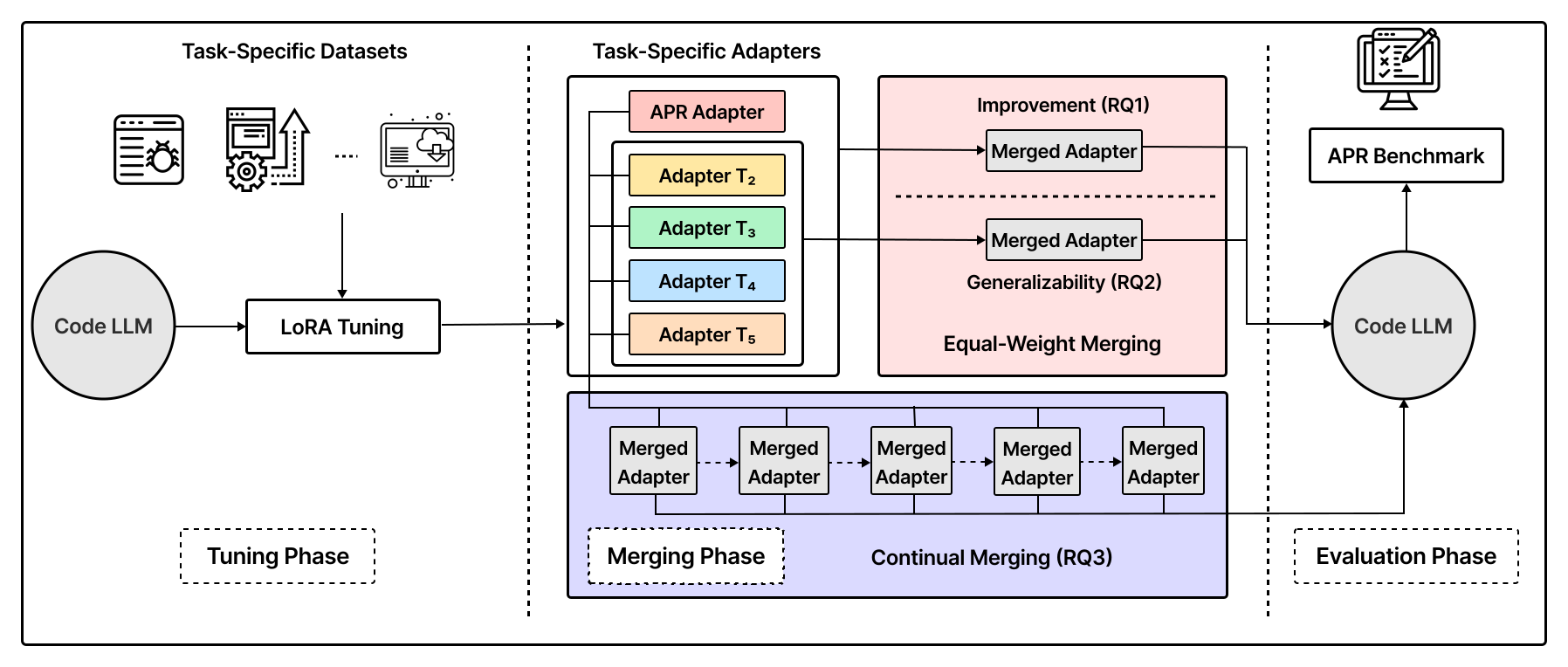}
\caption{An overview of MergeRepair.}
\label{fig:board}
\end{figure}

The main goal of this study is to investigate the performance of merged task-specific adapters on the APR task. 
Specifically, we use two main scenarios for merging adapters 1) equal-weight merging (RQ1 and RQ2) and 2) continual merging (RQ3), as shown in Figure~\ref{fig:board}. The explanation of each scenario is provided in Section~\ref{sec:exec}.
We design our experiments to answer the following research questions:

\textbf{RQ1: Does merging APR adapter with other task-specific adapters improve the performance of APR?} 
We are interested to investigate whether the performance of the APR task will be degraded or improved if we merge APR adapter with non-APR adapters. 

Prior studies have shown that merging adapters can improve the performance of single tasks in other domains \cite{mergekit, yang2024model}. In this RQ, we intend to investigate the potential improvement of merged adapters on the APR task, when APR is merged with four tasks of Development, Misc, Test \& QA, and Improvement. 
The alternative way to achieve better performance having an already fine-tuned model, is to continue the fine-tuning process with additional data and computation. The potential improvement using merging adapters is essentially interesting because the merging process of various adapters is computationally cheaper than further fine-tuning of models.
To study this RQ, we start by merging APR with only one task and evaluate it on the APR benchmark. Thereafter, we continue merging APR adapters with two, three, and all four tasks and assess the performance of the merged adapters on APR. In such a manner, we consider all subsets of four adapters merged with APR adapter.

\textbf{RQ2: Can merged adapters of other tasks generalize to APR?} 
Our main interest for this RQ is to understand if the performance of other task-specific adapters could be maintained for the APR task when merged together.

Training a Code LLM on a new task is costly, and one of the goals of merging adapters is to expand the model to new tasks (i.e., out of domain data) without additional training~\cite{mergekit}. 
Hence, we aim to analyze the generalizability of merged adapters to out-of-domain data. Particularly, we merge non-APR adapters together and evaluate them on the APR benchmark. For merging task-specific adapters, similar to RQ1, we start with adapters trained on a single task and {experiment with merging up to} four adapters. Each time, we test the merged model on the APR dataset. 

\textbf{RQ3: How does continual merging of other task-specific adapters with the APR adapter influence the performance of APR?} 
In real-world software projects, accessing several task-adapters at a time is less applicable; instead, new datasets and tasks will be available as the project evolves. In such situations, the ability to adapt the merged adapter to new data and tasks becomes crucial. 

To this end, we investigate the continual learning capacity of merged adapters in this RQ, using our proposed paradigm, \textit{continual merging}. Furthermore, the order in which different task-specific adapters will be merged is important as it determines their influence in the merged adapter
(Section \ref{sec:method} shows the weight contribution or influence of different task-adapters for this scenario).
We consider all permutations of the task-adapters to evaluate their order on APR performance. 

\subsection{Datasets}
\label{sec:data}

We utilize the CommitPackFT dataset released by the OctoPack study \cite{octopack}. 
This dataset is a refined version of the full CommitPack dataset, filtered by its publishers to make it suitable for instruction-tuning Code LLMs \cite{octopack}. CommitPackFT contains 277 programming languages and is 2GB of memory \cite{octopack}. These two datasets contain code samples from GitHub repositories, including the code before and after the commit change. In particular, the dataset includes three main fields: 1) commit messages, 2) old contents (the file content before the commit), and 3) new contents (the file content after the commit). The commit messages are the instructions used when instruction-tuning the models. Generally, these commits enhance the previous version of the code. A portion of the commits are related to bug-fixing commits tailored for the program repair task.

The Python split of the CommitPackFT dataset has been classified into five tasks by the publishers of the dataset. This classification shows the task information for each commit, which is done by using 1-shot prompting with the GPT-4 model.
The task classification for other programming languages is not available. 
Therefore, our experiments are limited to the Python split of the dataset.
We will conduct all the experiments on the Python split of the dataset, which contains $59,113$ samples. 
Table \ref{tab:dataset-stats} shows the available tasks in the Python dataset. This dataset is used to train the task-specific adapters. 
The tasks include Development, Bug Fixes, Misc, Test \& QA, and Improvement, where each one except Misc contains a main task and a small portion of the dataset as a closely related task. 
Bug Fixes is used for APR and includes bug fixes. Development mainly contains new features and 0.88\% of the whole data on user interfaces. Test \& QA includes mainly testing and a small subset on logging/instrumentation (0.62\% of the whole data). 
Improvement contains data mainly tagged as refactoring/code clean up, with small percentages labeled as formatting/linting (0.40\% of the whole data) and performance improvement (0.64\% of the whole data). 
The Misc subset contains a more variety of the tasks, including deprecation, build system/tooling, documentation, dependencies, configuration, and release management.

\begin{table}
    \centering
    \begin{tabular}{|l|c|}
    \hline
        \textbf{Task} & \# Records\\
        \hline
         Development & 15,656 (26.49\%)\\
         \hline
         Bug Fixes (APR)& 11,260 (19.02\%)\\
         \hline
         Misc & 11,617 (19.65\%)\\
         \hline
         Test \& QA & 8,255 (13.96\%)\\
         \hline
         Improvement & 12,325 (20.85\%)\\
         \hline
    \end{tabular}
    \caption{Task distribution of the CommitPackFT dataset.}
    \label{tab:dataset-stats}
\end{table}


\subsection{Experiment Design}
\label{sec:exec}

An overview of our approach, MergeRepair, is shown in Figure~\ref{fig:board}. MergeRepair consists of three main phases: tuning phase, merging phase and the evaluation phase. 

\begin{itemize}
\item \textbf{Tuning phase:} First, we train one instance of LoRA adapter for each task in the tuning phase. 
\item \textbf{Merging phase:} Then, we merge these task-specific adapters using two merging scenarios/paradigms: weight-space merging and continual merging. For the weight-space {merging}, we merge the parameters of all task-specific adapters with equivalent weights (i.e., influence) to investigate the potential improvement on APR (\textbf{RQ1}), generalizability to APR (\textbf{RQ2}), and the effect of continual merging (\textbf{RQ3}). This is done using three merging techniques. 
Subsequently, we merge non-APR adapters, again with equal weights.
\item \textbf{Evaluation phase:} we evaluate the merged adapter on the APR benchmark to study the above-mentioned RQs. 
\end{itemize}

For the continual merging scenario, we aim to simulate the continual learning capability of merged adapters in which we sequentially add new task-specific adapters to the previous merged adapter. In the following sections, we will explain the continual merging, merging methods, tasks, models, steps to answer each RQ, and evaluation metrics that we employ to study the above-mentioned RQs.


\subsubsection{Continual Merging}
\label{sec:method}

Let the parameters of the Code LLM be denoted by \(\theta_{M}\). We keep these parameters unchanged during instruction-tuning. For each target task, we have a separate LoRA adapter on top of the frozen model trained on that task. We refer to these trained adapters as task-specific adapters. Formally, having \(n\) available tasks denoted by \(T_{i}=\{T_{1}, T_{2}, ..., T_{n}\}\), we will also have \(n\) adapters, each with distinct weight parameter values. We define \(I_{T_{i}}\) as the influence of adapter \(T_{i}\) in the merged adapter.  

For continual merging, we maintain a single adapter as the merged adapter. 
Subsequently, we merge the parameters of the current merged adapter with one new task-specific adapter at a time. Therefore, starting from the first task, i.e., \(T_{1}\) we will have \(I_{T_{1}}=1\) in the first step. In the second step, we add the second task-specific adapter, i.e., \(T_{2}\), resulting in \(I_{T_{1}}=\frac{1}{2}\) and \(I_{T_{2}}=\frac{1}{2}\) in the merged adapter. 
In the final step, $n$, the influence of all previously added adapters to the merged adapter will be similar to a geometric progression with \(I_{T_{n}}=\frac{1}{2}\) for the last adapter and ratio of \(\frac{1}{2}\). More formally, the final influence of task-specific adapters in the merged adapter, starting from \(T_{1}\) and ending with \(T_{n}\), would be acquired by 

\begin{equation}
    I_{T_{i}}=\frac{1}{2^{n+1-i}}
\end{equation}

for \(i=\{2, 3, ..., n\}\) and \(I_{T_{i}}=\frac{1}{2^{n-i}}\) for \(i=1\), since we consider the first adapter as the merged adapter in the first step.

Consequently, in {this scenario}, the influence of the parameters of the previous task-adapters in the merged adapter will decrease, as we reach the final (i.e., $n$-th) adapter. Intuitively, the merged adapter attends to the new {task} while retaining less knowledge about the previous tasks.

For the continual merging scenario, we consider different orders of task-specific adapters in this study. The order of adding task-specific adapters is important as it will lead to different weight values for the parameters of the merged adapter. Having \(n\) different adapters, the total number of their permutations would be equal to \(n!\). 
Hence, to adhere to the computational resources available to us, for the continual merging scenario, we will merge APR adapter with up to three other task-specific adapters, totaling \(4!=24\) experiments required for each Code LLM.


\subsubsection{Merging Methods} We experiment with three merging methods, all compatible with LoRA adapter that we aim to employ. Weight-space averaging is selected as the baseline merging method. The other two methods Ties-merging \cite{ties} and DARE \cite{dare} propose more efficient merging techniques, which have been shown to perform better than weight-space averaging in prior research~\cite{mergekit}. 

\textbf{Weight-space averaging.} This method is the most straightforward approach to merge multiple models as it averages the weight parameters of the specialized models in order to obtain the parameters of the merged model \cite{modelsoup}. For weight-space averaging, the formal influence of task-specific adapter $T_i$, is defined as 
\begin{equation}
    I_{T_{i}}=\frac{1}{n}
\end{equation}
where $i=\{1,2,3,...,n\}$. This approach assigns equal importance to all adapters in the parameter space, creating a simple average of their weights.

\textbf{TIES-Merging.} This method uses a three-step approach to merge two or more models \cite{ties}. First, it trims redundant parameters that changed slightly during fine-tuning process. Next, it assigns positive or negative signs to each parameter by resolving parameter sign conflicts across all participating models. Finally, it averages the remained parameters that have the same sign values. The complete definition of TIES-Merging can be found in \cite{ties}.

\textbf{DARE.} DARE \cite{dare} uses a two-step method. First, a certain ratio of the parameters is dropped, and the remaining ones are rescaled to reconstruct an embedding vector similar to the initial parameter embeddings of the model. This process is done separately for each of the models. Ultimately, the parameters of all participating models are fused to construct the final merged model. The complete definition of DARE can be found in \cite{dare}.

\subsection{Tasks}

Table \ref{tab:dataset-stats} shows all the tasks, including Development, Misc, Test \& QA, and Improvement, along with the number of their samples. The ``Bug Fixes'' portion of this data is associated with the APR task and forms 19.02\% of the dataset. The Misc category includes code samples from different domains, including configuration, dependencies, and documentation~\cite{octopack}. 
Development includes records related to new features added to the programs along with a small portion of the user interface. 
Improvement contains samples from refactoring/code cleanup and performance improvement. The Test \& QA includes samples from testing and logging/instrumentation. For more details on the tasks refer to the OctoPack~\cite{octopack} study.    
For all these tasks, we used instruction-tuning data to fine-tune the Code LLMs using LoRA. This training provides the task-specific adapters for our merging experiments. 
The pre-trained models, without instruction-tuning, will generate open-ended outputs that are not suitable for production use. This process aims to adapt the pre-trained models to generate human-aligned outputs given an instruction and an input/output pair of code samples \cite{align}.

\subsection{Models}

We will use two state-of-the-art Code LLMs, StarCoder2 \cite{starcoder2} and Granite \cite{granite}, for our experiments. These models are chosen as they are more recent than popular Code LLMs like CodeLlama and Mistral. More importantly, StarCoder2 and Granite have the best performance for APR task at the time of this study~\cite{starcoder2, granite}. Both of these models have the same size, with three billion parameters each. Their checkpoints and source code are publicly available, allowing us to inject low-rank matrices of LoRA layers into the decoder blocks of these models and train only the added parameters. 
Another reason to choose StarCoder2 and Granite over other CodeLLMs is our restricted computational capacity. We ran experiments to fine-tune the two models using LoRA with a configuration that is reported by the models' authors to ensure the feasibility of our approach and its performance. This choice is not feasible for us with CodeLlama and Mistral.  

\textbf{starcoder2-3b.}
\starcoder \cite{starcoder2} is a category of models developed and released by the open-source community as part of the BigCode project. These models are published in three sizes: 3b, 7b, and 15b. They are trained on the Stack v2 dataset, which is built on top of the Software Heritage (SWH) repositories. They also utilize a comprehensive set of publicly available data sources such as GitHub issues, StackOverflow, Jupyter and Kaggle notebooks. The architecture of these models is based on the StarCodeBase \cite{starcoder} model, which uses a decoder-only architecture with further modifications. The 3b version of these models is trained on more than 3T tokens.
We chose this model as it has been developed by open source community and performs well on APR task.

\textbf{granite-3b-code-base.}
\granite models \cite{granite} are a group of four Code LLMs released by IBM Research, ranging from 3b to 34b models. These models use the decoder blocks of the Transformer model \cite{attention}. The pre-training process of these models consists of two phases. In the first phase, the models are trained on code data only, with the 3b model being trained on 4T of code tokens. In the second phase, the models are trained on 500B of tokens including both code and natural language tokens. In addition to casual language modeling objective, they use the Fill-In-the-Middle (FIM) \cite{fim} objective function during pre-training. To further align the model, they also train the models on various instruction-tuning datasets. 
We chose this model because it uses FIM objective during training phase it is the most competitive on program repair benchmarka compared to other models having the same of similar size.

The instruction-tuned version of these models are publicly available. However, these versions are trained on the entire dataset that we aim to use. Hence, we will select the base checkpoints (pre-trained only) of both models and train different task-specific adapters by further instruction-tuning the base models on each task-specific dataset. 

\subsection{Research method for each RQ}


For RQ1, we use merging with equal-weight merging (Figure~\ref{fig:board}) and add the other four tasks one by one to be merged with APR. As the order of tasks is not important in this part, we add tasks one by one to the APR adapter and compare the results with APR adapter on the benchmark dataset (see Section~\ref{sec:eval-metrics}). We apply this process for each of the three merging methods, weight-space averaging, TIES-Merging, and DARE, separately and report the results.

For RQ2, we use the same process as in RQ1. However, we do not add the APR adapter in the merged adapter. 
Instead, we test the merged adapter on APR to evaluate the generalizability of the merged adapter. We then report the results for each of the three merging methods and the combinations of the task-specific adapters.

For RQ3, we use the continual merging (Figure~\ref{fig:board} bottom of merging phase). For this purpose, we add the task-specific adapters one by one to build the merged adapter. In this case, the order of adding tasks is important and affects the results. 
We use four tasks, including APR, code development, code improvement, and test \& QA as identified in Table~\ref{tab:dataset-stats}. These tasks are all similar to each other in terms of being code-to-code tasks.
The experiments include models when two, three, or four task-specific adapters are used in the merged adapter, and all the permutations of the tasks. 
We opt to remove one of the tasks to keep the number of experiments manageable according to computational resources available to us.
Similar to previous RQs, we apply the same process for each of the merging techniques separately. This will show if continual merging helps improve the results compared to equal-weight averaging, and if so, it will be more effective on which merging method.

\subsection{Evaluation Metrics}
\label{sec:eval-metrics}

For all experiments we report the \passk scores of the models evaluated on the HumanEvalFix benchmark \cite{octopack}. This benchmark is created as part of the HumanEvalPack benchmarks released by the same research. They manually produced bugs in the extended version of the original HumanEval 
benchmark, which supports six programming languages, including Python. 

Compared to previous similarity-based metrics such as BLEU~\cite{bleu} and CodeBLEU~\cite{codebleu}, execution-based metrics, i.e., \passk are more reliable to capture the functional correctness of generated codes~\cite{codegen}. To calculate these scores, accessing to a dataset containing unit tests for all samples is required. In the \passk metric, to check the correctness of the generated code samples, one of the generated candidates should pass all the test cases of the corresponding sample in the dataset. More formally, by sampling \(k\) candidates for each record or problem of the dataset among all \(n\) generated candidates, the \passk metric could be calculated as follows: 

\begin{equation}
    pass@k:=\mathop{\mathbb{E}}_{Problems}\left[1-\frac{\binom{n-c}{k}}{\binom{n}{k}}\right]
\end{equation}

\noindent where \(c\) is the number of correct samples that pass all of the unit tests of each specific problem.
We report pass@1 and pass@10 similar to previous works~\cite{octopack}. 

Code LLMs have been shown to be sensitive to perturbations introduced in input prompts, which affects the reliability of the \passk metric~\cite{recode}. Therefore, \robustpass (RP\textsubscript{s}@k)~\cite{recode} was proposed to evaluate the code generation robustness of Code LLMs. Following \cite{recode}, we introduce random perturbations in the HumanEvalFix benchmark to evaluate the models in a more robust manner that reflects practical applications.
To calculate the \robustpass metric, $s$ random perturbations are introduced in the input prompt. $n$ candidates are generated for each input prompt, resulting in $n \times s$ generated candidates named \(f_i(x_j)\), where \(1\leq i\leq n\) and \(1\leq j\leq s\). Finally, the \robustpass metric is defined as below.
Here, \(rc_s(x) = \sum_{i=1}^{n}c_{i,s}(x)\) represents the worst-case correction for problem \(x\). For each \(i\), \(c_{i,s}(x)=1\) if and only if all \(f_i(x)\) generations are correct. Following the initial study~\cite{recode}, we report RP\textsubscript{5}@1 for our experiments. However, as creating perturbations is computationally expensive and beyond our computations capacity, we only apply \robustpass for a few cases, and we discuss them in the Discussions in Section~\ref{sec:disc}. 

\begin{equation}
    RobustPass_s@k:=\mathop{\mathbb{E}}_{x}\left[1-\frac{\binom{n-rc_s(x)}{k}}{\binom{n}{k}}\right]
\end{equation}



We apply pairwise t-test \cite{kim2015t} and report the cliff's delta effect size \cite{cliff1993dominance} to examine whether the results and their effects are statistically significant. Since the scores of the generated outputs are not normally distributed, with consultation with a Statistician professor with expertise in data science and machine learning, we randomly sampled the data with a large number, i.e., 400 times, to obtain approximately normal subsets suitable for applying the t-test.

\subsection{Configurations}
We conduct all experiments on one NVIDIA Tesla V100 32GB GPU. 
The models are trained on the CommitPackFT dataset for each task and then evaluated on HumanEvalFix benchmark. We set the temperature to 0.2 when evaluating the models and generate 20 samples to calculate pass@1 and pass@10 scores. The hyperparameters are set following the previous works~\cite{octopack, granite} for training each model. Table \ref{tab:hp} reports the values of these hyperparameters. For training the models using LoRA adapters, we insert LoRA layers into the decoders of the models and we apply them to q\_proj, o\_proj, k\_proj, v\_proj, c\_fc, c\_proj layers of the \starcoder and q\_proj, o\_proj, k\_proj, v\_proj, gate\_proj, up\_proj, down\_proj layers of \granite model.

\begin{table}[]
\centering
\caption{Values of different hyperparameters used for training the models. }
\label{tab:hp}
\begin{tabular}{@{}lcc@{}}
\toprule
\textbf{Hyperparameter}       & \textbf{\starcoder} & \textbf{\granite} \\ \midrule
learning\_rate                & 2e-5                             & 1e-5                                    \\
lr\_scheduler\_type           & cosine                           & cosine                                  \\
weight\_decay                 & 0.05                             & 0.05                                    \\
max\_steps                    & 10000                            & 10000                                   \\
warmup\_steps                 & 20                               & 250                                     \\
train\_micro\_batch\_size     & 4                                & 2                                       \\
gradient\_accumulation\_steps & 1                                & 8                                       \\
lora\_rank                    & 16                               & 16                                      \\ \bottomrule
\end{tabular}
\end{table}

To generate the HumanEvalFixPerturbed dataset used in the discussions, we applied the perturbations introduced by the ReCode~\cite{recode} study for the categories of \texttt{format}, \texttt{function}, and \texttt{syntax} and used five different seeds for generating each of the datasets following their approach. We did not use the \texttt{docstring} perturbation~\cite{recode}, as the bigcode-evaluation-harness library\footnote{https://github.com/bigcode-project/bigcode-evaluation-harness} we employed to evaluate the results -- used in the previous studies \cite{octopack} -- is not compatible with the prompt for the APR task.
Therefore, applying perturbations on this category is not providing a ground comparison among the two metrics, \passk and RobustPass@k.  


\section{Results}
\label{sec:res}

In this section, we report and discuss the results obtained for each RQ. Particularly, for each RQ, we report both \passone and \passten scores in a separate table. The tables with a light yellow background display \passone scores, while those with a light blue background present the \passten scores. Due to the large number of experiments for RQ3, we divide the results of the merged adapters with three and four base tasks in two categories. 
Additionally, for adapters having four tasks, we divide them into two tables for readability. 
In addition to \passone and \passten scores, we report the results of the statistical tests and effect size experiments conducted between each merged adapter and the APR adapter in order to investigate the importance of the results. For all experiments, we report the p-value of statistical tests with \texttt{*} for \texttt{p-value < 0.1}, \texttt{**} for \texttt{p-value < 0.05}, and \texttt{***} for \texttt{p-value < 0.01}. In the same vein, we report the \texttt{cliff\_delta} value of the effect size experiments using S for \texttt{|cliff\_delta| >= 0.11}, M for \texttt{|cliff\_delta| >= 0.28}, and L for \texttt{|cliff\_delta| >= 0.43}. 
For each model and merged adapter, the score of the best-performing merging method (row-wise) is shown in \textbf{bold}. Furthermore, the best-performing adapter for each merging method (column-wise) is \underline{underlined} for both models. Note that there might be multiple adapters with the exact same scores as the highest for each merging method or model. In such cases, all are shown in \textbf{bold} or are \underline{underlined}.

In all cases, we add the performance difference compared to the APR adapter in parenthesis. 
First, we provide the \passone and \passten scores of individual task-adapters on the APR benchmark as shown in Table \ref{tab:ind-passes}. Interestingly, T3, which is the Misc task has highest performance on APR, even surpassing the result of APR. Test \& QA on the other hand, have the lowest score among all tasks on the APR benchmark.

\begin{table}[]
\centering
\caption{Pass@1 and pass@10 scores of the individual task-specific adapters on the APR benchmark with \starcoder and \granite models. The highest scores are shown in \textbf{bold}. }
\label{tab:ind-passes}
\begin{tabular}{llccccc}
\hline
\multicolumn{2}{l}{\multirow{2}{*}{\textbf{Model/Task}}}          & \multicolumn{5}{c}{\textbf{Tasks}}                                                                                                   \\ \cline{3-7} 
\multicolumn{2}{l}{}                                              & \textbf{Program Repair (T1)} & \textbf{Improvement (T2)} & \textbf{Misc (T3)} & \textbf{Development (T4)} & \textbf{Test \& QA (T5)} \\ \hline
\multirow{2}{*}{\textbf{\starcoder}}        & \textbf{pass@1}  & 28.66\%                      & 27.90\%                   & \textbf{31.40\%}   & 24.66\%                   & 22.10\%                  \\
                                               & \textbf{pass@10} & 39.11\%                      & 37.99\%                   & \textbf{41.88\%}   & 32.66\%                   & 28.90\%                  \\ \hline
\multirow{2}{*}{\textbf{\granite}} & \textbf{pass@1}  & 16.16\%                      & 16.95\%                   & \textbf{18.63\%}   & 18.32\%                   & 13.20\%                  \\
                                               & \textbf{pass@10} & 19.84\%                      & 21.05\%                   & \textbf{22.35\%}   & 22.79\%                   & 17.68\%                  \\ \hline
\end{tabular}
\end{table}


\subsection{RQ1: Performance of Merged Adapters on APR}

\begin{table}[!htp]\centering
\caption{\passone scores of the merged adapters including T1 (APR) as one of their constituent tasks. The results are reported across three merging methods for \starcoder and \granite models. Note that the order of the tasks being merged does not affect the performance as they are all merged with equal weights. The tasks are listed as T1 (APR), T2 (Improvement), T3 (Misc), T4 (Development), and T5 (Test \& QA) }
\label{tab:rq1-pass1}
\scriptsize
\begin{tabular}{l|ccc|cccc}\toprule
\multirow{2}{*}{Merged Tasks} &\multicolumn{3}{c}{\starcoder} &\multicolumn{3}{c}{\granite} \\\cmidrule{2-7}
&dare-ties &ties &weight-averaging &dare-ties &ties &weight-averaging \\\midrule
\multirow{2}{*}{T1-T2} &28.41\% (-0.25) &\textbf{29.63\% (+0.97)} &28.93\% (+0.27) &\textbf{16.43\% (+0.27)} &16.28\% (+0.12) &16.25\% (+0.09) \\
&S** &S** &S* &S* &S* &S* \\
\multirow{2}{*}{T1-T3} &\cellcolor[HTML]{FFF2D7}30.18\% (+1.52) &\cellcolor[HTML]{FFF2D7}\textbf{31.01\% (+2.35)} &\cellcolor[HTML]{FFF2D7}30.15\% (+1.49) &\cellcolor[HTML]{FFF2D7}\textbf{17.41\% (+1.25)} &\cellcolor[HTML]{FFF2D7}17.07\% (+0.91) &\cellcolor[HTML]{FFF2D7}17.2\% (+1.04) \\
&\cellcolor[HTML]{FFF2D7}S** &\cellcolor[HTML]{FFF2D7}M*** &\cellcolor[HTML]{FFF2D7}M*** &\cellcolor[HTML]{FFF2D7}M*** &\cellcolor[HTML]{FFF2D7}M*** &\cellcolor[HTML]{FFF2D7}S* \\
\multirow{2}{*}{T1-T4} &\textbf{27.35\% (-1.31)} &26.89\% (-1.77) &27.13\% (-1.53) &\textbf{17.35\% (+1.19)} &16.86\% (+0.7) &\textbf{17.35\% (+1.19)} \\
&M*** &S*** &M*** &S** &S*** &S*** \\
\multirow{2}{*}{T1-T5} &\cellcolor[HTML]{FFF2D7}25.24\% (-3.42) &\cellcolor[HTML]{FFF2D7}\textbf{25.88\% (-2.78)} &\cellcolor[HTML]{FFF2D7}25.82\% (-2.84) &\cellcolor[HTML]{FFF2D7}\textbf{15.0\% (-1.16)} &\cellcolor[HTML]{FFF2D7}14.73\% (-1.43) &\cellcolor[HTML]{FFF2D7}14.73\% (-1.43) \\
&\cellcolor[HTML]{FFF2D7}L*** &\cellcolor[HTML]{FFF2D7}L*** &\cellcolor[HTML]{FFF2D7}L*** &\cellcolor[HTML]{FFF2D7}S*** &\cellcolor[HTML]{FFF2D7}S* &\cellcolor[HTML]{FFF2D7}L*** \\\midrule
\multirow{2}{*}{T1-T2-T3} &\ul{30.82\% (+2.16)} &\ul{\textbf{31.04\% (+2.38)}} &\ul{30.95\% (+2.29)} &17.29\% (+1.13) &17.29\% (+1.13) &\textbf{17.96\% (+1.8)} \\
&M*** &M*** &L*** &S*** &M*** &S*** \\
\multirow{2}{*}{T1-T2-T4} &\cellcolor[HTML]{FFF2D7}28.2\% (-0.46) &\cellcolor[HTML]{FFF2D7}28.26\% (-0.4) &\cellcolor[HTML]{FFF2D7}\textbf{28.99\% (+0.33)} &\cellcolor[HTML]{FFF2D7}\textbf{17.32\% (+1.16)} &\cellcolor[HTML]{FFF2D7}16.98\% (+0.82) &\cellcolor[HTML]{FFF2D7}\textbf{17.32\% (+1.16)} \\
&\cellcolor[HTML]{FFF2D7}S* &\cellcolor[HTML]{FFF2D7}S* &\cellcolor[HTML]{FFF2D7}S** &\cellcolor[HTML]{FFF2D7}S* &\cellcolor[HTML]{FFF2D7}S*** &\cellcolor[HTML]{FFF2D7}S*** \\
\multirow{2}{*}{T1-T2-T5} &\textbf{27.38\% (-1.28)} &27.26\% (-1.4) &26.49\% (-2.17) &\textbf{15.37\% (-0.79)} &15.15\% (-1.01) &15.34\% (-0.82) \\
&S*** &S** &M*** &S* &S* &S*** \\
\multirow{2}{*}{T1-T3-T4} &\cellcolor[HTML]{FFF2D7}\textbf{29.27\% (+0.61)} &\cellcolor[HTML]{FFF2D7}29.02\% (+0.36) &\cellcolor[HTML]{FFF2D7}28.51\% (-0.15) &\cellcolor[HTML]{FFF2D7}\ul{\textbf{17.93\% (+1.77)}} &\cellcolor[HTML]{FFF2D7}\ul{17.8\% (+1.64)} &\cellcolor[HTML]{FFF2D7}17.68\% (+1.52) \\
&\cellcolor[HTML]{FFF2D7}S* &\cellcolor[HTML]{FFF2D7}S* &\cellcolor[HTML]{FFF2D7}S* &\cellcolor[HTML]{FFF2D7}S*** &\cellcolor[HTML]{FFF2D7}M*** &\cellcolor[HTML]{FFF2D7}M*** \\
\multirow{2}{*}{T1-T3-T5} &28.05\% (-0.61) &\textbf{28.93\% (+0.27)} &27.99\% (-0.67) &16.65\% (+0.49) &16.22\% (+0.06) &\textbf{16.89\% (+0.73)} \\
&S*** &S* &S* &S* &S* &S*** \\
\multirow{2}{*}{T1-T4-T5} &\cellcolor[HTML]{FFF2D7}26.16\% (-2.5) &\cellcolor[HTML]{FFF2D7}\textbf{26.28\% (-2.38)} &\cellcolor[HTML]{FFF2D7}25.82\% (-2.84) &\cellcolor[HTML]{FFF2D7}15.88\% (-0.28) &\cellcolor[HTML]{FFF2D7}16.01\% (-0.15) &\cellcolor[HTML]{FFF2D7}\textbf{16.22\% (+0.06)} \\
&\cellcolor[HTML]{FFF2D7}L*** &\cellcolor[HTML]{FFF2D7}M*** &\cellcolor[HTML]{FFF2D7}M*** &\cellcolor[HTML]{FFF2D7}S* &\cellcolor[HTML]{FFF2D7}S* &\cellcolor[HTML]{FFF2D7}S* \\\midrule
\multirow{2}{*}{T1-T2-T3-T4} &30.21\% (+1.55) &30.06\% (+1.4) &\textbf{30.3\% (+1.64)} &17.47\% (+1.31) &17.71\% (+1.55) &\ul{\textbf{18.14\% (+1.98)}} \\
&M*** &S* &M*** &S** &M*** &M*** \\
\multirow{2}{*}{T1-T2-T3-T5} &\cellcolor[HTML]{FFF2D7}29.15\% (+0.49) &\cellcolor[HTML]{FFF2D7}\textbf{29.39\% (+0.73)} &\cellcolor[HTML]{FFF2D7}28.05\% (-0.61) &\cellcolor[HTML]{FFF2D7}15.91\% (-0.25) &\cellcolor[HTML]{FFF2D7}16.28\% (+0.12) &\cellcolor[HTML]{FFF2D7}\textbf{17.07\% (+0.91)} \\
&\cellcolor[HTML]{FFF2D7}S* &\cellcolor[HTML]{FFF2D7}S* &\cellcolor[HTML]{FFF2D7}S* &\cellcolor[HTML]{FFF2D7}S* &\cellcolor[HTML]{FFF2D7}S* &\cellcolor[HTML]{FFF2D7}S** \\
\multirow{2}{*}{T1-T2-T4-T5} &27.29\% (-1.37) &\textbf{27.53\% (-1.13)} &26.68\% (-1.98) &16.49\% (+0.33) &16.46\% (+0.3) &\textbf{17.1\% (+0.94)} \\
&S*** &S* &M*** &S* &S** &S* \\
\multirow{2}{*}{T1-T3-T4-T5} &\cellcolor[HTML]{FFF2D7}\textbf{28.69\% (+0.03)} &\cellcolor[HTML]{FFF2D7}27.56\% (-1.1) &\cellcolor[HTML]{FFF2D7}27.9\% (-0.76) &\cellcolor[HTML]{FFF2D7}16.65\% (+0.49) &\cellcolor[HTML]{FFF2D7}16.74\% (+0.58) &\cellcolor[HTML]{FFF2D7}\textbf{17.47\% (+1.31)} \\
&\cellcolor[HTML]{FFF2D7}S* &\cellcolor[HTML]{FFF2D7}S*** &\cellcolor[HTML]{FFF2D7}S* &\cellcolor[HTML]{FFF2D7}S* &\cellcolor[HTML]{FFF2D7}S* &\cellcolor[HTML]{FFF2D7}S*** \\\midrule
\multirow{2}{*}{T1-T2-T3-T4-T5} &27.8\% (-0.86) &\textbf{28.38\% (-0.28)} &27.32\% (-1.34) &\textbf{17.29\% (+1.13)} &16.65\% (+0.49) &17.13\% (+0.97) \\
&M*** &S* &S* &S** &S** &S*** \\
\bottomrule
\end{tabular}
\end{table}

\begin{table}[!htp]\centering
\caption{\passten scores of the merged adapters including T1 (APR) as one of their constituent tasks. The results are reported across three merging methods for \starcoder and \granite models. Note that the order of the tasks being merged does not affect the performance as they are all merged with equal weights. The tasks are listed as T1 (APR), T2 (Improvement), T3 (Misc), T4 (Development), and T5 (Test \& QA). }
\label{tab:rq1-pass10}
\scriptsize
\begin{tabular}{l|ccc|cccc}\toprule
\multirow{2}{*}{Merged Tasks} &\multicolumn{3}{c}{\starcoder} &\multicolumn{3}{c}{\granite} \\\cmidrule{2-7}
&dare-ties &ties &weight-averaging &dare-ties &ties &weight-averaging \\\midrule
\multirow{2}{*}{T1-T2} &40.54\% (+1.43) &39.88\% (+0.77) &41.46\% (+2.35) &19.37\% (-0.47) &\textbf{19.97\% (+0.13)} &19.51\% (-0.33) \\
&S*** &S* &S* &S* &S* &S* \\
\multirow{2}{*}{T1-T3} &\cellcolor[HTML]{D9EAFD}\ul{41.72\% (+2.61)} &\cellcolor[HTML]{D9EAFD}42.24\% (+3.13) &\cellcolor[HTML]{D9EAFD}\textbf{42.37\% (+3.26)} &\cellcolor[HTML]{D9EAFD}\textbf{21.32\% (+1.48)} &\cellcolor[HTML]{D9EAFD}20.52\% (+0.68) &\cellcolor[HTML]{D9EAFD}21.29\% (+1.45) \\
&\cellcolor[HTML]{D9EAFD}L*** &\cellcolor[HTML]{D9EAFD}L*** &\cellcolor[HTML]{D9EAFD}M*** &\cellcolor[HTML]{D9EAFD}M*** &\cellcolor[HTML]{D9EAFD}S*** &\cellcolor[HTML]{D9EAFD}M*** \\
\multirow{2}{*}{T1-T4} &\textbf{38.94\% (-0.17)} &37.78\% (-1.33) &38.61\% (-0.5) &21.57\% (+1.73) &21.15\% (+1.31) &\textbf{21.65\% (+1.81)} \\
&S* &M*** &S*** &M*** &S*** &M*** \\
\multirow{2}{*}{T1-T5} &\cellcolor[HTML]{D9EAFD}\textbf{35.18\% (-3.93)} &\cellcolor[HTML]{D9EAFD}34.43\% (-4.68) &\cellcolor[HTML]{D9EAFD}34.76\% (-4.35) &\cellcolor[HTML]{D9EAFD}19.29\% (-0.55) &\cellcolor[HTML]{D9EAFD}19.26\% (-0.58) &\cellcolor[HTML]{D9EAFD}\textbf{19.49\% (-0.35)} \\
&\cellcolor[HTML]{D9EAFD}L*** &\cellcolor[HTML]{D9EAFD}L*** &\cellcolor[HTML]{D9EAFD}L*** &\cellcolor[HTML]{D9EAFD}S* &\cellcolor[HTML]{D9EAFD}S* &\cellcolor[HTML]{D9EAFD}S* \\\midrule
\multirow{2}{*}{T1-T2-T3} &41.53\% (+2.42) &\ul{42.94\% (+3.83)} &\ul{\textbf{43.12\% (+4.01)}} &20.44\% (+0.6) &21.27\% (+1.43) &\textbf{21.65\% (+1.81)} \\
&M*** &L*** &M*** &S* &M*** &M*** \\
\multirow{2}{*}{T1-T2-T4} &\cellcolor[HTML]{D9EAFD}39.64\% (+0.53) &\cellcolor[HTML]{D9EAFD}40.22\% (+1.11) &\cellcolor[HTML]{D9EAFD}\textbf{41.57\% (+2.46)} &\cellcolor[HTML]{D9EAFD}20.44\% (+0.6) &\cellcolor[HTML]{D9EAFD}21.41\% (+1.57) &\cellcolor[HTML]{D9EAFD}\textbf{21.53\% (+1.69)} \\
&\cellcolor[HTML]{D9EAFD}S* &\cellcolor[HTML]{D9EAFD}S* &\cellcolor[HTML]{D9EAFD}S*** &\cellcolor[HTML]{D9EAFD}S* &\cellcolor[HTML]{D9EAFD}M*** &\cellcolor[HTML]{D9EAFD}M*** \\
\multirow{2}{*}{T1-T2-T5} &37.91\% (-1.2) &\textbf{39.81\% (+0.7)} &37.8\% (-1.31) &19.84\% (+0.0) &19.48\% (-0.36) &\textbf{19.97\% (+0.13)} \\
&S* &S* &M*** &S* &S* &S* \\
\multirow{2}{*}{T1-T3-T4} &\cellcolor[HTML]{D9EAFD}\textbf{40.49\% (+1.38)} &\cellcolor[HTML]{D9EAFD}39.39\% (+0.28) &\cellcolor[HTML]{D9EAFD}39.47\% (+0.36) &\cellcolor[HTML]{D9EAFD}\ul{\textbf{21.96\% (+2.12)}} &\cellcolor[HTML]{D9EAFD}\ul{21.58\% (+1.74)} &\cellcolor[HTML]{D9EAFD}21.16\% (+1.32) \\
&\cellcolor[HTML]{D9EAFD}S** &\cellcolor[HTML]{D9EAFD}S* &\cellcolor[HTML]{D9EAFD}S* &\cellcolor[HTML]{D9EAFD}M*** &\cellcolor[HTML]{D9EAFD}M*** &\cellcolor[HTML]{D9EAFD}S** \\
\multirow{2}{*}{T1-T3-T5} &38.24\% (-0.87) &\textbf{39.34\% (+0.23)} &37.25\% (-1.86) &20.45\% (+0.61) &19.85\% (+0.01) &\textbf{20.52\% (+0.68)} \\
&S* &S* &L*** &S* &S* &S** \\
\multirow{2}{*}{T1-T4-T5} &\cellcolor[HTML]{D9EAFD}37.23\% (-1.88) &\cellcolor[HTML]{D9EAFD}\textbf{38.29\% (-0.82)} &\cellcolor[HTML]{D9EAFD}36.18\% (-2.93) &\cellcolor[HTML]{D9EAFD}19.76\% (-0.08) &\cellcolor[HTML]{D9EAFD}\textbf{20.51\% (+0.67)} &\cellcolor[HTML]{D9EAFD}20.15\% (+0.31) \\
&\cellcolor[HTML]{D9EAFD}S*** &\cellcolor[HTML]{D9EAFD}S* &\cellcolor[HTML]{D9EAFD}L*** &\cellcolor[HTML]{D9EAFD}S* &\cellcolor[HTML]{D9EAFD}S** &\cellcolor[HTML]{D9EAFD}S* \\\midrule
\multirow{2}{*}{T1-T2-T3-T4} &39.76\% (+0.65) &41.14\% (+2.03) &\textbf{41.74\% (+2.63)} &20.83\% (+0.99) &20.68\% (+0.84) &\ul{\textbf{22.45\% (+2.61)}} \\
&S*** &S** &S** &S* &S*** &L*** \\
\multirow{2}{*}{T1-T2-T3-T5} &\cellcolor[HTML]{D9EAFD}40.61\% (+1.5) &\cellcolor[HTML]{D9EAFD}40.35\% (+1.24) &\cellcolor[HTML]{D9EAFD}\textbf{40.77\% (+1.66)} &\cellcolor[HTML]{D9EAFD}19.18\% (-0.66) &\cellcolor[HTML]{D9EAFD}19.91\% (+0.07) &\cellcolor[HTML]{D9EAFD}\textbf{20.92\% (+1.08)} \\
&\cellcolor[HTML]{D9EAFD}S*** &\cellcolor[HTML]{D9EAFD}S** &\cellcolor[HTML]{D9EAFD}S* &\cellcolor[HTML]{D9EAFD}S* &\cellcolor[HTML]{D9EAFD}S* &\cellcolor[HTML]{D9EAFD}S*** \\
\multirow{2}{*}{T1-T2-T4-T5} &38.3\% (-0.81) &\textbf{39.09\% (-0.02)} &38.82\% (-0.29) &\textbf{21.37\% (+1.53)} &20.88\% (+1.04) &21.11\% (+1.27) \\
&S* &S** &S*** &S*** &S*** &S*** \\
\multirow{2}{*}{T1-T3-T4-T5} &\cellcolor[HTML]{D9EAFD}37.46\% (-1.65) &\cellcolor[HTML]{D9EAFD}38.23\% (-0.88) &\cellcolor[HTML]{D9EAFD}\textbf{38.33\% (-0.78)} &\cellcolor[HTML]{D9EAFD}20.55\% (+0.71) &\cellcolor[HTML]{D9EAFD}20.68\% (+0.84) &\cellcolor[HTML]{D9EAFD}\textbf{21.07\% (+1.23)} \\
&\cellcolor[HTML]{D9EAFD}S** &\cellcolor[HTML]{D9EAFD}S** &\cellcolor[HTML]{D9EAFD}S*** &\cellcolor[HTML]{D9EAFD}S*** &\cellcolor[HTML]{D9EAFD}S*** &\cellcolor[HTML]{D9EAFD}S*** \\\midrule
\multirow{2}{*}{T1-T2-T3-T4-T5} &39.01\% (-0.1) &\textbf{39.84\% (+0.73)} &37.31\% (-1.8) &20.82\% (+0.98) &19.69\% (-0.15) &\textbf{20.89\% (+1.05)} \\
&S* &S* &M*** &S* &S* &S*** \\
\bottomrule
\end{tabular}
\end{table}

In this RQ, we merge the APR adapter with all the other task-specific adapters and evaluate the performance of the merged adapter on the APR benchmark. Tables \ref{tab:rq1-pass1} and \ref{tab:rq1-pass10} report the \passone and \passten scores of these experiments, respectively. 
As \passone scores consider the model performance with one generation of code, we mainly discuss this score for the results. However, for the completeness of the work, we also included \passten scores and some related discussions.
Comparing the two models, \starcoder and \granite, the results for \starcoder are higher, which could reflect the ability of the \starcoder model for APR task.

According to the \passone scores in Table \ref{tab:rq1-pass1}, among the 15 merged adapters with different combinations, we can observe the performance improvement over the APR adapter in less than half of the merged adapters (6 using dare-ties method, 7 using ties method, and 5 using weight-averaging method) for the first model (StarCoder2). Whereas, for the second model, this figure rises to more than half of the merged adapters (11 using dare-ties method, 12 using ties method, and 13 using weight-averaging method). 
In other words, a considerable number of the merged adapters evaluated with \granite model have higher scores than the APR task that is evaluated with the same model. 
However, for the \starcoder model, fewer number of the merged adapters have led to improved performance compared to the APR task. 
One of the reasons for this difference between different models can be rooted in the scores of other individual task-specific adapters with each model (refer to the scores reported in Table \ref{tab:ind-passes}). 
Particularly, other task-specific adapters except T5, perform better than APR using the \granite model, while APR is the second best task when evaluated with the \starcoder model. \textit{As a result, merging APR with other tasks is more probable to be effective when applied in \granite instead of the \starcoder model.} This finding is valid across all merging methods according to the \passone and \passten scores of these methods.

Comparing different merged adapters (column-wise comparison), according to the \passone scores of the merged adapters, T1-T2-T3 performs the best among all the other merged adapters with each of the three merging methods when evaluated with the \starcoder model. 
The performance of the T1-T2-T3 adapter improved the APR performance, which is statistically significant and effect size is Medium to Large.
However, for the \granite model, T1-T3-T4 adapter performs better than others using two merging methods, i.e., dare-ties and ties. 
In this case, the performance is improved compared to APR score, which is statistically significant, but the effect size is Small for dare-ties and Medium for the ties.
With the weight-averaging method, T1-T2-T3-T4 performs better than other adapters, and the improvement over APR is statistically significant with Medium effect size. 
These results are in agreement with the results of the individual task-specific adapters, as T3 alone reaches the highest performance on the APR benchmark across both models (see Table \ref{tab:ind-passes}). As seen above, T3 is also included in all the best-performing merged adapters.

It is also worth mentioning that while some merged adapters improve the performance of the APR, the \passone scores of all the merged adapters on the APR benchmark are less than the scores of the highest performing individual task-specific adapters (i.e., T3), which are $31.40\%$ and $18.63\%$ for \starcoder and \granite, respectively. 
That being said, it does not affect our discussions about the merging technique to improve the results, compared to the APR individual adapter. We associate the reason of high performance for the T3 adapter to the diversity of the records in its training dataset. When we investigated the results, we observed that some samples were close to program repair task. Thus, it could affect the performance of T3 on APR task in a positive way. 

For the \passten scores, the same tasks, i.e., T1-T2-T3 and T1-T3-T4, perform better than others for \starcoder and \granite, respectively. Interestingly, the \passten scores of the T1-T2-T3 and T1-T3 merged adapters using ties and weight-averaging methods are even higher than the scores of their constituent individual task-adapters (i.e., T1, T2, and T3). Specifically, the \passten score of the best performing individual task-specific adapter which is $41.88\%$ is lower than the \passten scores of the above-mentioned merged adapters. Despite being rare when considering the total number of the experiments conducted for different combinations of the tasks, \textit{these results show that merging different task-specific adapters could lead to even higher performance for a specific downstream task like APR.}

We experimented with merging different \textit{number} of task-specific adapters to construct the merge adapters as shown in Tables \ref{tab:rq1-pass1} and \ref{tab:rq1-pass10}. However, we do not observe any specific pattern on the score of the merged adapters with different number of tasks involved in their merging process. 
In other words, increasing or decreasing the number of task-specific adapters used to build the merged adapters does not necessarily lead to performance degradation or improvement over the APR benchmark. For example, for the \starcoder model, the scores of the T1-T2-T3-T4-T5 adapter is higher than the scores of the T1-T5 adapter for all merging techniques, while having more constituent tasks. 
Conversely, the scores of the T1-T2 adapter is higher than the scores of T1-T2-T3-T4-T5 adapter in the same model for all the merging techniques. 
\textit{As discussed above, we find that the type of the tasks involved in the merged adapters has a higher affect on the performance of the merged adapter rather than the number of the tasks in the merged adapter.}  

When comparing the results of the merging methods in terms of the \passone scores (i.e., row-wise), ties method performs better than the other two methods applied on the \starcoder model whereas dare-ties and weight-averaging perform better if \granite is used as the base model. Surprisingly, ties performs the worst for the latter model and dare-ties and weight-averaging perform close to each other with weight-averaging being slightly better. 
For \passten, a mixed results is observed for the \starcoder model, where weight-averaging and ties have better scores. For the \granite model, almost the same pattern exists for the results of the \passten scores. More specifically, weight-averaging performs the best for both models with ties being close to weight-averaging for \starcoder and the worst for \granite. 

By analyzing the merged adapters that are trained on \starcoder model and merged using two adapters, T1-T3 performs better than others, and after that T1-T2 is the second best-performing adapter with T1-T4 and T1-T5 coming afterwards. This suggests that T3 helps improve the performance of the APR task while the others either degrade the performance or increase it slightly (T1-T2). 
In the same vein, in the category of merged adapters that are created using three adapters, merging with T3 and T2 tasks lead to the highest performance improvement over other merged adapters. 
For the merged adapters that are built by four task-specific adapters, the combination of T2, T3, and T4 leads the highest improvement as expected. Lastly, the merged adapter of all task-specific adapters, performs as one of the middlemost adapters. These results are valid for both the \passone and \passten scores as reported in Tables \ref{tab:rq1-pass1} and \ref{tab:rq1-pass10}, respectively.
Again, this observation supports our discussion that the merged task plays an important role in improving the performance compared to APR adapter. 

For the \granite model, almost the same pattern exists with some few changes. Specifically, since T4 performs better than T2 using this model as reported in Table \ref{tab:ind-passes}, the merged adapter T1-T4 performs better than T1-T2. 
In addition, as T3 and T5 have the best and worst scores on APR, respectively, T1-T3 has the highest score and T1-T5 has the lower score among all adapters that merge two adapters.
Similarly, merging T1 with T3 and T4 performs better than all merged adapters created from three tasks and merging with T2, T3, and T4 is the best-performing merged adapter among all merged adapters consisting of four tasks. Finally, the last adapters, i.e., T1-T2-T3-T4-T5 performs relatively as one of the middlemost adapters.   


\begin{tcolorbox}[colback=green!5!white,colframe=boxheader,title=RQ1 Summary]

    Among all the experiments conducted using different combinations of task-specific adapters, we can observe performance improvement on the APR benchmark in some experiments. Particularly, in 19 and 36 experiments out of 45, the \passone scores on the APR has improved for the \starcoder and \granite models, respectively. However, this highly depend on the type of the base model used and the constituent tasks of the merged adapter. Merging adapters using \granite as the base model is more effective in improving the performance. In addition, the individual task with higher performance than APR can lead to higher performance in the merged adapter on APR benchmark while the ones with lower performance degrade the performance of the merged adapter. 
    Additionally, the performance of the merged adapter depends more on the task-adapter being used in the merging than the number of task-adapters being merged.

\end{tcolorbox}

\subsection{RQ2: Generalizability of Merged Adapters to APR}

\begin{table}[!htp]\centering
\caption{\passone scores of the merged adapters consisting of non-APR task adapters. The results are reported across three merging methods for \starcoder and \granite models. Note that the order of the tasks being merged does not affect the performance as they are all merged with equal weights. The tasks are listed as T1 (APR), T2 (Improvement), T3 (Misc), T4 (Development), and T5 (Test \& QA). }
\label{tab:rq2-pass1}
\scriptsize
\begin{tabular}{l|ccc|cccc}\toprule
\multirow{2}{*}{Meged Tasks} &\multicolumn{3}{c}{\starcoder} &\multicolumn{3}{c}{\granite} \\\cmidrule{2-7}
&dare-ties &ties &weight-averaging &dare-ties &ties &weight-averaging \\\midrule
\multirow{2}{*}{T2-T3} &\ul{30.3\% (+1.64)} &\ul{\textbf{31.95\% (+3.29)}} &\ul{30.67\% (+2.01)} &\textbf{18.32\% (+2.16)} &17.96\% (+1.8) &18.23\% (+2.07) \\
&S*** &L*** &L*** &M*** &M*** &M*** \\
\multirow{2}{*}{T2-T4} &\cellcolor[HTML]{FFF2D7}\textbf{27.26\% (-1.4)} &\cellcolor[HTML]{FFF2D7}27.2\% (-1.46) &\cellcolor[HTML]{FFF2D7}26.62\% (-2.04) &\cellcolor[HTML]{FFF2D7}18.26\% (+2.1) &\cellcolor[HTML]{FFF2D7}18.05\% (+1.89) &\cellcolor[HTML]{FFF2D7}\textbf{18.84\% (+2.68)} \\
&\cellcolor[HTML]{FFF2D7}M*** &\cellcolor[HTML]{FFF2D7}S*** &\cellcolor[HTML]{FFF2D7}S*** &\cellcolor[HTML]{FFF2D7}L*** &\cellcolor[HTML]{FFF2D7}L*** &\cellcolor[HTML]{FFF2D7}L*** \\
\multirow{2}{*}{T2-T5} &25.85\% (-2.81) &\textbf{26.28\% (-2.38)} &25.82\% (-2.84) &14.85\% (-1.31) &14.91\% (-1.25) &\textbf{14.97\% (-1.19)} \\
&L*** &L*** &M*** &M*** &M*** &M*** \\
\multirow{2}{*}{T3-T4} &\cellcolor[HTML]{FFF2D7}27.26\% (-1.4) &\cellcolor[HTML]{FFF2D7}\textbf{27.9\% (-0.76)} &\cellcolor[HTML]{FFF2D7}28.99\% (+0.33) &\cellcolor[HTML]{FFF2D7}\ul{\textbf{19.3\% (+3.14)}} &\cellcolor[HTML]{FFF2D7}\ul{18.84\% (+2.68)} &\cellcolor[HTML]{FFF2D7}\ul{19.09\% (+2.93)} \\
&\cellcolor[HTML]{FFF2D7}S** &\cellcolor[HTML]{FFF2D7}S* &\cellcolor[HTML]{FFF2D7}S*** &\cellcolor[HTML]{FFF2D7}L*** &\cellcolor[HTML]{FFF2D7}L*** &\cellcolor[HTML]{FFF2D7}L*** \\
\multirow{2}{*}{T3-T5} &26.01\% (-2.65) &\textbf{26.83\% (-1.83)} &26.37\% (-2.29) &16.77\% (+0.61) &16.55\% (+0.39) &\textbf{17.41\% (+1.25)} \\
&M*** &M*** &M*** &S** &S* &S*** \\
\multirow{2}{*}{T4-T5} &\cellcolor[HTML]{FFF2D7}24.42\% (-4.24) &\cellcolor[HTML]{FFF2D7}24.79\% (-3.87) &\cellcolor[HTML]{FFF2D7}\textbf{24.91\% (-3.75)} &\cellcolor[HTML]{FFF2D7}16.34\% (+0.18) &\cellcolor[HTML]{FFF2D7}15.82\% (-0.34) &\cellcolor[HTML]{FFF2D7}\textbf{16.43\% (+0.27)} \\
&\cellcolor[HTML]{FFF2D7}L*** &\cellcolor[HTML]{FFF2D7}L*** &\cellcolor[HTML]{FFF2D7}L*** &\cellcolor[HTML]{FFF2D7}S* &\cellcolor[HTML]{FFF2D7}S* &\cellcolor[HTML]{FFF2D7}S* \\\midrule
\multirow{2}{*}{T2-T3-T4} &\textbf{29.27\% (+0.61)} &29.12\% (+0.46) &29.09\% (+0.43) &18.48\% (+2.32) &18.78\% (+2.62) &\textbf{18.99\% (+2.83)} \\
&S* &S* &S** &L*** &L*** &L*** \\
\multirow{2}{*}{T2-T3-T5} &\cellcolor[HTML]{FFF2D7}\textbf{29.02\% (+0.36)} &\cellcolor[HTML]{FFF2D7}28.48\% (-0.18) &\cellcolor[HTML]{FFF2D7}27.77\% (-0.89) &\cellcolor[HTML]{FFF2D7}\textbf{17.04\% (+0.88)} &\cellcolor[HTML]{FFF2D7}16.55\% (+0.39) &\cellcolor[HTML]{FFF2D7}16.8\% (+0.64) \\
&\cellcolor[HTML]{FFF2D7}S* &\cellcolor[HTML]{FFF2D7}S* &\cellcolor[HTML]{FFF2D7}S* &\cellcolor[HTML]{FFF2D7}S** &\cellcolor[HTML]{FFF2D7}S** &\cellcolor[HTML]{FFF2D7}S* \\
\multirow{2}{*}{T2-T4-T5} &25.09\% (-3.57) &25.88\% (-2.78) &\textbf{26.16\% (-2.5)} &16.74\% (+0.58) &16.25\% (+0.09) &\textbf{17.5\% (+1.34)} \\
&L*** &L*** &M*** &S* &S* &M*** \\
\multirow{2}{*}{T3-T4-T5} &\cellcolor[HTML]{FFF2D7}\textbf{27.8\% (-0.86)} &\cellcolor[HTML]{FFF2D7}27.1\% (-1.56) &\cellcolor[HTML]{FFF2D7}26.37\% (-2.29) &\cellcolor[HTML]{FFF2D7}17.23\% (+1.07) &\cellcolor[HTML]{FFF2D7}17.26\% (+1.1) &\cellcolor[HTML]{FFF2D7}\textbf{17.8\% (+1.64)} \\
&\cellcolor[HTML]{FFF2D7}S* &\cellcolor[HTML]{FFF2D7}M*** &\cellcolor[HTML]{FFF2D7}M*** &\cellcolor[HTML]{FFF2D7}S*** &\cellcolor[HTML]{FFF2D7}M*** &\cellcolor[HTML]{FFF2D7}M*** \\\midrule
\multirow{2}{*}{T2-T3-T4-T5} &27.5\% (-1.16) &\textbf{27.84\% (-0.82)} &27.29\% (-1.37) &16.92\% (+0.76) &17.47\% (+1.31) &\textbf{17.77\% (+1.61)} \\
&S*** &S* &S*** &S** &M*** &M*** \\
\bottomrule
\end{tabular}
\end{table}

\begin{table}[!htp]\centering
\caption{\passten scores of the merged adapters consisting of non-APR task adapters. The results are reported across three merging methods for \starcoder and \granite models. Note that the order of the tasks being merged does not affect the performance as they are all merged with equal weights. The tasks are listed as T1 (APR), T2 (Improvement), T3 (Misc), T4 (Development), and T5 (Test \& QA). }
\label{tab:rq2-pass10}
\scriptsize
\begin{tabular}{l|ccc|cccc}\toprule
\multirow{2}{*}{Merged Tasks} &\multicolumn{3}{c}{\starcoder} &\multicolumn{3}{c}{\granite} \\\cmidrule{2-7}
&dare-ties &ties &weight-averaging &dare-ties &ties &weight-averaging \\\midrule
\multirow{2}{*}{T2-T3} &39.48\% (+0.37) &\ul{\textbf{41.79\% (+2.68)}} &\ul{40.79\% (+1.68)} &21.53\% (+1.69) &20.86\% (+1.02) &\textbf{21.78\% (+1.94)} \\
&S* &M*** &M*** &S*** &S** &M*** \\
\multirow{2}{*}{T2-T4} &\cellcolor[HTML]{D9EAFD}37.09\% (-2.02) &\cellcolor[HTML]{D9EAFD}\textbf{37.89\% (-1.22)} &\cellcolor[HTML]{D9EAFD}35.94\% (-3.17) &\cellcolor[HTML]{D9EAFD}22.27\% (+2.43) &\cellcolor[HTML]{D9EAFD}\ul{21.98\% (+2.14)} &\cellcolor[HTML]{D9EAFD}\ul{\textbf{22.63\% (+2.79)}} \\
&\cellcolor[HTML]{D9EAFD}L*** &\cellcolor[HTML]{D9EAFD}S** &\cellcolor[HTML]{D9EAFD}M*** &\cellcolor[HTML]{D9EAFD}M*** &\cellcolor[HTML]{D9EAFD}M*** &\cellcolor[HTML]{D9EAFD}L*** \\
\multirow{2}{*}{T2-T5} &35.62\% (-3.49) &\textbf{36.49\% (-2.62)} &35.17\% (-3.94) &\textbf{19.58\% (-0.26)} &18.78\% (-1.06) &18.83\% (-1.01) \\
&L*** &M*** &L*** &S* &M*** &S* \\
\multirow{2}{*}{T3-T4} &\cellcolor[HTML]{D9EAFD}37.5\% (-1.61) &\cellcolor[HTML]{D9EAFD}39.21\% (+0.1) &\cellcolor[HTML]{D9EAFD}\textbf{40.47\% (+1.36)} &\cellcolor[HTML]{D9EAFD}21.88\% (+2.04) &\cellcolor[HTML]{D9EAFD}21.9\% (+2.06) &\cellcolor[HTML]{D9EAFD}\textbf{22.1\% (+2.26)} \\
&\cellcolor[HTML]{D9EAFD}M*** &\cellcolor[HTML]{D9EAFD}S** &\cellcolor[HTML]{D9EAFD}S*** &\cellcolor[HTML]{D9EAFD}S*** &\cellcolor[HTML]{D9EAFD}M*** &\cellcolor[HTML]{D9EAFD}L*** \\
\multirow{2}{*}{T3-T5} &35.46\% (-3.65) &\textbf{37.0\% (-2.11)} &35.78\% (-3.33) &21.4\% (+1.56) &20.49\% (+0.65) &\textbf{21.58\% (+1.74)} \\
&L*** &S*** &M*** &S*** &S* &M*** \\
\multirow{2}{*}{T4-T5} &\cellcolor[HTML]{D9EAFD}32.54\% (-6.57) &\cellcolor[HTML]{D9EAFD}\textbf{34.33\% (-4.78)} &\cellcolor[HTML]{D9EAFD}33.25\% (-5.86) &\cellcolor[HTML]{D9EAFD}\textbf{20.82\% (+0.98)} &\cellcolor[HTML]{D9EAFD}19.5\% (-0.34) &\cellcolor[HTML]{D9EAFD}20.58\% (+0.74) \\
&\cellcolor[HTML]{D9EAFD}L*** &\cellcolor[HTML]{D9EAFD}L*** &\cellcolor[HTML]{D9EAFD}L*** &\cellcolor[HTML]{D9EAFD}S* &\cellcolor[HTML]{D9EAFD}S* &\cellcolor[HTML]{D9EAFD}S*** \\\midrule
\multirow{2}{*}{T2-T3-T4} &39.45\% (+0.34) &\textbf{41.02\% (+1.91)} &40.71\% (+1.6) &\ul{\textbf{22.41\% (+2.57)}} &21.59\% (+1.75) &22.2\% (+2.36) \\
&S* &M*** &M*** &L*** &M*** &L*** \\
\multirow{2}{*}{T2-T3-T5} &\cellcolor[HTML]{D9EAFD}\ul{\textbf{39.6\% (+0.49)}} &\cellcolor[HTML]{D9EAFD}39.17\% (+0.06) &\cellcolor[HTML]{D9EAFD}39.28\% (+0.17) &\cellcolor[HTML]{D9EAFD}\textbf{20.52\% (+0.68)} &\cellcolor[HTML]{D9EAFD}19.79\% (-0.05) &\cellcolor[HTML]{D9EAFD}20.08\% (+0.24) \\
&\cellcolor[HTML]{D9EAFD}S* &\cellcolor[HTML]{D9EAFD}S* &\cellcolor[HTML]{D9EAFD}S** &\cellcolor[HTML]{D9EAFD}S* &\cellcolor[HTML]{D9EAFD}S* &\cellcolor[HTML]{D9EAFD}S* \\
\multirow{2}{*}{T2-T4-T5} &33.4\% (-5.71) &35.64\% (-3.47) &\textbf{35.96\% (-3.15)} &20.83\% (+0.99) &20.33\% (+0.49) &\textbf{21.41\% (+1.57)} \\
&L*** &M*** &M*** &S* &S* &M*** \\
\multirow{2}{*}{T3-T4-T5} &\cellcolor[HTML]{D9EAFD}37.29\% (-1.82) &\cellcolor[HTML]{D9EAFD}\textbf{37.58\% (-1.53)} &\cellcolor[HTML]{D9EAFD}36.01\% (-3.1) &\cellcolor[HTML]{D9EAFD}20.89\% (+1.05) &\cellcolor[HTML]{D9EAFD}20.76\% (+0.92) &\cellcolor[HTML]{D9EAFD}\textbf{21.42\% (+1.58)} \\
&\cellcolor[HTML]{D9EAFD}M*** &\cellcolor[HTML]{D9EAFD}S** &\cellcolor[HTML]{D9EAFD}M*** &\cellcolor[HTML]{D9EAFD}S** &\cellcolor[HTML]{D9EAFD}S** &\cellcolor[HTML]{D9EAFD}M*** \\\midrule
\multirow{2}{*}{T2-T3-T4-T5} &\textbf{38.27\% (-0.84)} &38.2\% (-0.91) &36.93\% (-2.18) &20.5\% (+0.66) &20.56\% (+0.72) &\textbf{21.63\% (+1.79)} \\
&M*** &S* &S* &S* &S* &M*** \\
\bottomrule
\end{tabular}
\end{table}

In this RQ, we exclude the T1 (APR) adapter when merging different combinations of task-specific adapters and evaluate the merged adapters on the APR benchmark. This will result in having merged adapters consisting of two, three, and four non-APR adapters which will ultimately lead to a set of 11 different merged adapters. The \passone and \passten scores of these merged adapters for both models via three merging methods are reported in Tables~\ref{tab:rq2-pass1} and \ref{tab:rq2-pass10}, respectively.

Despite of excluding the APR task when merging different task-specific adapters, we do not observe a significant performance drops on the APR benchmark when comparing with the results of RQ1. In particular, for the first model, \starcoder, the overall performance on the APR benchmark is decreased with some merged adapters increasing the results.
For the second model, \granite, the performance is increased in more cases. As mentioned earlier, the main reason of this change is related to the high performance of the other individual task-specific adapters on APR benchmark with the \granite model. Hence, the performance of the individual tasks should be considered when investigating the root causes of the performance change of the merged adapters. 

According to the \passone scores reported in Table \ref{tab:rq2-pass1}, the best performing merged adapters are T2-T3 and T3-T4 for the \starcoder and \granite models, respectively. For these merged adapters on both models, the results are improved compared to the APR adapter, and the difference is statistically significant with Large effect size in almost all of the merging techniques. 
Such results are in compliance with the scores of the individual task-specific adapters. In fact, T3 is the best performing individual task, and the performance of T2 is slightly lower than APR with the \starcoder model as shown in Table \ref{tab:ind-passes}. Therefore, their merged adapter has the highest score compared to all other adapter combinations. Similarly, T3 and T4 perform better than the rest of individual adapters when used with \granite model (see Table \ref{tab:ind-passes}). These results applies for all merging methods across both models. However, for the \passten scores as reported in Table \ref{tab:rq2-pass10}, although the same individual tasks perform better than the rest for each model, their merged adapters do not always perform the best. Specifically, the merged adapter of T2-T3-T5 performs the best with dare-ties method for the \starcoder model and T2-T3-T4 performs the best with dare-ties method with \granite model. This is interesting as T5 alone has the worst performance among all task-specific adapters for \starcoder model and yet leads to the highest performance when merged with T2 and T3 tasks. 

Considering the effectiveness of the merging methods, while ties performs slightly better than others for \starcoder model, it is the worst merging method when applied with \granite model. Dare-ties and weight-averaging methods perform worse than ties for the \starcoder model, but for the \granite model, weight-averaging is the most effective method as it performs better than others regarding both \passone and \passten scores.

Similar to the previous RQ, we do not observe much performance difference with merged adapters derived from different \textit{number} of tasks-specific adapters. In fact, the performance of the merged adapters prone to be related with the performance of the individual tasks than the number of the tasks contributed in the merged adapters. In particular, as T3 and T2 are performing well with \starcoder model, their merged adapter is also the best adapter among all adapters. Similarly, for \granite, the merged adapter of T3 and T4 performs better than other adapters as expected. Note that in this RQ, merging four adapters reduces the performance compared to most of the merged adapter with two or three adapters.

\begin{tcolorbox}[colback=green!5!white,colframe=boxheader,title=RQ2 Summary]

   Similar to RQ1, the performance of the merged adapters, whether it rises or falls, depends on the performance of the individual task-specific adapters on the APR benchmark. The individual adapters' performance has the greatest influence on the final score of the merged adapter. In other words, if the individual task adapters perform well, the merged adapter is unlikely to experience a significant performance drop. Consequently, we observe that merged adapters originating from non-APR tasks can still perform well on the APR benchmark, provided their constituent adapters achieve strong APR scores. In some cases, the merged adapter even outperforms the APR-specific adapter, demonstrating the generalizability of the merging approach. This performance improvement over APR-specific adapter is observed in 8 and 29 \passone scores out of 33 experiments for \starcoder and \granite models, respectively. 
    
\end{tcolorbox}

\subsection{RQ3: Performance of Continual Merging }

In this RQ, we are specifically interested in investigating the effect of the order of the adapters added in the merged adapter on the performance of the merged adapter on the APR benchmark. As aforementioned, 
the sequential addition of task-specific adapters will lead to higher weight of the adapters which were merged lastly. As such, we can argue that the performance of the merged adapter is close to that of the last adapters in the continual merging process. Please note that due to resource restrictions, we only experiment with 4 tasks (i.e., APR, Development, Improvement, and Test \& QA) to be merged with varying orders. We opted for excluding the Misc adapter as after investigating the dataset, we found that this category did not represent one single type of code change and could not help much in terms of the interpretation of the results.
The \passone results of this RQ are shown in Tables~\ref{tab:rq3-pass1-3tasks} and \ref{tab:rq3-pass1-4tasks-1}. Tables~\ref{tab:rq3-pass10-3tasks} and \ref{tab:rq3-pass10-4tasks-1} represent the results for \passten. 
Note that for \passone, most results are statistically significant, with mainly effect sizes of Medium or Large for \starcoder and varying effect sizes for \granite.

Among these four tasks, excluding T3, APR (T1) is the best performing task with \starcoder and Development (T4) with the \granite model (see Table \ref{tab:ind-passes}). Based on this, we observe that the scores of the sequentially merged adapters are aligned with the performance of the individual task-specific adapters. Looking at the \passone scores of merging 3 tasks as reported in Table \ref{tab:rq3-pass1-3tasks}, we can verify that the last adapters (the rightmost ones) have a significant effect on the performance of the merged adapter. For example, for the base tasks of T1, T2, T4, the merged adapters with T2-T4-T1 and T4-T2-T1 orders perform better than the rest via all merging methods for the \starcoder model. Likewise, for \granite, the scores of the merged tasks with T1-T2-T4 and T2-T1-T4 orders are higher than those of the others with the same set of tasks (T1, T2, and T4). 

In this vein, the performance of the merged tasks consisting of three task-specific adapters built from other base tasks, follows a similar pattern. Particularly, for the base tasks of T1, T2, and T5, the merged adapters with either T1 or T2 as the last added task perform better than those two with T5 as the last task (for the \starcoder model). This happens as the score of the individual task-specific adapter of T5 is the lowest on the APR benchmark via \starcoder model. Likewise, for the merged adapters of T1, T4, and T5 the same results are obtained when T5 is added in the last step. These trends are valid for the \passten scores of the merged adapters as reported in Table \ref{tab:rq3-pass10-3tasks}.

\begin{table}[!htp]\centering
\caption{\passone scores of the merged adapters using continual merging approach built from three task-specific adapters. The results are reported across three merging methods for \starcoder and \granite models. Note that the order of the tasks being merged is important in this RQ as it affects the weight of each adapter used in the merged adapters. The rightmost adapters have higher weights. The tasks are listed as T1 (APR), T2 (Improvement), T3 (Misc), T4 (Development), and T5 (Test \& QA). }
\label{tab:rq3-pass1-3tasks}
\scriptsize
\begin{tabular}{lc|ccc|cccc}\toprule
\multirow{2}{*}{\rotatebox{90}{Tasks}} &\multirow{2}{*}{Order} &\multicolumn{3}{c}{\starcoder} &\multicolumn{3}{c}{\granite} \\\cmidrule{3-8}
& &dare-ties &ties &weight-averaging &dare-ties &ties &weight-averaging \\\midrule
\multirow{12}{*}{\rotatebox{90}{T1,T2,T4}} &\multirow{2}{*}{T1-T2-T4} &27.32\% (-1.34) &27.35\% (-1.31) &\textbf{27.5\% (-1.16)} &\ul{17.84\% (+1.68)} &\ul{17.47\% (+1.31)} &\ul{\textbf{17.99\% (+1.83)}} \\
& &S* &S** &S*** &M*** &M*** &S*** \\
&\multirow{2}{*}{T1-T4-T2} &\cellcolor[HTML]{FFF2D7}28.29\% (-0.37) &\cellcolor[HTML]{FFF2D7}\textbf{28.81\% (+0.15)} &\cellcolor[HTML]{FFF2D7}\ul{28.78\% (+0.12)} &\cellcolor[HTML]{FFF2D7}\textbf{17.47\% (+1.31)} &\cellcolor[HTML]{FFF2D7}17.32\% (+1.16) &\cellcolor[HTML]{FFF2D7}17.23\% (+1.07) \\
& &\cellcolor[HTML]{FFF2D7}S* &\cellcolor[HTML]{FFF2D7}S* &\cellcolor[HTML]{FFF2D7}S* &\cellcolor[HTML]{FFF2D7}M*** &\cellcolor[HTML]{FFF2D7}S*** &\cellcolor[HTML]{FFF2D7}S*** \\
&\multirow{2}{*}{T2-T1-T4} &\textbf{28.17\% (-0.49)} &27.35\% (-1.31) &27.5\% (-1.16) &17.71\% (+1.55) &\ul{17.47\% (+1.31)} &\ul{\textbf{17.99\% (+1.83)}} \\
& &S* &S* &S*** &M*** &S** &M*** \\
&\multirow{2}{*}{T2-T4-T1} &\cellcolor[HTML]{FFF2D7}\textbf{29.21\% (+0.55)} &\cellcolor[HTML]{FFF2D7}\ul{28.96\% (+0.3)} &\cellcolor[HTML]{FFF2D7}\ul{28.78\% (+0.12)} &\cellcolor[HTML]{FFF2D7}\textbf{17.07\% (+0.91)} &\cellcolor[HTML]{FFF2D7}16.55\% (+0.39) &\cellcolor[HTML]{FFF2D7}16.86\% (+0.7) \\
& &\cellcolor[HTML]{FFF2D7}S* &\cellcolor[HTML]{FFF2D7}S* &\cellcolor[HTML]{FFF2D7}S* &\cellcolor[HTML]{FFF2D7}S* &\cellcolor[HTML]{FFF2D7}S* &\cellcolor[HTML]{FFF2D7}S* \\
&\multirow{2}{*}{T4-T1-T2} &28.11\% (-0.55) &\textbf{28.81\% (+0.15)} &\ul{28.78\% (+0.12)} &\textbf{17.47\% (+1.31)} &17.32\% (+1.16) &17.23\% (+1.07) \\
& &S** &S* &S* &M*** &S*** &S*** \\
&\multirow{2}{*}{T4-T2-T1} &\cellcolor[HTML]{FFF2D7}\ul{\textbf{29.3\% (+0.64)}} &\cellcolor[HTML]{FFF2D7}\ul{28.96\% (+0.3)} &\cellcolor[HTML]{FFF2D7}\ul{28.78\% (+0.12)} &\cellcolor[HTML]{FFF2D7}\textbf{17.01\% (+0.85)} &\cellcolor[HTML]{FFF2D7}16.55\% (+0.39) &\cellcolor[HTML]{FFF2D7}16.86\% (+0.7) \\
& &\cellcolor[HTML]{FFF2D7}S** &\cellcolor[HTML]{FFF2D7}S** &\cellcolor[HTML]{FFF2D7}S* &\cellcolor[HTML]{FFF2D7}S*** &\cellcolor[HTML]{FFF2D7}S* &\cellcolor[HTML]{FFF2D7}S** \\\midrule
\multirow{12}{*}{\rotatebox{90}{T1,T2,T5}} &\multirow{2}{*}{T1-T2-T5} &25.73\% (-2.93) &\textbf{25.88\% (-2.78)} &25.37\% (-3.29) &15.21\% (-0.95) &14.24\% (-1.92) &\textbf{15.24\% (-0.92)} \\
& &L*** &M*** &L*** &S*** &M*** &S*** \\
&\multirow{2}{*}{T1-T5-T2} &\cellcolor[HTML]{FFF2D7}27.9\% (-0.76) &\cellcolor[HTML]{FFF2D7}\textbf{28.11\% (-0.55)} &\cellcolor[HTML]{FFF2D7}27.59\% (-1.07) &\cellcolor[HTML]{FFF2D7}\textbf{16.28\% (+0.12)} &\cellcolor[HTML]{FFF2D7}15.61\% (-0.55) &\cellcolor[HTML]{FFF2D7}15.91\% (-0.25) \\
& &\cellcolor[HTML]{FFF2D7}S* &\cellcolor[HTML]{FFF2D7}S* &\cellcolor[HTML]{FFF2D7}S*** &\cellcolor[HTML]{FFF2D7}S* &\cellcolor[HTML]{FFF2D7}S** &\cellcolor[HTML]{FFF2D7}S* \\
&\multirow{2}{*}{T2-T1-T5} &25.52\% (-3.14) &\textbf{25.88\% (-2.78)} &25.37\% (-3.29) &14.36\% (-1.8) &14.24\% (-1.92) &\textbf{15.24\% (-0.92)} \\
& &L*** &M*** &L*** &S*** &M*** &S** \\
&\multirow{2}{*}{T2-T5-T1} &\cellcolor[HTML]{FFF2D7}27.68\% (-0.98) &\cellcolor[HTML]{FFF2D7}\textbf{28.2\% (-0.46)} &\cellcolor[HTML]{FFF2D7}27.41\% (-1.25) &\cellcolor[HTML]{FFF2D7}\textbf{15.7\% (-0.46)} &\cellcolor[HTML]{FFF2D7}15.58\% (-0.58) &\cellcolor[HTML]{FFF2D7}15.58\% (-0.58) \\
& &\cellcolor[HTML]{FFF2D7}S*** &\cellcolor[HTML]{FFF2D7}S* &\cellcolor[HTML]{FFF2D7}M*** &\cellcolor[HTML]{FFF2D7}S* &\cellcolor[HTML]{FFF2D7}S* &\cellcolor[HTML]{FFF2D7}S* \\
&\multirow{2}{*}{T5-T1-T2} &27.84\% (-0.82) &\textbf{28.11\% (-0.55)} &27.59\% (-1.07) &15.82\% (-0.34) &15.61\% (-0.55) &\textbf{15.91\% (-0.25)} \\
& &S* &S* &S*** &S* &S* &S* \\
&\multirow{2}{*}{T5-T2-T1} &\cellcolor[HTML]{FFF2D7}28.11\% (-0.55) &\cellcolor[HTML]{FFF2D7}\textbf{28.2\% (-0.46)} &\cellcolor[HTML]{FFF2D7}27.41\% (-1.25) &\cellcolor[HTML]{FFF2D7}\textbf{15.73\% (-0.43)} &\cellcolor[HTML]{FFF2D7}15.58\% (-0.58) &\cellcolor[HTML]{FFF2D7}15.58\% (-0.58) \\
& &\cellcolor[HTML]{FFF2D7}S* &\cellcolor[HTML]{FFF2D7}S* &\cellcolor[HTML]{FFF2D7}S*** &\cellcolor[HTML]{FFF2D7}S* &\cellcolor[HTML]{FFF2D7}S** &\cellcolor[HTML]{FFF2D7}S* \\\midrule
\multirow{12}{*}{\rotatebox{90}{T1,T4,T5}} &\multirow{2}{*}{T1-T4-T5} &\textbf{25.43\% (-3.23)} &24.63\% (-4.03) &\textbf{25.43\% (-3.23)} &15.49\% (-0.67) &15.34\% (-0.82) &\textbf{15.55\% (-0.61)} \\
& &L*** &L*** &L*** &S* &S* &S** \\
&\multirow{2}{*}{T1-T5-T4} &\cellcolor[HTML]{FFF2D7}\textbf{26.07\% (-2.59)} &\cellcolor[HTML]{FFF2D7}25.09\% (-3.57) &\cellcolor[HTML]{FFF2D7}25.95\% (-2.71) &\cellcolor[HTML]{FFF2D7}16.92\% (+0.76) &\cellcolor[HTML]{FFF2D7}16.55\% (+0.39) &\cellcolor[HTML]{FFF2D7}\textbf{17.26\% (+1.1)} \\
& &\cellcolor[HTML]{FFF2D7}L*** &\cellcolor[HTML]{FFF2D7}L*** &\cellcolor[HTML]{FFF2D7}L*** &\cellcolor[HTML]{FFF2D7}S*** &\cellcolor[HTML]{FFF2D7}S* &\cellcolor[HTML]{FFF2D7}S*** \\
&\multirow{2}{*}{T4-T1-T5} &24.73\% (-3.93) &24.63\% (-4.03) &\textbf{25.43\% (-3.23)} &\textbf{15.7\% (-0.46)} &15.34\% (-0.82) &15.55\% (-0.61) \\
& &L*** &L*** &L*** &S* &S** &S* \\
&\multirow{2}{*}{T4-T5-T1} &\cellcolor[HTML]{FFF2D7}\textbf{27.07\% (-1.59)} &\cellcolor[HTML]{FFF2D7}25.73\% (-2.93) &\cellcolor[HTML]{FFF2D7}26.46\% (-2.2) &\cellcolor[HTML]{FFF2D7}\textbf{16.49\% (+0.33)} &\cellcolor[HTML]{FFF2D7}16.07\% (-0.09) &\cellcolor[HTML]{FFF2D7}16.1\% (-0.06) \\
& &\cellcolor[HTML]{FFF2D7}S*** &\cellcolor[HTML]{FFF2D7}M*** &\cellcolor[HTML]{FFF2D7}M*** &\cellcolor[HTML]{FFF2D7}S** &\cellcolor[HTML]{FFF2D7}S* &\cellcolor[HTML]{FFF2D7}S* \\
&\multirow{2}{*}{T5-T1-T4} &\textbf{26.28\% (-2.38)} &25.09\% (-3.57) &25.95\% (-2.71) &16.55\% (+0.39) &16.55\% (+0.39) &\textbf{17.26\% (+1.1)} \\
& &L*** &L*** &L*** &S*** &S* &S* \\
&\multirow{2}{*}{T5-T4-T1} &\cellcolor[HTML]{FFF2D7}\textbf{26.8\% (-1.86)} &\cellcolor[HTML]{FFF2D7}25.73\% (-2.93) &\cellcolor[HTML]{FFF2D7}26.46\% (-2.2) &\cellcolor[HTML]{FFF2D7}\textbf{16.37\% (+0.21)} &\cellcolor[HTML]{FFF2D7}16.07\% (-0.09) &\cellcolor[HTML]{FFF2D7}16.1\% (-0.06) \\
& &\cellcolor[HTML]{FFF2D7}S*** &\cellcolor[HTML]{FFF2D7}L*** &\cellcolor[HTML]{FFF2D7}M*** &\cellcolor[HTML]{FFF2D7}S* &\cellcolor[HTML]{FFF2D7}S* &\cellcolor[HTML]{FFF2D7}S* \\
\bottomrule
\end{tabular}
\end{table}
    
\begin{table}[!htp]\centering
\caption{\passone scores of the merged adapters using continual merging approach built from four task-specific adapters. The results are reported across three merging methods for \starcoder and \granite models. Note that the order of the tasks being merged is important in this RQ as it affects the weight of each adapter used in the merged adapters. The rightmost adapters have higher weights.The tasks are listed as T1 (APR), T2 (Improvement), T3 (Misc), T4 (Development), and T5 (Test \& QA).}
\label{tab:rq3-pass1-4tasks-1}
\scriptsize
\begin{tabular}{lc|ccc|cccc}\toprule
\multirow{2}{*}{\rotatebox{90}{Tasks}} &\multirow{2}{*}{Order} &\multicolumn{3}{c}{\starcoder} &\multicolumn{3}{c}{\granite} \\\cmidrule{3-8}
& &dare-ties &ties &weight-averaging &dare-ties &ties &weight-averaging \\\midrule
\multirow{48}{*}{\rotatebox{90}{T1,T2,T4,T5}} &\multirow{2}{*}{T1-T2-T4-T5} &25.03\% (-3.63) &24.33\% (-4.33) &\textbf{25.73\% (-2.93)} &15.85\% (-0.31) &15.4\% (-0.76) &\textbf{16.28\% (+0.12)} \\
& &L*** &L*** &L*** &S* &S* &S* \\
&\multirow{2}{*}{T1-T2-T5-T4} &\cellcolor[HTML]{FFF2D7}25.21\% (-3.45) &\cellcolor[HTML]{FFF2D7}25.55\% (-3.11) &\cellcolor[HTML]{FFF2D7}\textbf{25.67\% (-2.99)} &\cellcolor[HTML]{FFF2D7}17.04\% (+0.88) &\cellcolor[HTML]{FFF2D7}16.43\% (+0.27) &\cellcolor[HTML]{FFF2D7}\textbf{17.1\% (+0.94)} \\
& &\cellcolor[HTML]{FFF2D7}L*** &\cellcolor[HTML]{FFF2D7}M*** &\cellcolor[HTML]{FFF2D7}L*** &\cellcolor[HTML]{FFF2D7}S*** &\cellcolor[HTML]{FFF2D7}S* &\cellcolor[HTML]{FFF2D7}S* \\
&\multirow{2}{*}{T1-T4-T2-T5} &25.43\% (-3.23) &25.49\% (-3.17) &\textbf{25.73\% (-2.93)} &14.66\% (-1.5) &14.82\% (-1.34) &\textbf{15.82\% (-0.34)} \\
& &L*** &L*** &L*** &S** &M*** &S* \\
&\multirow{2}{*}{T1-T4-T5-T2} &\cellcolor[HTML]{FFF2D7}26.31\% (-2.35) &\cellcolor[HTML]{FFF2D7}\textbf{27.29\% (-1.37)} &\cellcolor[HTML]{FFF2D7}27.16\% (-1.5) &\cellcolor[HTML]{FFF2D7}\textbf{17.13\% (+0.97)} &\cellcolor[HTML]{FFF2D7}16.13\% (-0.03) &\cellcolor[HTML]{FFF2D7}16.37\% (+0.21) \\
& &\cellcolor[HTML]{FFF2D7}M*** &\cellcolor[HTML]{FFF2D7}S* &\cellcolor[HTML]{FFF2D7}M*** &\cellcolor[HTML]{FFF2D7}S** &\cellcolor[HTML]{FFF2D7}S* &\cellcolor[HTML]{FFF2D7}S* \\
&\multirow{2}{*}{T1-T5-T2-T4} &25.85\% (-2.81) &\textbf{26.46\% (-2.2)} &26.43\% (-2.23) &17.65\% (+1.49) &16.95\% (+0.79) &\ul{\textbf{17.87\% (+1.71)}} \\
& &L*** &M*** &L*** &L*** &S* &M*** \\
&\multirow{2}{*}{T1-T5-T4-T2} &\cellcolor[HTML]{FFF2D7}27.47\% (-1.19) &\cellcolor[HTML]{FFF2D7}\textbf{27.96\% (-0.7)} &\cellcolor[HTML]{FFF2D7}27.38\% (-1.28) &\cellcolor[HTML]{FFF2D7}\textbf{17.56\% (+1.4)} &\cellcolor[HTML]{FFF2D7}16.77\% (+0.61) &\cellcolor[HTML]{FFF2D7}17.01\% (+0.85) \\
& &\cellcolor[HTML]{FFF2D7}S* &\cellcolor[HTML]{FFF2D7}S* &\cellcolor[HTML]{FFF2D7}M*** &\cellcolor[HTML]{FFF2D7}M*** &\cellcolor[HTML]{FFF2D7}S*** &\cellcolor[HTML]{FFF2D7}S*** \\
&\multirow{2}{*}{T2-T1-T4-T5} &25.46\% (-3.2) &24.33\% (-4.33) &\textbf{25.73\% (-2.93)} &15.85\% (-0.31) &15.4\% (-0.76) &\textbf{16.28\% (+0.12)} \\
& &L*** &L*** &L*** &S* &S*** &S* \\
&\multirow{2}{*}{T2-T1-T5-T4} &\cellcolor[HTML]{FFF2D7}\textbf{25.76\% (-2.9)} &\cellcolor[HTML]{FFF2D7}25.55\% (-3.11) &\cellcolor[HTML]{FFF2D7}25.67\% (-2.99) &\cellcolor[HTML]{FFF2D7}\textbf{17.23\% (+1.07)} &\cellcolor[HTML]{FFF2D7}16.43\% (+0.27) &\cellcolor[HTML]{FFF2D7}17.1\% (+0.94) \\
& &\cellcolor[HTML]{FFF2D7}L*** &\cellcolor[HTML]{FFF2D7}M*** &\cellcolor[HTML]{FFF2D7}L*** &\cellcolor[HTML]{FFF2D7}S*** &\cellcolor[HTML]{FFF2D7}S* &\cellcolor[HTML]{FFF2D7}M*** \\
&\multirow{2}{*}{T2-T4-T1-T5} &25.4\% (-3.26) &25.3\% (-3.36) &\textbf{25.55\% (-3.11)} &14.7\% (-1.46) &14.7\% (-1.46) &\textbf{15.43\% (-0.73)} \\
& &L*** &L*** &L*** &S*** &M*** &S* \\
&\multirow{2}{*}{T2-T4-T5-T1} &\cellcolor[HTML]{FFF2D7}27.04\% (-1.62) &\cellcolor[HTML]{FFF2D7}\textbf{27.35\% (-1.31)} &\cellcolor[HTML]{FFF2D7}27.13\% (-1.53) &\cellcolor[HTML]{FFF2D7}15.79\% (-0.37) &\cellcolor[HTML]{FFF2D7}15.46\% (-0.7) &\cellcolor[HTML]{FFF2D7}\textbf{16.13\% (-0.03)} \\
& &\cellcolor[HTML]{FFF2D7}M*** &\cellcolor[HTML]{FFF2D7}S** &\cellcolor[HTML]{FFF2D7}M*** &\cellcolor[HTML]{FFF2D7}S* &\cellcolor[HTML]{FFF2D7}S* &\cellcolor[HTML]{FFF2D7}S* \\
&\multirow{2}{*}{T2-T5-T1-T4} &26.77\% (-1.89) &\textbf{27.16\% (-1.5)} &26.52\% (-2.14) &\textbf{17.44\% (+1.28)} &\ul{17.26\% (+1.1)} &17.16\% (+1.0) \\
& &M*** &S* &M*** &M*** &S*** &S*** \\
&\multirow{2}{*}{T2-T5-T4-T1} &\cellcolor[HTML]{FFF2D7}\textbf{27.59\% (-1.07)} &\cellcolor[HTML]{FFF2D7}27.2\% (-1.46) &\cellcolor[HTML]{FFF2D7}27.32\% (-1.34) &\cellcolor[HTML]{FFF2D7}16.62\% (+0.46) &\cellcolor[HTML]{FFF2D7}16.43\% (+0.27) &\cellcolor[HTML]{FFF2D7}\textbf{16.8\% (+0.64)} \\
& &\cellcolor[HTML]{FFF2D7}S* &\cellcolor[HTML]{FFF2D7}S* &\cellcolor[HTML]{FFF2D7}S*** &\cellcolor[HTML]{FFF2D7}S** &\cellcolor[HTML]{FFF2D7}S* &\cellcolor[HTML]{FFF2D7}S* \\

&\multirow{2}{*}{T4-T1-T2-T5} &\textbf{25.73\% (-2.93)} &25.49\% (-3.17) &\textbf{25.73\% (-2.93)} &14.94\% (-1.22) &14.82\% (-1.34) &\textbf{15.82\% (-0.34)} \\
& &L*** &M*** &L*** &S** &M*** &S* \\
&\multirow{2}{*}{T4-T1-T5-T2} &\cellcolor[HTML]{FFF2D7}27.01\% (-1.65) &\cellcolor[HTML]{FFF2D7}\textbf{27.29\% (-1.37)} &\cellcolor[HTML]{FFF2D7}27.16\% (-1.5) &\cellcolor[HTML]{FFF2D7}\textbf{16.46\% (+0.3)} &\cellcolor[HTML]{FFF2D7}16.13\% (-0.03) &\cellcolor[HTML]{FFF2D7}16.37\% (+0.21) \\
& &\cellcolor[HTML]{FFF2D7}M*** &\cellcolor[HTML]{FFF2D7}S*** &\cellcolor[HTML]{FFF2D7}S*** &\cellcolor[HTML]{FFF2D7}S** &\cellcolor[HTML]{FFF2D7}S* &\cellcolor[HTML]{FFF2D7}S* \\
&\multirow{2}{*}{T4-T2-T1-T5} &25.4\% (-3.26) &25.3\% (-3.36) &\textbf{25.55\% (-3.11)} &15.37\% (-0.79) &14.7\% (-1.46) &\textbf{15.43\% (-0.73)} \\
& &L*** &L*** &L*** &S* &M*** &S* \\
&\multirow{2}{*}{T4-T2-T5-T1} &\cellcolor[HTML]{FFF2D7}\textbf{28.02\% (-0.64)} &\cellcolor[HTML]{FFF2D7}27.35\% (-1.31) &\cellcolor[HTML]{FFF2D7}27.13\% (-1.53) &\cellcolor[HTML]{FFF2D7}15.98\% (-0.18) &\cellcolor[HTML]{FFF2D7}15.46\% (-0.7) &\cellcolor[HTML]{FFF2D7}\textbf{16.13\% (-0.03)} \\
& &\cellcolor[HTML]{FFF2D7}S* &\cellcolor[HTML]{FFF2D7}S* &\cellcolor[HTML]{FFF2D7}M*** &\cellcolor[HTML]{FFF2D7}S* &\cellcolor[HTML]{FFF2D7}S** &\cellcolor[HTML]{FFF2D7}S* \\
&\multirow{2}{*}{T4-T5-T1-T2} &\ul{\textbf{28.69\% (+0.03)}} &28.38\% (-0.28) &28.26\% (-0.4) &\textbf{16.89\% (+0.73)} &16.34\% (+0.18) &16.86\% (+0.7) \\
& &S* &S* &S* &S*** &S* &S** \\
&\multirow{2}{*}{T4-T5-T2-T1} &\cellcolor[HTML]{FFF2D7}28.57\% (-0.09) &\cellcolor[HTML]{FFF2D7}\ul{28.69\% (+0.03)} &\cellcolor[HTML]{FFF2D7}\ul{\textbf{29.02\% (+0.36)}} &\cellcolor[HTML]{FFF2D7}15.95\% (-0.21) &\cellcolor[HTML]{FFF2D7}16.13\% (-0.03) &\cellcolor[HTML]{FFF2D7}\textbf{16.62\% (+0.46)} \\
& &\cellcolor[HTML]{FFF2D7}S* &\cellcolor[HTML]{FFF2D7}S* &\cellcolor[HTML]{FFF2D7}S* &\cellcolor[HTML]{FFF2D7}S* &\cellcolor[HTML]{FFF2D7}S* &\cellcolor[HTML]{FFF2D7}S* \\
&\multirow{2}{*}{T5-T1-T2-T4} &\textbf{27.26\% (-1.4)} &26.46\% (-2.2) &26.43\% (-2.23) &\ul{\textbf{17.87\% (+1.71)}} &16.95\% (+0.79) &\ul{\textbf{17.87\% (+1.71)}} \\
& &S*** &M*** &L*** &L*** &S* &M*** \\
&\multirow{2}{*}{T5-T1-T4-T2} &\cellcolor[HTML]{FFF2D7}\textbf{28.08\% (-0.58)} &\cellcolor[HTML]{FFF2D7}27.96\% (-0.7) &\cellcolor[HTML]{FFF2D7}27.38\% (-1.28) &\cellcolor[HTML]{FFF2D7}\textbf{17.5\% (+1.34)} &\cellcolor[HTML]{FFF2D7}16.77\% (+0.61) &\cellcolor[HTML]{FFF2D7}17.01\% (+0.85) \\
& &\cellcolor[HTML]{FFF2D7}S* &\cellcolor[HTML]{FFF2D7}S* &\cellcolor[HTML]{FFF2D7}S* &\cellcolor[HTML]{FFF2D7}M*** &\cellcolor[HTML]{FFF2D7}S** &\cellcolor[HTML]{FFF2D7}S* \\
&\multirow{2}{*}{T5-T2-T1-T4} &\textbf{27.41\% (-1.25)} &27.16\% (-1.5) &26.52\% (-2.14) &\textbf{17.68\% (+1.52)} &\ul{17.26\% (+1.1)} &17.16\% (+1.0) \\
& &S*** &S** &L*** &M*** &S*** &S*** \\
&\multirow{2}{*}{T5-T2-T4-T1} &\cellcolor[HTML]{FFF2D7}\textbf{27.62\% (-1.04)} &\cellcolor[HTML]{FFF2D7}27.2\% (-1.46) &\cellcolor[HTML]{FFF2D7}27.32\% (-1.34) &\cellcolor[HTML]{FFF2D7}16.52\% (+0.36) &\cellcolor[HTML]{FFF2D7}16.43\% (+0.27) &\cellcolor[HTML]{FFF2D7}\textbf{16.8\% (+0.64)} \\
& &\cellcolor[HTML]{FFF2D7}S** &\cellcolor[HTML]{FFF2D7}S** &\cellcolor[HTML]{FFF2D7}S** &\cellcolor[HTML]{FFF2D7}S* &\cellcolor[HTML]{FFF2D7}S* &\cellcolor[HTML]{FFF2D7}S* \\
&\multirow{2}{*}{T5-T4-T1-T2} &28.32\% (-0.34) &\textbf{28.38\% (-0.28)} &28.26\% (-0.4) &16.74\% (+0.58) &16.34\% (+0.18) &\textbf{16.86\% (+0.7)} \\
& &S* &S* &S* &S*** &S* &S** \\
&\multirow{2}{*}{T5-T4-T2-T1} &\cellcolor[HTML]{FFF2D7}28.41\% (-0.25) &\cellcolor[HTML]{FFF2D7}\ul{28.69\% (+0.03)} &\cellcolor[HTML]{FFF2D7}\ul{\textbf{29.02\% (+0.36)}} &\cellcolor[HTML]{FFF2D7}16.01\% (-0.15) &\cellcolor[HTML]{FFF2D7}16.13\% (-0.03) &\cellcolor[HTML]{FFF2D7}\textbf{16.62\% (+0.46)} \\
& &\cellcolor[HTML]{FFF2D7}S* &\cellcolor[HTML]{FFF2D7}S** &\cellcolor[HTML]{FFF2D7}S* &\cellcolor[HTML]{FFF2D7}S* &\cellcolor[HTML]{FFF2D7}S* &\cellcolor[HTML]{FFF2D7}S** \\

\bottomrule
\end{tabular}
\end{table}

\begin{table}[!htp]\centering
\caption{\passten scores of the merged adapters using continual merging approach built from three task-specific adapters. The results are reported across three merging methods for \starcoder and \granite models. Note that the order of the tasks being merged is important in this RQ as it affects the weight of each adapter used in the merged adapters. The rightmost adapters have higher weights.The tasks are listed as T1 (APR), T2 (Improvement), T3 (Misc), T4 (Development), and T5 (Test \& QA). }
\label{tab:rq3-pass10-3tasks}
\scriptsize
\begin{tabular}{lc|ccc|cccc}\toprule
\multirow{2}{*}{\rotatebox{90}{Tasks}} &\multirow{2}{*}{Order} &\multicolumn{3}{c}{\starcoder} &\multicolumn{3}{c}{\granite} \\\cmidrule{3-8}
& &dare-ties &ties &weight-averaging &dare-ties &ties &weight-averaging \\\midrule
\multirow{12}{*}{\rotatebox{90}{T1,T2,T4}} &\multirow{2}{*}{T1-T2-T4} &\textbf{40.25\% (+1.14)} &39.12\% (+0.01) &38.5\% (-0.61) &\textbf{22.31\% (+2.47)} &21.38\% (+1.54) &21.57\% (+1.73) \\
& &M*** &S* &S* &L*** &S* &M*** \\
&\multirow{2}{*}{T1-T4-T2} &\cellcolor[HTML]{D9EAFD}39.32\% (+0.21) &\cellcolor[HTML]{D9EAFD}\textbf{39.57\% (+0.46)} &\cellcolor[HTML]{D9EAFD}39.21\% (+0.1) &\cellcolor[HTML]{D9EAFD}21.84\% (+2.0) &\cellcolor[HTML]{D9EAFD}\ul{\textbf{22.09\% (+2.25)}} &\cellcolor[HTML]{D9EAFD}\ul{21.73\% (+1.89)} \\
& &\cellcolor[HTML]{D9EAFD}S** &\cellcolor[HTML]{D9EAFD}S* &\cellcolor[HTML]{D9EAFD}S*** &\cellcolor[HTML]{D9EAFD}M*** &\cellcolor[HTML]{D9EAFD}M*** &\cellcolor[HTML]{D9EAFD}M*** \\
&\multirow{2}{*}{T2-T1-T4} &\textbf{39.15\% (+0.04)} &39.12\% (+0.01) &38.5\% (-0.61) &21.56\% (+1.72) &21.38\% (+1.54) &\textbf{21.57\% (+1.73)} \\
& &S* &S* &S* &M*** &S** &M*** \\
&\multirow{2}{*}{T2-T4-T1} &\cellcolor[HTML]{D9EAFD}39.78\% (+0.67) &\cellcolor[HTML]{D9EAFD}39.98\% (+0.87) &\cellcolor[HTML]{D9EAFD}\ul{\textbf{40.75\% (+1.64)}} &\cellcolor[HTML]{D9EAFD}\ul{\textbf{22.38\% (+2.54)}} &\cellcolor[HTML]{D9EAFD}20.6\% (+0.76) &\cellcolor[HTML]{D9EAFD}21.25\% (+1.41) \\
& &\cellcolor[HTML]{D9EAFD}S*** &\cellcolor[HTML]{D9EAFD}S* &\cellcolor[HTML]{D9EAFD}M*** &\cellcolor[HTML]{D9EAFD}M*** &\cellcolor[HTML]{D9EAFD}S* &\cellcolor[HTML]{D9EAFD}M*** \\
&\multirow{2}{*}{T4-T1-T2} &\textbf{39.81\% (+0.7)} &39.57\% (+0.46) &39.21\% (+0.1) &21.41\% (+1.57) &\ul{\textbf{22.09\% (+2.25)}} &\ul{21.73\% (+1.89)} \\
& &S* &S* &S* &M*** &S*** &M*** \\
&\multirow{2}{*}{T4-T2-T1} &\cellcolor[HTML]{D9EAFD}\ul{\textbf{41.51\% (+2.4)}} &\cellcolor[HTML]{D9EAFD}39.98\% (+0.87) &\cellcolor[HTML]{D9EAFD}\ul{40.75\% (+1.64)} &\cellcolor[HTML]{D9EAFD}\textbf{22.11\% (+2.27)} &\cellcolor[HTML]{D9EAFD}20.6\% (+0.76) &\cellcolor[HTML]{D9EAFD}21.25\% (+1.41) \\
& &\cellcolor[HTML]{D9EAFD}L*** &\cellcolor[HTML]{D9EAFD}S* &\cellcolor[HTML]{D9EAFD}M*** &\cellcolor[HTML]{D9EAFD}M*** &\cellcolor[HTML]{D9EAFD}S* &\cellcolor[HTML]{D9EAFD}S*** \\\midrule
\multirow{12}{*}{\rotatebox{90}{T1,T2,T5}} &\multirow{2}{*}{T1-T2-T5} &34.9\% (-4.21) &34.99\% (-4.12) &\textbf{35.05\% (-4.06)} &19.13\% (-0.71) &18.99\% (-0.85) &\textbf{19.44\% (-0.4)} \\
& &L*** &L*** &L*** &S* &S*** &S* \\
&\multirow{2}{*}{T1-T5-T2} &\cellcolor[HTML]{D9EAFD}39.26\% (+0.15) &\cellcolor[HTML]{D9EAFD}39.49\% (+0.38) &\cellcolor[HTML]{D9EAFD}\textbf{39.74\% (+0.63)} &\cellcolor[HTML]{D9EAFD}\textbf{20.43\% (+0.59)} &\cellcolor[HTML]{D9EAFD}19.47\% (-0.37) &\cellcolor[HTML]{D9EAFD}20.19\% (+0.35) \\
& &\cellcolor[HTML]{D9EAFD}S* &\cellcolor[HTML]{D9EAFD}S* &\cellcolor[HTML]{D9EAFD}S*** &\cellcolor[HTML]{D9EAFD}S* &\cellcolor[HTML]{D9EAFD}S*** &\cellcolor[HTML]{D9EAFD}S* \\
&\multirow{2}{*}{T2-T1-T5} &34.71\% (-4.4) &34.99\% (-4.12) &\textbf{35.05\% (-4.06)} &\textbf{19.45\% (-0.39)} &18.99\% (-0.85) &19.44\% (-0.4) \\
& &L*** &L*** &L*** &S* &S*** &S* \\
&\multirow{2}{*}{T2-T5-T1} &\cellcolor[HTML]{D9EAFD}40.32\% (+1.21) &\cellcolor[HTML]{D9EAFD}\ul{\textbf{41.06\% (+1.95)}} &\cellcolor[HTML]{D9EAFD}39.89\% (+0.78) &\cellcolor[HTML]{D9EAFD}19.91\% (+0.07) &\cellcolor[HTML]{D9EAFD}19.84\% (+0.0) &\cellcolor[HTML]{D9EAFD}\textbf{20.3\% (+0.46)} \\
& &\cellcolor[HTML]{D9EAFD}S*** &\cellcolor[HTML]{D9EAFD}M*** &\cellcolor[HTML]{D9EAFD}S*** &\cellcolor[HTML]{D9EAFD}S* &\cellcolor[HTML]{D9EAFD}S** &\cellcolor[HTML]{D9EAFD}S* \\
&\multirow{2}{*}{T5-T1-T2} &38.82\% (-0.29) &39.49\% (+0.38) &\textbf{39.74\% (+0.63)} &\textbf{20.68\% (+0.84)} &19.47\% (-0.37) &20.19\% (+0.35) \\
& &S* &S* &S*** &S* &S*** &S** \\
&\multirow{2}{*}{T5-T2-T1} &\cellcolor[HTML]{D9EAFD}39.88\% (+0.77) &\cellcolor[HTML]{D9EAFD}\ul{\textbf{41.06\% (+1.95)}} &\cellcolor[HTML]{D9EAFD}39.89\% (+0.78) &\cellcolor[HTML]{D9EAFD}20.02\% (+0.18) &\cellcolor[HTML]{D9EAFD}19.84\% (+0.0) &\cellcolor[HTML]{D9EAFD}\textbf{20.3\% (+0.46)} \\
& &\cellcolor[HTML]{D9EAFD}S*** &\cellcolor[HTML]{D9EAFD}M*** &\cellcolor[HTML]{D9EAFD}S*** &\cellcolor[HTML]{D9EAFD}S* &\cellcolor[HTML]{D9EAFD}S* &\cellcolor[HTML]{D9EAFD}S** \\\midrule
\multirow{12}{*}{\rotatebox{90}{T1,T4,T5}} &\multirow{2}{*}{T1-T4-T5} &\textbf{35.16\% (-3.95)} &33.71\% (-5.4) &35.12\% (-3.99) &\textbf{20.22\% (+0.38)} &19.78\% (-0.06) &19.84\% (+0.0) \\
& &L*** &L*** &L*** &S* &S* &S* \\
&\multirow{2}{*}{T1-T5-T4} &\cellcolor[HTML]{D9EAFD}36.34\% (-2.77) &\cellcolor[HTML]{D9EAFD}34.11\% (-5.0) &\cellcolor[HTML]{D9EAFD}\textbf{37.06\% (-2.05)} &\cellcolor[HTML]{D9EAFD}21.25\% (+1.41) &\cellcolor[HTML]{D9EAFD}20.68\% (+0.84) &\cellcolor[HTML]{D9EAFD}\textbf{21.31\% (+1.47)} \\
& &\cellcolor[HTML]{D9EAFD}S*** &\cellcolor[HTML]{D9EAFD}L*** &\cellcolor[HTML]{D9EAFD}S*** &\cellcolor[HTML]{D9EAFD}M*** &\cellcolor[HTML]{D9EAFD}S* &\cellcolor[HTML]{D9EAFD}M*** \\
&\multirow{2}{*}{T4-T1-T5} &33.7\% (-5.41) &33.71\% (-5.4) &\textbf{35.12\% (-3.99)} &\textbf{20.38\% (+0.54)} &19.78\% (-0.06) &19.84\% (+0.0) \\
& &L*** &L*** &M*** &S* &S* &S* \\
&\multirow{2}{*}{T4-T5-T1} &\cellcolor[HTML]{D9EAFD}\textbf{37.88\% (-1.23)} &\cellcolor[HTML]{D9EAFD}37.16\% (-1.95) &\cellcolor[HTML]{D9EAFD}37.19\% (-1.92) &\cellcolor[HTML]{D9EAFD}\textbf{21.03\% (+1.19)} &\cellcolor[HTML]{D9EAFD}20.07\% (+0.23) &\cellcolor[HTML]{D9EAFD}19.77\% (-0.07) \\
& &\cellcolor[HTML]{D9EAFD}S* &\cellcolor[HTML]{D9EAFD}S*** &\cellcolor[HTML]{D9EAFD}S* &\cellcolor[HTML]{D9EAFD}S** &\cellcolor[HTML]{D9EAFD}S* &\cellcolor[HTML]{D9EAFD}S* \\
&\multirow{2}{*}{T5-T1-T4} &35.19\% (-3.92) &34.11\% (-5.0) &\textbf{37.06\% (-2.05)} &20.81\% (+0.97) &20.68\% (+0.84) &\textbf{21.31\% (+1.47)} \\
& &M*** &L*** &S** &S* &S* &M*** \\
&\multirow{2}{*}{T5-T4-T1} &\cellcolor[HTML]{D9EAFD}\textbf{38.03\% (-1.08)} &\cellcolor[HTML]{D9EAFD}37.16\% (-1.95) &\cellcolor[HTML]{D9EAFD}37.19\% (-1.92) &\cellcolor[HTML]{D9EAFD}19.83\% (-0.01) &\cellcolor[HTML]{D9EAFD}\textbf{20.07\% (+0.23)} &\cellcolor[HTML]{D9EAFD}19.77\% (-0.07) \\
& &\cellcolor[HTML]{D9EAFD}S* &\cellcolor[HTML]{D9EAFD}M*** &\cellcolor[HTML]{D9EAFD}S** &\cellcolor[HTML]{D9EAFD}S* &\cellcolor[HTML]{D9EAFD}S* &\cellcolor[HTML]{D9EAFD}S* \\
\bottomrule
\end{tabular}
\end{table}

\begin{table}[!htp]\centering
\caption{\passten scores of the merged adapters using continual merging approach built from four task-specific adapters. The results are reported across three merging methods for \starcoder and \granite models. Note that the order of the tasks being merged is important in this RQ as it affects the weight of each adapter used in the merged adapters. The rightmost adapters have higher weights. The tasks are listed as T1 (APR), T2 (Improvement), T3 (Misc), T4 (Development), and T5 (Test \& QA).}
\label{tab:rq3-pass10-4tasks-1}
\scriptsize
\begin{tabular}{lc|ccc|ccccc}\toprule
\multirow{2}{*}{\rotatebox{90}{Tasks}} &\multirow{2}{*}{Order} &\multicolumn{3}{c}{\starcoder} &\multicolumn{3}{c}{\granite} \\\cmidrule{3-8}
& &dare-ties &ties &weight-averaging &dare-ties &ties &weight-averaging \\\midrule
\multirow{48}{*}{\rotatebox{90}{T1,T2,T4,T5}} &\multirow{2}{*}{T1-T2-T4-T5} &34.03\% (-5.08) &34.07\% (-5.04) &\textbf{35.46\% (-3.65)} &20.51\% (+0.67) &19.92\% (+0.08) &\textbf{20.56\% (+0.72)} \\
& &L*** &L*** &L*** &S** &S* &S** \\
&\multirow{2}{*}{T1-T2-T5-T4} &\cellcolor[HTML]{D9EAFD}35.09\% (-4.02) &\cellcolor[HTML]{D9EAFD}35.79\% (-3.32) &\cellcolor[HTML]{D9EAFD}\textbf{36.31\% (-2.8)} &\cellcolor[HTML]{D9EAFD}20.96\% (+1.12) &\cellcolor[HTML]{D9EAFD}20.36\% (+0.52) &\cellcolor[HTML]{D9EAFD}\textbf{21.11\% (+1.27)} \\
& &\cellcolor[HTML]{D9EAFD}L*** &\cellcolor[HTML]{D9EAFD}L*** &\cellcolor[HTML]{D9EAFD}S*** &\cellcolor[HTML]{D9EAFD}S*** &\cellcolor[HTML]{D9EAFD}S* &\cellcolor[HTML]{D9EAFD}S*** \\
&\multirow{2}{*}{T1-T4-T2-T5} &34.17\% (-4.94) &34.92\% (-4.19) &\textbf{35.77\% (-3.34)} &19.58\% (-0.26) &19.14\% (-0.7) &\textbf{20.51\% (+0.67)} \\
& &L*** &L*** &L*** &S* &S* &S** \\
&\multirow{2}{*}{T1-T4-T5-T2} &\cellcolor[HTML]{D9EAFD}35.96\% (-3.15) &\cellcolor[HTML]{D9EAFD}37.97\% (-1.14) &\cellcolor[HTML]{D9EAFD}\textbf{38.15\% (-0.96)} &\cellcolor[HTML]{D9EAFD}20.7\% (+0.86) &\cellcolor[HTML]{D9EAFD}\textbf{20.98\% (+1.14)} &\cellcolor[HTML]{D9EAFD}20.28\% (+0.44) \\
& &\cellcolor[HTML]{D9EAFD}L*** &\cellcolor[HTML]{D9EAFD}S*** &\cellcolor[HTML]{D9EAFD}S* &\cellcolor[HTML]{D9EAFD}S*** &\cellcolor[HTML]{D9EAFD}S* &\cellcolor[HTML]{D9EAFD}S* \\
&\multirow{2}{*}{T1-T5-T2-T4} &36.99\% (-2.12) &\textbf{37.62\% (-1.49)} &35.7\% (-3.41) &\ul{21.59\% (+1.75)} &20.98\% (+1.14) &\ul{\textbf{21.99\% (+2.15)}} \\
& &S** &S** &M*** &M*** &S* &L*** \\
&\multirow{2}{*}{T1-T5-T4-T2} &\cellcolor[HTML]{D9EAFD}\textbf{38.48\% (-0.63)} &\cellcolor[HTML]{D9EAFD}37.87\% (-1.24) &\cellcolor[HTML]{D9EAFD}37.68\% (-1.43) &\cellcolor[HTML]{D9EAFD}\textbf{21.46\% (+1.62)} &\cellcolor[HTML]{D9EAFD}20.69\% (+0.85) &\cellcolor[HTML]{D9EAFD}20.86\% (+1.02) \\
& &\cellcolor[HTML]{D9EAFD}S* &\cellcolor[HTML]{D9EAFD}S*** &\cellcolor[HTML]{D9EAFD}S* &\cellcolor[HTML]{D9EAFD}S*** &\cellcolor[HTML]{D9EAFD}S* &\cellcolor[HTML]{D9EAFD}S** \\
&\multirow{2}{*}{T2-T1-T4-T5} &34.34\% (-4.77) &34.07\% (-5.04) &\textbf{35.46\% (-3.65)} &20.53\% (+0.69) &19.92\% (+0.08) &\textbf{20.56\% (+0.72)} \\
& &L*** &L*** &L*** &S* &S* &S** \\
&\multirow{2}{*}{T2-T1-T5-T4} &\cellcolor[HTML]{D9EAFD}35.99\% (-3.12) &\cellcolor[HTML]{D9EAFD}35.79\% (-3.32) &\cellcolor[HTML]{D9EAFD}\textbf{36.31\% (-2.8)} &\cellcolor[HTML]{D9EAFD}20.86\% (+1.02) &\cellcolor[HTML]{D9EAFD}20.36\% (+0.52) &\cellcolor[HTML]{D9EAFD}\textbf{21.11\% (+1.27)} \\
& &\cellcolor[HTML]{D9EAFD}M*** &\cellcolor[HTML]{D9EAFD}L*** &\cellcolor[HTML]{D9EAFD}S*** &\cellcolor[HTML]{D9EAFD}S** &\cellcolor[HTML]{D9EAFD}S* &\cellcolor[HTML]{D9EAFD}S*** \\
&\multirow{2}{*}{T2-T4-T1-T5} &\textbf{35.54\% (-3.57)} &34.9\% (-4.21) &35.22\% (-3.89) &\textbf{20.02\% (+0.18)} &19.0\% (-0.84) &19.85\% (+0.01) \\
& &L*** &L*** &L*** &S* &M*** &S* \\
&\multirow{2}{*}{T2-T4-T5-T1} &\cellcolor[HTML]{D9EAFD}37.78\% (-1.33) &\cellcolor[HTML]{D9EAFD}\textbf{38.94\% (-0.17)} &\cellcolor[HTML]{D9EAFD}37.26\% (-1.85) &\cellcolor[HTML]{D9EAFD}19.97\% (+0.13) &\cellcolor[HTML]{D9EAFD}19.75\% (-0.09) &\cellcolor[HTML]{D9EAFD}\textbf{20.76\% (+0.92)} \\
& &\cellcolor[HTML]{D9EAFD}S* &\cellcolor[HTML]{D9EAFD}S* &\cellcolor[HTML]{D9EAFD}S** &\cellcolor[HTML]{D9EAFD}S* &\cellcolor[HTML]{D9EAFD}S* &\cellcolor[HTML]{D9EAFD}S** \\
&\multirow{2}{*}{T2-T5-T1-T4} &38.17\% (-0.94) &\textbf{38.9\% (-0.21)} &36.61\% (-2.5) &\textbf{21.45\% (+1.61)} &\ul{21.29\% (+1.45)} &21.09\% (+1.25) \\
& &S* &S* &S*** &M*** &S* &S*** \\
&\multirow{2}{*}{T2-T5-T4-T1} &\cellcolor[HTML]{D9EAFD}39.83\% (+0.72) &\cellcolor[HTML]{D9EAFD}\textbf{40.27\% (+1.16)} &\cellcolor[HTML]{D9EAFD}38.46\% (-0.65) &\cellcolor[HTML]{D9EAFD}\textbf{20.76\% (+0.92)} &\cellcolor[HTML]{D9EAFD}20.57\% (+0.73) &\cellcolor[HTML]{D9EAFD}21.23\% (+1.39) \\
& &\cellcolor[HTML]{D9EAFD}S*** &\cellcolor[HTML]{D9EAFD}S* &\cellcolor[HTML]{D9EAFD}S* &\cellcolor[HTML]{D9EAFD}S* &\cellcolor[HTML]{D9EAFD}S* &\cellcolor[HTML]{D9EAFD}S*** \\

&\multirow{2}{*}{T4-T1-T2-T5} &33.94\% (-5.17) &34.92\% (-4.19) &\textbf{35.77\% (-3.34)} &19.02\% (-0.82) &19.14\% (-0.7) &\textbf{20.51\% (+0.67)} \\
& &L*** &L*** &M*** &S* &S*** &S* \\
&\multirow{2}{*}{T4-T1-T5-T2} &\cellcolor[HTML]{D9EAFD}\textbf{38.76\% (-0.35)} &\cellcolor[HTML]{D9EAFD}37.97\% (-1.14) &\cellcolor[HTML]{D9EAFD}38.15\% (-0.96) &\cellcolor[HTML]{D9EAFD}20.56\% (+0.72) &\cellcolor[HTML]{D9EAFD}\textbf{20.98\% (+1.14)} &\cellcolor[HTML]{D9EAFD}20.28\% (+0.44) \\
& &\cellcolor[HTML]{D9EAFD}S* &\cellcolor[HTML]{D9EAFD}S* &\cellcolor[HTML]{D9EAFD}S* &\cellcolor[HTML]{D9EAFD}S* &\cellcolor[HTML]{D9EAFD}S* &\cellcolor[HTML]{D9EAFD}S* \\
&\multirow{2}{*}{T4-T2-T1-T5} &34.12\% (-4.99) &34.9\% (-4.21) &\textbf{35.22\% (-3.89)} &\textbf{20.44\% (+0.6)} &19.0\% (-0.84) &19.85\% (+0.01) \\
& &L*** &L*** &L*** &M*** &S*** &S* \\
&\multirow{2}{*}{T4-T2-T5-T1} &\cellcolor[HTML]{D9EAFD}\textbf{40.1\% (+0.99)} &\cellcolor[HTML]{D9EAFD}38.94\% (-0.17) &\cellcolor[HTML]{D9EAFD}37.26\% (-1.85) &\cellcolor[HTML]{D9EAFD}20.7\% (+0.86) &\cellcolor[HTML]{D9EAFD}19.75\% (-0.09) &\cellcolor[HTML]{D9EAFD}\textbf{20.76\% (+0.92)} \\
& &\cellcolor[HTML]{D9EAFD}S*** &\cellcolor[HTML]{D9EAFD}S* &\cellcolor[HTML]{D9EAFD}S* &\cellcolor[HTML]{D9EAFD}S** &\cellcolor[HTML]{D9EAFD}S** &\cellcolor[HTML]{D9EAFD}S*** \\
&\multirow{2}{*}{T4-T5-T1-T2} &\ul{\textbf{41.28\% (+2.17)}} &39.16\% (+0.05) &40.23\% (+1.12) &20.59\% (+0.75) &19.98\% (+0.14) &\textbf{21.15\% (+1.31)} \\
& &M*** &S* &M*** &S** &S* &M*** \\
&\multirow{2}{*}{T4-T5-T2-T1} &\cellcolor[HTML]{D9EAFD}40.73\% (+1.62) &\cellcolor[HTML]{D9EAFD}\ul{\textbf{42.12\% (+3.01)}} &\cellcolor[HTML]{D9EAFD}\ul{41.87\% (+2.76)} &\cellcolor[HTML]{D9EAFD}20.51\% (+0.67) &\cellcolor[HTML]{D9EAFD}20.13\% (+0.29) &\cellcolor[HTML]{D9EAFD}\textbf{20.52\% (+0.68)} \\
& &\cellcolor[HTML]{D9EAFD}M*** &\cellcolor[HTML]{D9EAFD}M*** &\cellcolor[HTML]{D9EAFD}L*** &\cellcolor[HTML]{D9EAFD}S* &\cellcolor[HTML]{D9EAFD}S* &\cellcolor[HTML]{D9EAFD}S* \\
&\multirow{2}{*}{T5-T1-T2-T4} &37.24\% (-1.87) &\textbf{37.62\% (-1.49)} &35.7\% (-3.41) &21.52\% (+1.68) &20.98\% (+1.14) &\ul{\textbf{21.99\% (+2.15)}} \\
& &S* &S*** &L*** &M*** &S* &M*** \\
&\multirow{2}{*}{T5-T1-T4-T2} &\cellcolor[HTML]{D9EAFD}\textbf{39.48\% (+0.37)} &\cellcolor[HTML]{D9EAFD}37.87\% (-1.24) &\cellcolor[HTML]{D9EAFD}37.68\% (-1.43) &\cellcolor[HTML]{D9EAFD}\textbf{21.44\% (+1.6)} &\cellcolor[HTML]{D9EAFD}20.69\% (+0.85) &\cellcolor[HTML]{D9EAFD}20.86\% (+1.02) \\
& &\cellcolor[HTML]{D9EAFD}S* &\cellcolor[HTML]{D9EAFD}S*** &\cellcolor[HTML]{D9EAFD}S* &\cellcolor[HTML]{D9EAFD}S** &\cellcolor[HTML]{D9EAFD}S* &\cellcolor[HTML]{D9EAFD}M*** \\
&\multirow{2}{*}{T5-T2-T1-T4} &38.68\% (-0.43) &\textbf{38.9\% (-0.21)} &36.61\% (-2.5) &\textbf{21.43\% (+1.59)} &\ul{21.29\% (+1.45)} &21.09\% (+1.25) \\
& &S* &S* &M*** &M*** &S** &S*** \\
&\multirow{2}{*}{T5-T2-T4-T1} &\cellcolor[HTML]{D9EAFD}40.1\% (+0.99) &\cellcolor[HTML]{D9EAFD}\textbf{40.27\% (+1.16)} &\cellcolor[HTML]{D9EAFD}38.46\% (-0.65) &\cellcolor[HTML]{D9EAFD}20.7\% (+0.86) &\cellcolor[HTML]{D9EAFD}20.57\% (+0.73) &\cellcolor[HTML]{D9EAFD}\textbf{21.23\% (+1.39)} \\
& &\cellcolor[HTML]{D9EAFD}M*** &\cellcolor[HTML]{D9EAFD}S** &\cellcolor[HTML]{D9EAFD}S* &\cellcolor[HTML]{D9EAFD}S*** &\cellcolor[HTML]{D9EAFD}S* &\cellcolor[HTML]{D9EAFD}M*** \\
&\multirow{2}{*}{T5-T4-T1-T2} &\textbf{40.39\% (+1.28)} &39.16\% (+0.05) &40.23\% (+1.12) &20.56\% (+0.72) &19.98\% (+0.14) &\textbf{21.15\% (+1.31)} \\
& &S*** &S* &M*** &S* &S* &M*** \\
&\multirow{2}{*}{T5-T4-T2-T1} &\cellcolor[HTML]{D9EAFD}41.05\% (+1.94) &\cellcolor[HTML]{D9EAFD}\ul{\textbf{42.12\% (+3.01)}} &\cellcolor[HTML]{D9EAFD}\ul{41.87\% (+2.76)} &\cellcolor[HTML]{D9EAFD}\textbf{20.78\% (+0.94)} &\cellcolor[HTML]{D9EAFD}20.13\% (+0.29) &\cellcolor[HTML]{D9EAFD}20.52\% (+0.68) \\
& &\cellcolor[HTML]{D9EAFD}M*** &\cellcolor[HTML]{D9EAFD}L*** &\cellcolor[HTML]{D9EAFD}L*** &\cellcolor[HTML]{D9EAFD}S* &\cellcolor[HTML]{D9EAFD}S* &\cellcolor[HTML]{D9EAFD}S* \\

\bottomrule
\end{tabular}
\end{table}

When merging four task-specific adapters, we can still observe that the low-performing tasks such as T5 degrade the performance of the merged adapters if they are merged in the last steps of the continual merging. Particularly, almost all of the merged adapters having T5 as their last or second last adapter, have the \passone score of close to 25\% via all merging methods with \starcoder model. Similar poor results are obtained for the \granite model as reported in Table \ref{tab:rq3-pass1-4tasks-1}. On the other hand, the merged adapters having T1 and T2 as the rightmost adapters, perform better than the others in most cases. For example, for \starcoder model, T4-T5-T1-T2, T4-T5-T2-T1, and T5-T4-T2-T1 are the best performing tasks. For the \granite model, the best performing tasks are T5-T1-T2-T4, T5-T2-T1-T4, T2-T5-T1-T4, and T1-T5-T2-T4 in all of which the T4 task exists as the last one. Recall that for this model, T4 has the best \passone score compared to the rest of the individual task-specific adapters excluding T3. Almost the same tasks performed the best based on the \passten scores across both models and all three merging methods as observed from Table \ref{tab:rq3-pass10-4tasks-1}.

Interestingly, we still observe that in several cases the scores of the merged adapters are higher than the individual scores of their base tasks. In fact, for the \starcoder model, all the merged adapters that have positive performance difference as reported in parenthesis in front of their real scores, have higher performance than that of their individual task adapters. Please note that this performance difference reported in all tables is against T1 (APR) task; hence, as T1 has the highest performance among all the individual task adapters (excluding T3) for the \starcoder model, the previous interpretation is derived. This pattern happens for both \passone and \passten scores of the first model. For the second model, however, as T4 is the best performing task instead of T1, there exist more considerable number of merged adapters that have higher performance compared to T1 task for both \passone and \passten scores.

\begin{figure}[h]

\begin{subfigure}{0.32\textwidth}
    \includegraphics[width=\linewidth, height=280pt]{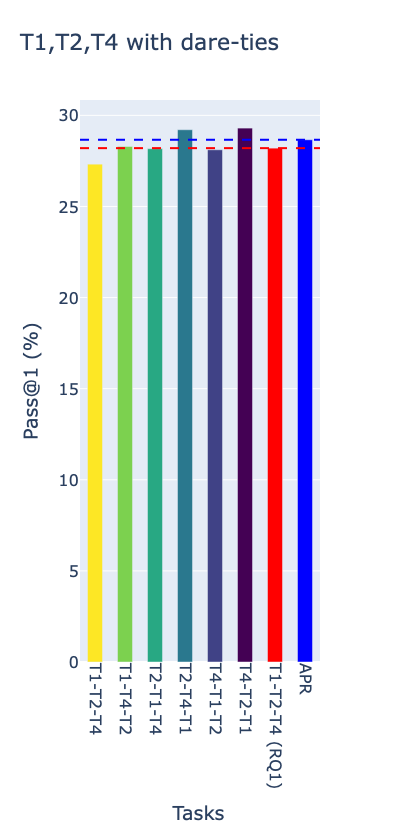} 
    \label{fig:fsub1}
\end{subfigure}
\begin{subfigure}{0.32\textwidth}
    \includegraphics[width=\linewidth, height=280pt]{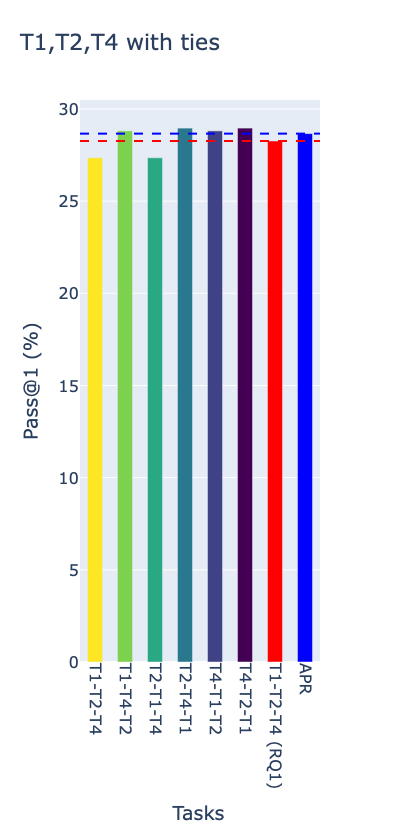}
    \label{fig:fsub2}
\end{subfigure}
\begin{subfigure}{0.32\textwidth}
    \includegraphics[width=\linewidth, height=280pt]{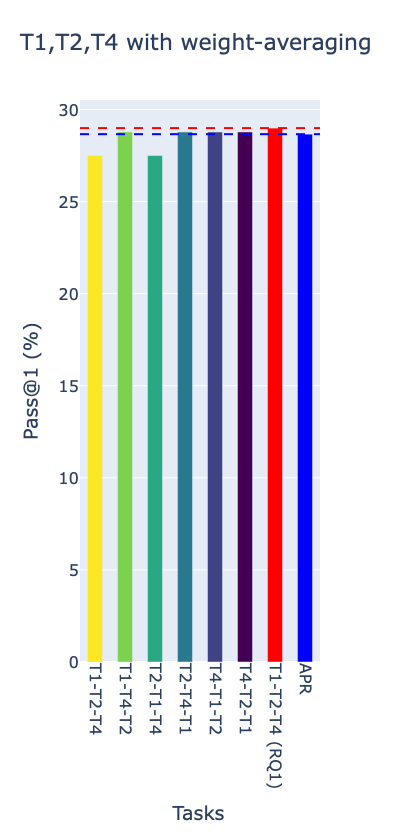}
    \label{fig:fsub3}
\end{subfigure}

\caption{Bar plot of \passone scores obtained by merged adapters of T1, T2, and T4 using continual merging (RQ3) and equal-weight merging (RQ1) approaches using \starcoder model. The tasks are listed as T1 (APR), T2 (Improvement), T3 (Misc), T4 (Development), and T5 (Test \& QA). } 
\label{fig:first-bars-1}

\end{figure}

\begin{figure}[h]

\begin{subfigure}{0.32\textwidth}
    \includegraphics[width=\linewidth, height=280pt]{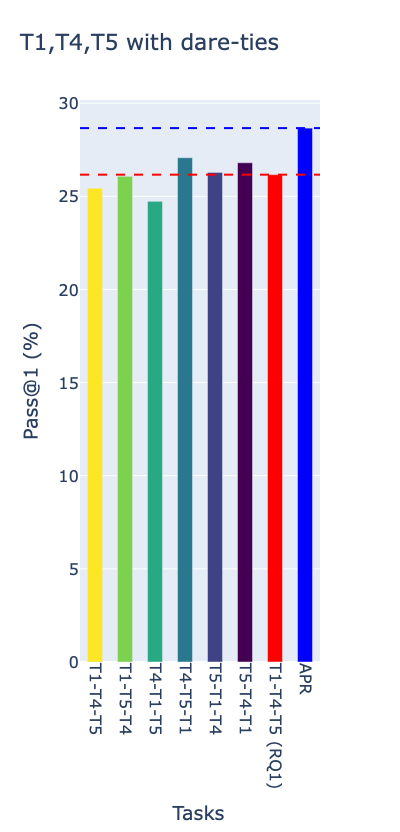} 
    \label{fig:fsub1}
\end{subfigure}
\begin{subfigure}{0.32\textwidth}
    \includegraphics[width=\linewidth, height=280pt]{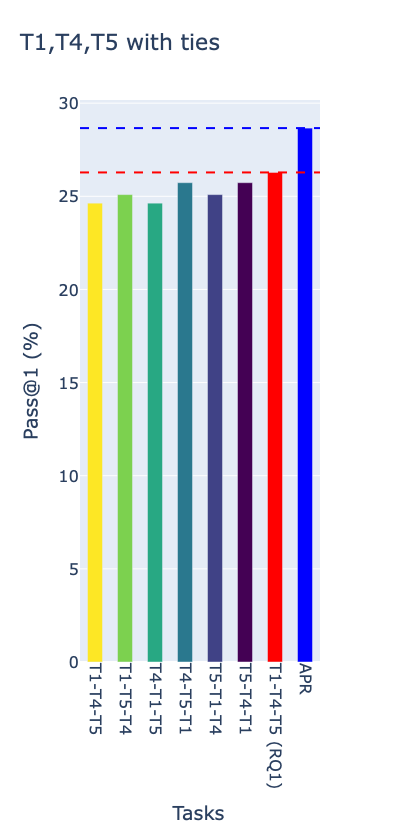}
    \label{fig:fsub2}
\end{subfigure}
\begin{subfigure}{0.32\textwidth}
    \includegraphics[width=\linewidth, height=280pt]{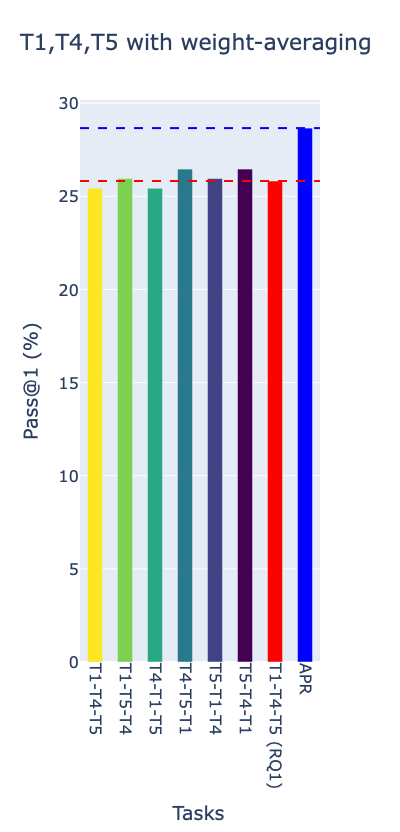}
    \label{fig:fsub3}
\end{subfigure}

\caption{Bar plot of \passone scores obtained by merged adapters of T1, T4, and T5 using continual merging (RQ3) and equal-weight merging (RQ1) approaches using \starcoder model. The tasks are listed as T1 (APR), T2 (Improvement), T3 (Misc), T4 (Development), and T5 (Test \& QA). }
\label{fig:first-bars-2}

\end{figure}


\begin{figure}[h]

\begin{subfigure}{0.32\textwidth}
    \includegraphics[width=\linewidth, height=280pt]{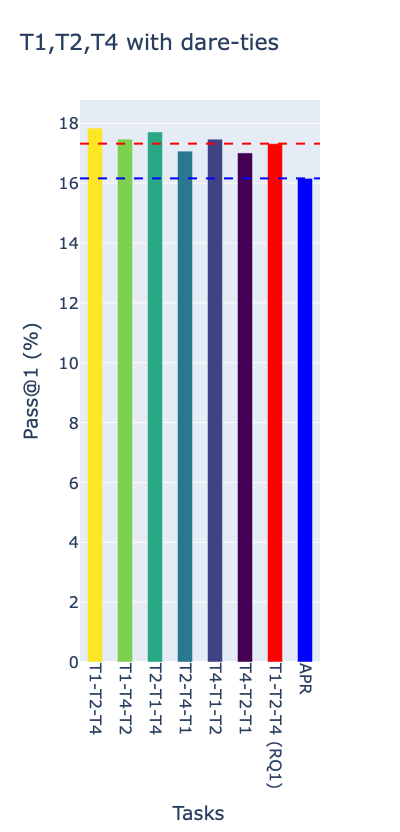} 
    \label{fig:ssub1}
\end{subfigure}
\begin{subfigure}{0.32\textwidth}
    \includegraphics[width=\linewidth, height=280pt]{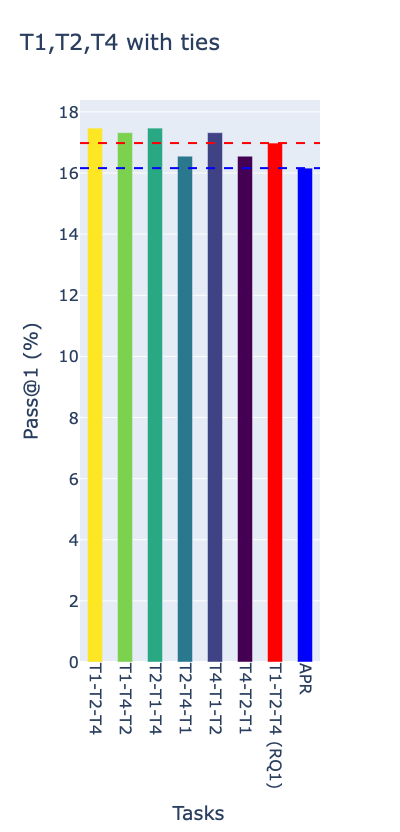}
    \label{fig:ssub2}
\end{subfigure}
\begin{subfigure}{0.32\textwidth}
    \includegraphics[width=\linewidth, height=280pt]{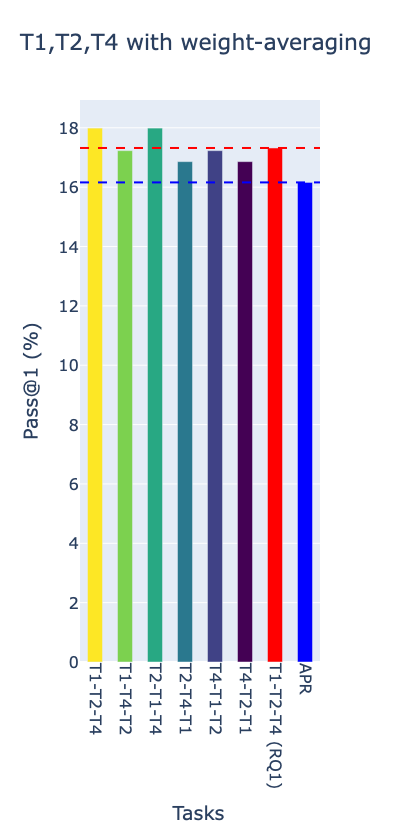}
    \label{fig:ssub3}
\end{subfigure}

\caption{Bar plot of \passone scores obtained by merged adapters of T1, T2, and T4 using continual merging (RQ3) and equal-weight merging (RQ1) approaches using \granite model. The tasks are listed as T1 (APR), T2 (Improvement), T3 (Misc), T4 (Development), and T5 (Test \& QA). }
\label{fig:second-bars-1}

\end{figure}

\begin{figure}[h]

\begin{subfigure}{0.32\textwidth}
    \includegraphics[width=\linewidth, height=280pt]{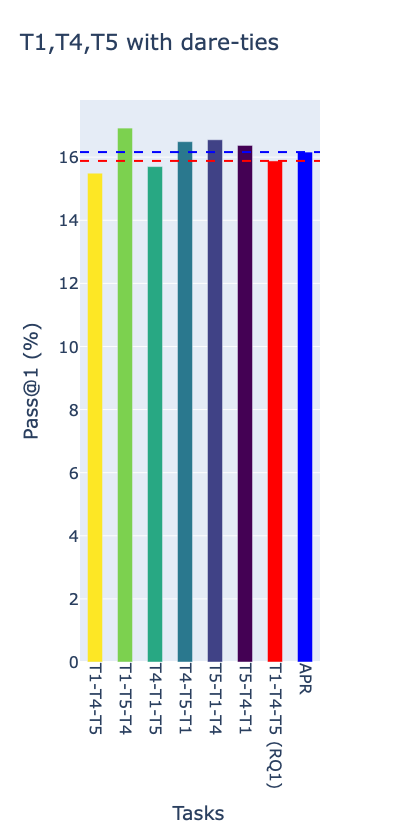} 
    \label{fig:ssub1}
\end{subfigure}
\begin{subfigure}{0.32\textwidth}
    \includegraphics[width=\linewidth, height=280pt]{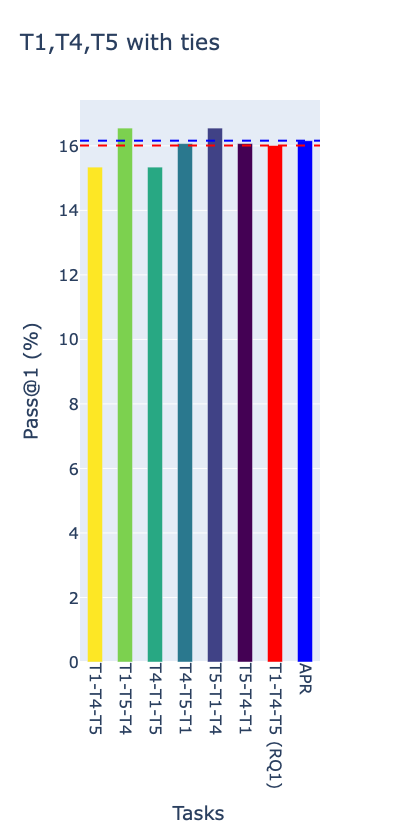}
    \label{fig:ssub2}
\end{subfigure}
\begin{subfigure}{0.32\textwidth}
    \includegraphics[width=\linewidth, height=280pt]{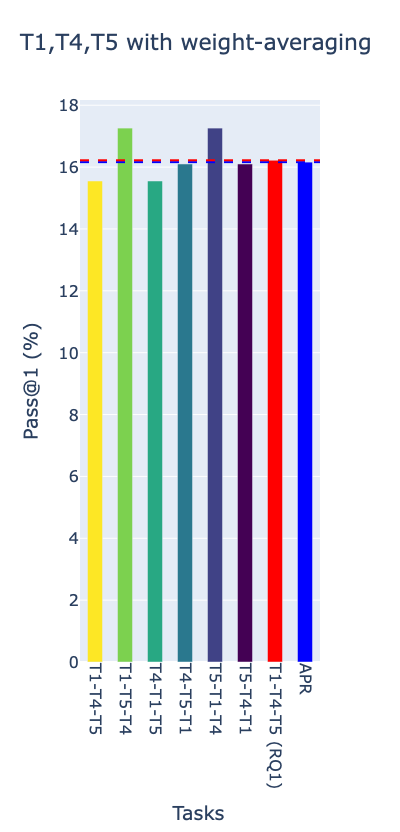}
    \label{fig:ssub3}
\end{subfigure}

\caption{Bar plot of \passone scores obtained by merged adapters of T1, T4, and T5 using continual merging (RQ3) and equal-weight merging (RQ1) approaches using \granite model. The tasks are listed as T1 (APR), T2 (Improvement), T3 (Misc), T4 (Development), and T5 (Test \& QA). }
\label{fig:second-bars-2}

\end{figure}

Comparing the results of this RQ with the results of the first RQ, we can observe that continual merging can lead to better performance on APR benchmark if the order of the tasks are selected carefully. Figure \ref{fig:first-bars-1} illustrates the comparison between \passone scores of the T1, T2, and T4 tasks merged using equal-weight merging (RQ1) and continual merging along with the APR adapter, using \starcoder model. 
In all plots, the APR plot is shown in \textcolor{blue}{blue} and the \textcolor{red}{red} plot refers to the equal-weight merging adapter (RQ1). 
As it is observed, when using dare-ties or ties merging methods, we can achieve better performance when the order of the merged adapters is optimal. 
The performance of the equal-weight merged adapter of these adapters is slightly higher than the performance of all the merged adapters using continual merging.

Figure \ref{fig:first-bars-2} shows \passone score for the adapters derived by T1, T4, and T5 tasks on top of \starcoder model. Unlike the previous tasks, here we can observe that the equal-weight merging approach only performs better than continual merging with the ties method. With the two other merging methods, we can again achieve better performance in continual merging using two orders that contain the APR (T1) task in the last position.

Similarly, with the \granite model, we achieve better performance using continual merging compared to equal-weight merging when an efficient order in continual merging is used. Figures~\ref{fig:second-bars-1} and \ref{fig:second-bars-2} represent the associated plots. Using \granite model, the performance is improved with continual merging with any of the merging methods. 


As the order of the base tasks plays a significant role in the results of the merged adapter, knowing with which order we can obtain a superior result can help reduce the trials to find the best order of merged adapters. According to the results reported in Table \ref{tab:rq3-pass1-3tasks}, having a set of tasks, a greedy approach to find the best order of tasks that can perform better than others is to sort the best performing tasks based on their individual scores (reported in Table \ref{tab:ind-passes}) and merge them in ascending order; meaning that the tasks with higher scores should be merged in the last steps. For example, for the tasks of T1, T2, and T4, their ascending order based on their individual performances is T4-T2-T1 for the \starcoder model. Consequently, the merged adapter of T4-T2-T1 performs the best among all the merged adapters of the same base tasks. Similarly, for the tasks of T1, T2, and T5, the best order is T5-T2-T1 which has the highest performance over the merged adapters of the same tasks. Also, for T1, T4, and T5 the best order is T5-T4-T1, which is again the best performing merged adapter in its category. Note that these results are consistent across all merging methods. However, when merging four tasks together, this greedy approach does not always lead to the best performance of the adapters. As it is observed in Table \ref{tab:rq3-pass1-4tasks-1}, the best performing adapters are T4-T5-T1-T2 and T4-T5-T2-T1 and none of them has the best order based on the individual results.

For the second model, i.e., granite-2b-code-base, the best order of tasks is T5-T1-T2-T4, but this order does not always perform the best when merging all these adapters together except for the both \passone and \passten scores of the weight-averaging method. Unlike the previous model, for the merged adapters with three base adapters the greedy approach does not always produce merged adapters with the highest scores. For example, when merging T1, T2, and T5 adapters, the best performing adapter is T1-T5-T2 which does not have the best order based on their individual performances.


\begin{tcolorbox}[colback=green!5!white,colframe=boxheader,title=RQ3 Summary]

    In the continual merging approach, the order of merging task-specific adapters affects the performance of the merged adapters significantly. It is consistently observed that merging the best performing individual adapters in the last iterations lead to higher overall performance of the merged adapter on the APR benchmark. This is related with the weight influence of adapters as the lastly added adapters have higher weights in the merged adapter. Such observation happens among all merging methods and both models. On the other hand, merging the low-performance adapters in the last iterations can degrade the performance of the merged adapter. \\
    Additionally, for most of the merged task-specific adapters, we observe an improved performance when using continual merging, considering the best order of the adapter, compared to their equal-weight averaging used in RQ1. This shows that having different weights for the task-specific adapters and considering the order of the adapters in the merged adapter can affect the performance of the merged adapter. 
    
\end{tcolorbox}

\section{Discussion}
\label{sec:disc}

This section provides a more detailed analysis of the experiments conducted on merging task-specific adapters. Specifically, we present examples from the HumanEvalFix benchmark, highlighting cases where individual adapters either succeeded or failed to solve the problems. Additionally, we compare the efficiency of the continual merging approach with the equal-weight merging approach, and evaluate the robustness of both individual and merged adapters under input perturbations. Finally, we analyze the Fraction of Sign Difference in the weight parameters of the adapters.






\subsection{Example Cases}

\begin{table}[]
\centering
\caption{An example of the HumanEvalFix benchmark for which the individual task-specific adapters are not able to solve the problem while the merged version of them using the weight-averaging method solves the problem successfully.}
\label{tab:s24-case}
\begin{tabular}{@{}ll@{}}
\toprule
\begin{tabular}[c]{@{}l@{}}The input prompt consisting of the question, buggy\\ code, test case, and the initial part of the answer that the\\ model is expected to complete.\end{tabular}                       & \begin{tabular}[c]{@{}l@{}}
\begin{lstlisting}
Question: 
Fix bugs in largest_divisor 

def largest_divisor(n: int) -> int: 
    for i in reversed(range(n)):
        if n - i == 0:
            return i

def check(largest_divisor):
    assert largest_divisor(3) == 1
    assert largest_divisor(7) == 1
    assert largest_divisor(10) == 5
    assert largest_divisor(100) == 50
    assert largest_divisor(49) == 7

check(largest_divisor)

Answer:
def largest_divisor(n: int) -> int:
\end{lstlisting}
\end{tabular} \\ \midrule
\begin{tabular}[c]{@{}l@{}}The wrong code generated by each of the task-specific adapters of\\ T1, T2, and T3. Note that all of these adapters have generated the same\\ code for the given prompt.\end{tabular} & \begin{tabular}[c]{@{}l@{}}
\begin{lstlisting}
def largest_divisor(n: int) -> int:
    for i in reversed(range(1, n + 1)):
        if n % i == 0:
            return i
\end{lstlisting}
\end{tabular}                                                                                                                                        \\ \midrule
\begin{tabular}[c]{@{}l@{}}The correct code generated by the merged version of the constituent\\ adapters (T1, T2, and T3) derived by weight-averaging method.\end{tabular}                                      & \begin{tabular}[c]{@{}l@{}}
\begin{lstlisting}
def largest_divisor(n: int) -> int:
    for i in reversed(range(n)):
        if n % i == 0:
            return i
    return 1
\end{lstlisting}
\end{tabular}                                                                                                                              \\ \bottomrule
\end{tabular}
\end{table}

As mentioned earlier, in some cases, we observed that merged adapters could outperform the individual task-specific adapters on the APR benchmark. For example, the \passten score of the T1-T2-T3 merged adapter is higher than that of any individual task-specific adapter. While there were many samples in the evaluation dataset that all these adapters could solve, there were a few examples that only the merged version was able to solve. As shown in Figure \ref{tab:s24-case}, the problem is solved exclusively by the merged adapter and remains unsolved when using the individual task adapters.
This suggests that merging different task adapters can be beneficial for solving certain samples.

\begin{table}[]
\centering
\caption{Different perturbation types on problem 15 of the HumanEvalFix dataset. Note that only the first part of the input prompt is included and the rest are eliminated as those are the same for all samples.}
\label{tab:perturbed-examples}
\begin{tabular}{ll}
\hline
non-perturbed code                                                                           & \begin{tabular}[c]{@{}l@{}}Question: Fix bugs in largest\_divisor.\\ \begin{lstlisting}
def largest_divisor(n: int) -> int:
    """For a given number n, find the largest number that divides n evenly, 
    smaller than n
    >>> largest_divisor(15)
    5
    """
    for i in reversed(range(n)):
        if n - i == 0:
            return i
\end{lstlisting}\end{tabular}                                                                                                                              \\ \hline
\begin{tabular}[c]{@{}l@{}}perturbed by \texttt{format} \\ (doc2comment)\end{tabular}                    & \begin{tabular}[c]{@{}l@{}}Question: Fix bugs in largest\_divisor.\\ \begin{lstlisting}[escapechar=\%]
def largest_divisor(n: int) -> int:
    %\reduline{\#}% For a given number n, find the largest number that divides n evenly, 
    %\reduline{\#}% smaller than n
    %\reduline{\#}% >>> largest_divisor(15)
    %\reduline{\#}% 5
    for i in reversed(range(n)):
        if n - i == 0:
            return i
\end{lstlisting}\end{tabular}                                                                                                                                 \\ \hline
\begin{tabular}[c]{@{}l@{}}perturbed by \texttt{syntax} \\ (DeadCodeInserter)\end{tabular}               & \begin{tabular}[c]{@{}l@{}}Question: Fix bugs in largest\_divisor.\\ \begin{lstlisting}[escapechar=\%]
def largest_divisor(n: int) -> int:
    """For a given number n, find the largest number that divides n evenly, 
    smaller than n
    >>> largest_divisor(15)
    5
    """
    %\reduline{for i in reversed(range(n)):}%
        %\reduline{while False:}%
            %\reduline{if n \% i == 0:}%
                %\reduline{return i}%

    for i in reversed(range(n)):
        if n - i == 0:
            return i
\end{lstlisting}\end{tabular} \\ \hline

\begin{tabular}[c]{@{}l@{}}perturbed by \texttt{function} \\ (FuncRenameButterFinger)\end{tabular}     & \begin{tabular}[c]{@{}l@{}}Question: Fix bugs in \reduline{larhest\_divisor}.\\ 
\begin{lstlisting}[escapechar=\%]
def %\reduline{larhest\_divisor}%(n: int) -> int:
    """ For a given number n, find the largest number that divides n evenly, 
    smaller than n
    >>> %\reduline{larhest\_divisor}%(15)
    5
    """
    for i in reversed(range(n)):
        if n - i == 0:
            return i    
\end{lstlisting}\end{tabular}                                                                                                                             \\ \hline
                                                                        
\end{tabular}
\end{table}

\subsection{Robustness Evaluation of Task-Specific Adapters}

Code LLMs have been shown to be sensitive to small perturbations in input prompts. In real-world applications, such issues with prompts are common. However, most current benchmarks used to evaluate Code LLMs do not adequately capture the robustness capabilities of these models. A recent study \cite{recode} proposed a framework for introducing intentional perturbations to the input prompts, thereby enabling the evaluation of model robustness, primarily for the code generation task.

Following this work, we extend their framework to the automated program repair task by applying their perturbations to the HumanEvalFix dataset, resulting in a new dataset called \textbf{HumanEvalFixPerturbed}. This dataset is used to evaluate model robustness and compute the \robustpass scores, which are then compared with the original \passk scores.
Following the previous study, we apply all perturbations spanning over four categories of \texttt{format}, \texttt{syntax},  and \texttt{function} to HumanEvalFix dataset, resulting in $1640$, $2296$ and $4264$  samples in \texttt{format}, \texttt{syntax}, and \texttt{function} categories, respectively. 
Table~\ref{tab:perturbed-examples} shows examples of these perturbations. Note that for each category, there are multiple perturbations that are applied by the library.


For each of these categories, we apply the perturbations using 5 different seeds, following the methodology of the previous work \cite{recode}. We then compute the \robustpass scores with $k=1$ for the individual task-specific adapters and two selected merged adapters.
It is important to note that evaluating \robustpass scores for all merged adapters is computationally intensive and not feasible given our available resources. Specifically, evaluating a single adapter across all four variants of the perturbed dataset takes at least 8--28x time, which is significantly more than the approximate 1 hour needed to compute the \passk score for a single model.
As a result, we limit this experiment to a subset of adapters. 

\begin{table}[]
\centering
\caption{RobustPass@1 scores of individual task-specific adapters vs. two merged adapters trained using \textbackslash{}textbackslash\{\}starcoder as the backbone model evaluated across four categories of perturbations.}
\label{tab:rpk}
\begin{tabular}{@{}lllcccc@{}}
\toprule
                                               & \multicolumn{2}{l}{\textbf{Tasks \& Merging Methods}} & \multicolumn{1}{l}{\textbf{Format}} & \textbf{Syntax}  & \textbf{Function} & \multicolumn{1}{l}{\textbf{Docstring}} \\ \midrule
\multirow{3}{*}{\textbf{Individual Adapters}}  & \multicolumn{2}{l}{\textbf{Development (T4)}}         & 13.01\%                             & 13.01\%          & 03.66\%           & 24.81\%                                \\
                                               & \multicolumn{2}{l}{\textbf{Improvement (T2)}}         & 13.11\%                             & 11.79\%          & 04.47\%           & 26.23\%                                \\
                                               & \multicolumn{2}{l}{\textbf{Program Repair (T1)}}      & \textbf{14.33\%}                    & \textbf{13.11\%} & \textbf{05.28\%}  & \textbf{29.01\%}                       \\ \midrule
\multirow{3}{*}{\textbf{Merged Adapter (RQ3)}} & \multirow{3}{*}{\textbf{T4-T2-T1}} & weight-averaging & \textbf{14.33\%}                    & 12.80\%          & 04.88\%           & \textbf{29.01\%}                       \\
                                               &                                    & ties             & \textbf{14.33\%}                    & 12.50\%          & 04.57\%           & \textbf{29.01\%}                       \\
                                               &                                    & dare-ties        & 14.13\%                             & 12.70\%          & 04.88\%           & 27.59\%                                \\ \bottomrule
\end{tabular}
\end{table}

Table \ref{tab:rpk} reports the RobustPass@1 scores for individual task-specific adapters, along with results for merged adapter, T4-T2-T1, as this one showed the best performance when adapter order mattered (RQ3). 
When comparing the RobustPass@1 results of the merged adapter with those of the individual adapters, we do not observe significant differences. However, when comparing the RobustPass@1 scores with the original Pass@1 scores for the same adapters, we observe a general decline. These drops are significant across all categories, with the exception of the \texttt{function} category, which shows a substantial decrease.  

The \texttt{function} category yields the lowest RobustPass@1 scores among all categories, highlighting the importance of having correct and meaningful function names in the input prompts to the model. It is important to note that in this category, function names appear in multiple parts of the input prompt, such as in the function definition and in the test cases. In our adapted code, we apply the same change all over and modify all occurrences of the function names. 
Among the remaining categories, \texttt{format} and \texttt{syntax} have similar scores, the latter being a bit lower. 
Among the different merging techniques, all three have similar scores, and there is no specific pattern showing either of them is more robust.


\subsection{Continual Merging vs. Equal-Weight Merging}

One of the main advantages of the continual merging approach is its memory efficiency compared to equal-weight merging. Specifically, the continual merging strategy requires storing only one merged adapter at a time along with the next adapter to be merged. Adapters are merged sequentially, eliminating the need to load and store all of them simultaneously. This property becomes especially important when dealing with a large number of task-specific adapters or when the adapter sizes are substantial.
Notably, this memory efficiency is achieved without a significant drop in the performance of the merged adapters on the APR task. In fact, the continual merging approach can still yield high performance on the APR benchmark.

As mentioned earlier, when comparing the results of continual merging with equal-weight merging, we do not observe a significant difference in overall performance. Nevertheless, since the order in which task-specific adapters are merged affects the final performance, better results can be achieved by carefully selecting the merging sequence.
Consider the Pass@1 scores of the merged adapters involving three task-specific adapters, as reported in Tables \ref{tab:rq1-pass1} and \ref{tab:rq3-pass1-3tasks} for equal-weight and continual merging, respectively. Interestingly, for every adapter combination merged using the equal-weight approach, we can identify a variant of continual merging with the same base tasks that outperforms it. This trend also holds for merged adapters involving four task-specific adapters.
These observations suggest that continual merging can be more effective, provided that an optimal order of merging is identified.

\subsection{Fraction of Sign Difference in Adapter Weight Parameters}

To further analyze the results, we consider the internal adapter parameters of the models using the fraction of sign difference~\cite{generaliz}, i.e., the portion of parameters having opposite signs, of the participating adapters in the merging process. 
Most existing merging methods take into account the sign of the parameters when merging different models or adapters. For instance, in the TIES method, only parameters with matching sign values are averaged in the final step. Therefore, analyzing the Fraction of Sign Difference (FSD) is crucial to understanding whether there is a relation between the performance of merged adapters and the similarity of their sign patterns (low or high FSD).
A higher FSD value indicates that the adapters have more parameters with differing sign values, whereas a lower FSD suggests fewer such differences. Intuitively, adapters with a lower FSD can be considered more similar to each other, and thus their performance is expected to be more aligned compared to adapters with a higher FSD.

\begin{figure}
    \centering
    \begin{minipage}{0.49\textwidth}
        \centering
        \includegraphics[width=\linewidth]{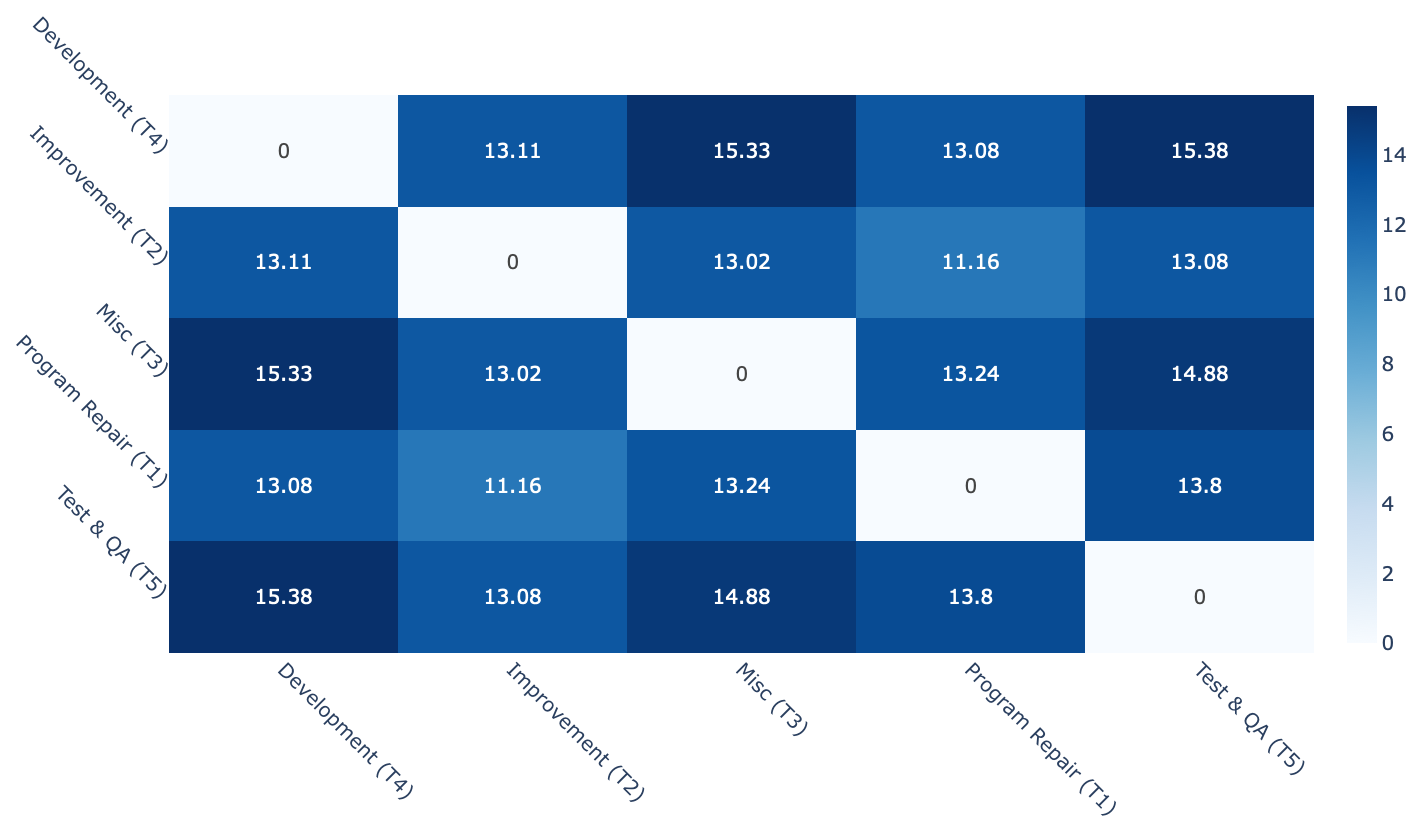}
        \subcaption{task-specific adapters trained on \starcoder}
        \label{fig:left}
    \end{minipage}
    \begin{minipage}{0.49\textwidth}
        \centering
        \includegraphics[width=\linewidth]{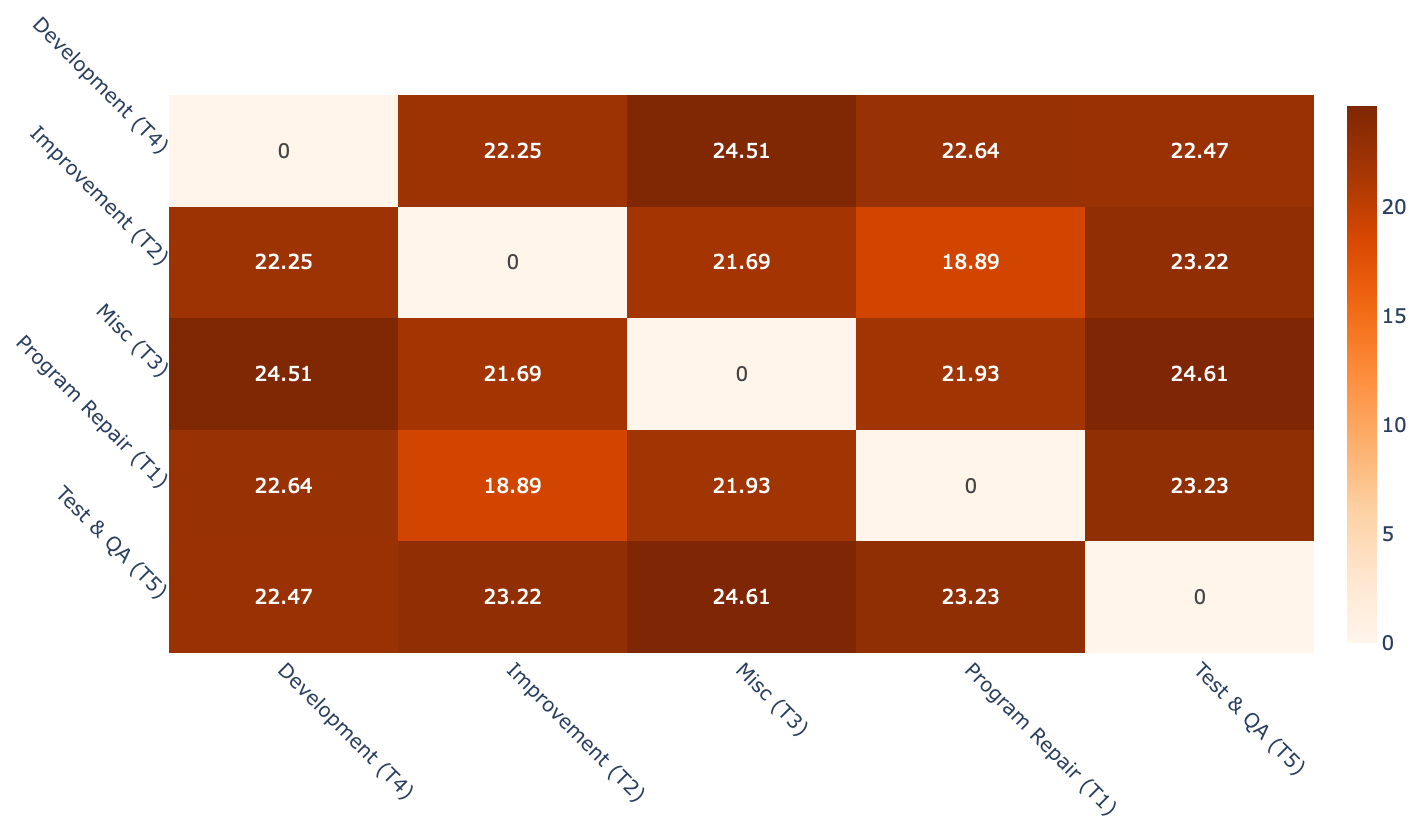}
        \subcaption{task-specific adapters trained on \granite}
        \label{fig:right}
    \end{minipage}
    
    \caption{The Fraction of Sign Difference (FSD) between individual task-specific adapters trained on top of \starcoder and \granite models. The values show the percentage of parameters having different signs.}
    \label{fig:fsd-ind}
\end{figure}

Figure \ref{fig:fsd-ind} shows the FSD of individual task-specific adapters that were trained on top of \starcoder model compared to other adapters. By analyzing this figure with the performance of each adapter reported in Table \ref{tab:ind-passes}, we observe that the tasks with a lower FSD, perform relatively closer to each other in terms of \passk score. For example, T1 and T2 have the lowest FSD and their performance is also very close to each other being 28.66\% and 27.90\%, respectively. Similarly, T5 has the highest FSD with T1 and their performance difference is also larger than all the other tasks compared with T1. However, the same observation does not hold for T4 and T3 in comparison with T1. While T3 performs closer to T1, it has a higher FSD which shows that there is no general correlation between the FSD and the performance of the individual adapters. 

On the other hand, having the FSD values between different task adapters, we cannot understand which adapters will perform better or worse compared to a specific adapter like T1. For instance, while T3 performs better than T1, it has a higher FSD compared to the FSD value of T1 and T2. In fact, the FSD can only be helpful to understand the relative similarity of task-adapters and therefore, we can conclude which adapters will lead to a smaller change after applying merging. 

As it is observed from Figure \ref{fig:right}, the same pattern exists for the \granite model. As mentioned before, the FSD values do not represent the performance drop or improvement of the adapters on T1. But, they represent the similarity of task-specific adapters, and therefore their performance scores' closeness to each other. As such, we observe that the FSD of T1 and T2 is still lower than the FSD of T1 and other adapters, although for this model, the performance of the T2 task is better than T1.

In summary, when having the FSD value of APR task relative to other task-specific adapters, we can decide which task-specific adapters will have a higher impact on the APR task if they are merged together, in terms of having a closer performance to APR. In other words, when merging APR adapter with task-specific adapters having a high FSD value, the resulting adapter will perform significantly different (could be higher or lower) compared to the case where we merge APR with a task-specific adapter having a low FSD. This is supported by the differences in the performance that we observed in Table~\ref{tab:rq1-pass1}, specifically when on adapter is merged with APR (T1). There is however, no specific conclusion when more adapters are merged with APR, because it depends on other adapters that are added as third or fourth ones.

\begin{figure}
\includegraphics[width=\textwidth]{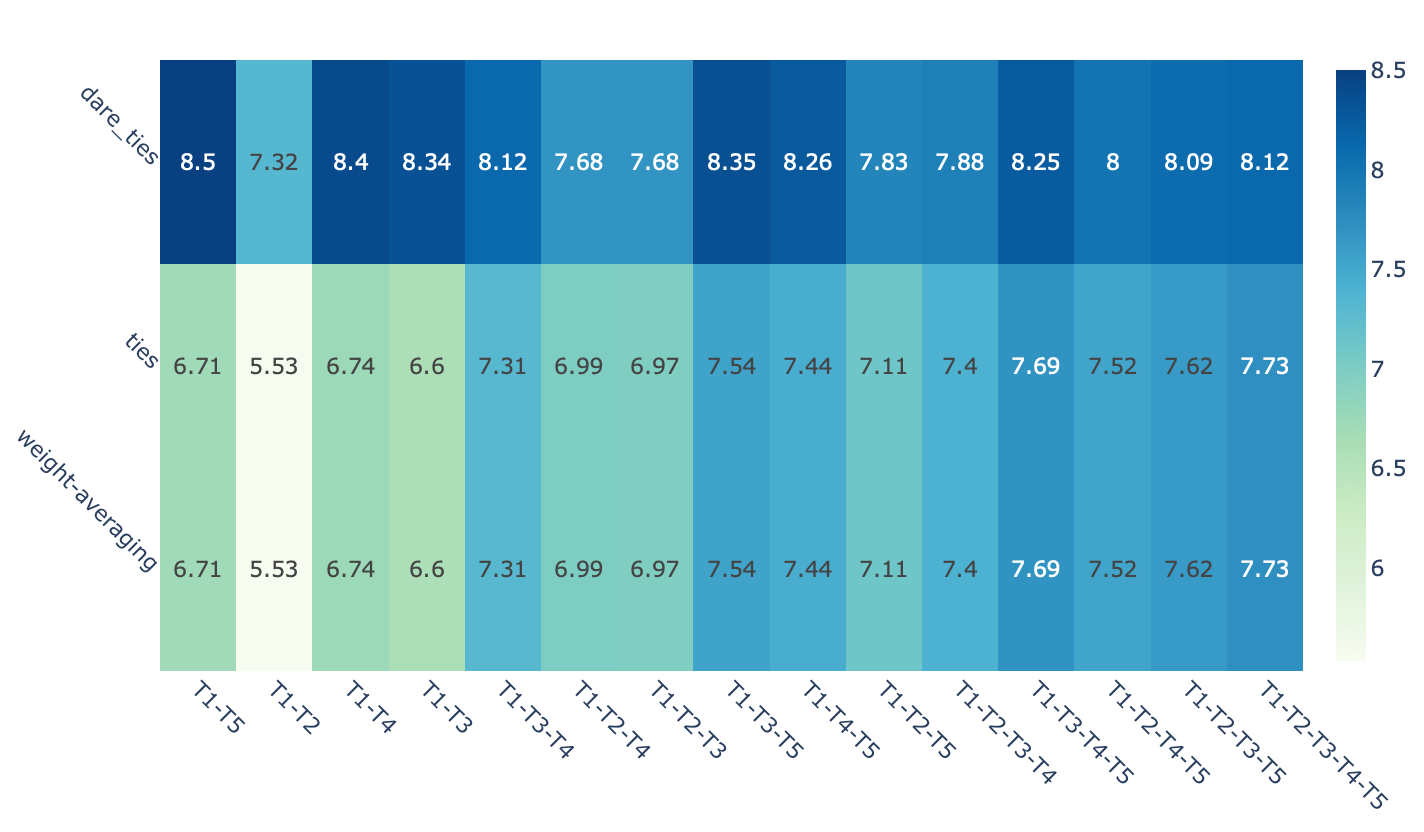}
\caption{The Fraction of Sign Difference (FSD) between the merged adapters of RQ1 vs. APR adapter trained on \starcoder model.}
\label{fig:fsd-rq1}
\end{figure}

We also compute the FSD values between the merged adapters of each RQ and APR adapter. 
As the FSD differences of the adapters trained on \starcoder and its performance on all RQs are better than Granite, in the following, we only consider the \starcoder model. 
Figure \ref{fig:fsd-rq1} shows these values for RQ1 adapters and APR. The first observation is that the FSD value of adapters merged with ties method are all similar to the ones merged with weight-averaging method. This happens due to the fact that ties merging technique is similar to weight-averaging. In ties merging, after assigning the sign vectors, the final parameters are averaged, similar to weight-averaging. 

Moreover, when comparing the values of merged adapters with different number of adapters, it can be observed that the adapters with lower constituent task-adapters have a lower FSD for ties and weight-averaging methods. As the number of adapters increases in these methods, the FSD values get higher with the last merged adapter having the highest FSD. Almost the opposite trend can be seen for the dare\_ties method meaning that the merged adapters with more constituent adapters have relatively lower FSD values. Therefore, we cannot correlate the number of adapters in each merged adapter with their FSD values with respect to the APR adapter.

Additionally, the overall values of FSD for the adapters of RQ1 vs. APR adapter are lower than the FSD values of single task-specific adapters vs. APR adapter across all methods and merged adapters. This suggests that after merging APR with other similar adapters, the resulting adapters become more similar to APR in terms of their FSD. 

It is worth mentioning that such results do not correlate with the performance of the merged adapters. For example, a lower FSD value for the merged adapters does not necessarily show a high performance for that adapter. More specifically, this can be observed when considering the results of T1-T2 and T1-T3 tasks. Although T1-T2 has a lower FSD value, it performs worse than T1-T3 merged adapter. This happens across all merging methods.    

\begin{figure}
\includegraphics[width=\textwidth]{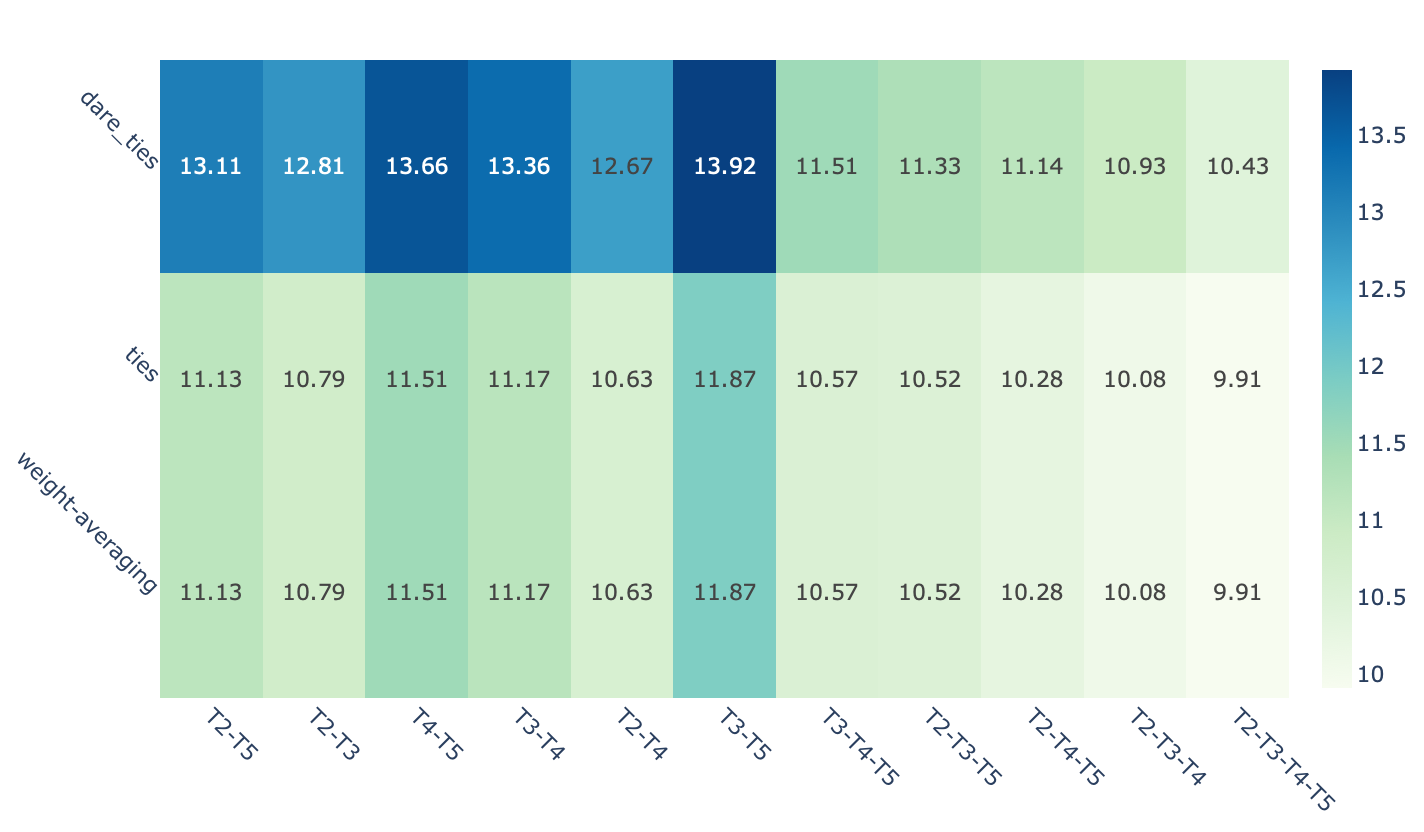}
\caption{The Fraction of Sign Difference (FSD) between the merged adapters of RQ2 vs. APR adapter trained on \starcoder model.}
\label{fig:fsd-rq2}
\end{figure}

Figure \ref{fig:fsd-rq2} reports the FSD values of the merged adapters in the second RQ again with respect to the APR adapter. As it is observed the overall values are higher than the values of the first RQ. This could be valid and expected as in this RQ, we eliminate the APR adapter form the merged adapter. Hence, the merged adapter of other task-specific adapters should have more parameters with different sign values compared with APR. This unveils the reason behind having a lower performance for the merged adapters of non-APR task-specific adapters compared to the first RQ where all the merged adapters include the APR adapter and their performance are higher in general.  

Unlike the previous RQ, the FSD values of the merged adapters get lower as the number of the tasks involved in the merged adapter increases. This trend is consistent for all three merging methods. 

Similar to previous observations, the low/high score of FSD is not related to the performance of the merged adapters. For example, the merged adapter T2-T3-T4-T5 has the lowest FSD in Figure~\ref{fig:fsd-rq2}, but, also the a relatively lower performance compared to when two adapters are merged (See Table~\ref{tab:rq2-pass1}).

\begin{figure}
\includegraphics[width=\textwidth]{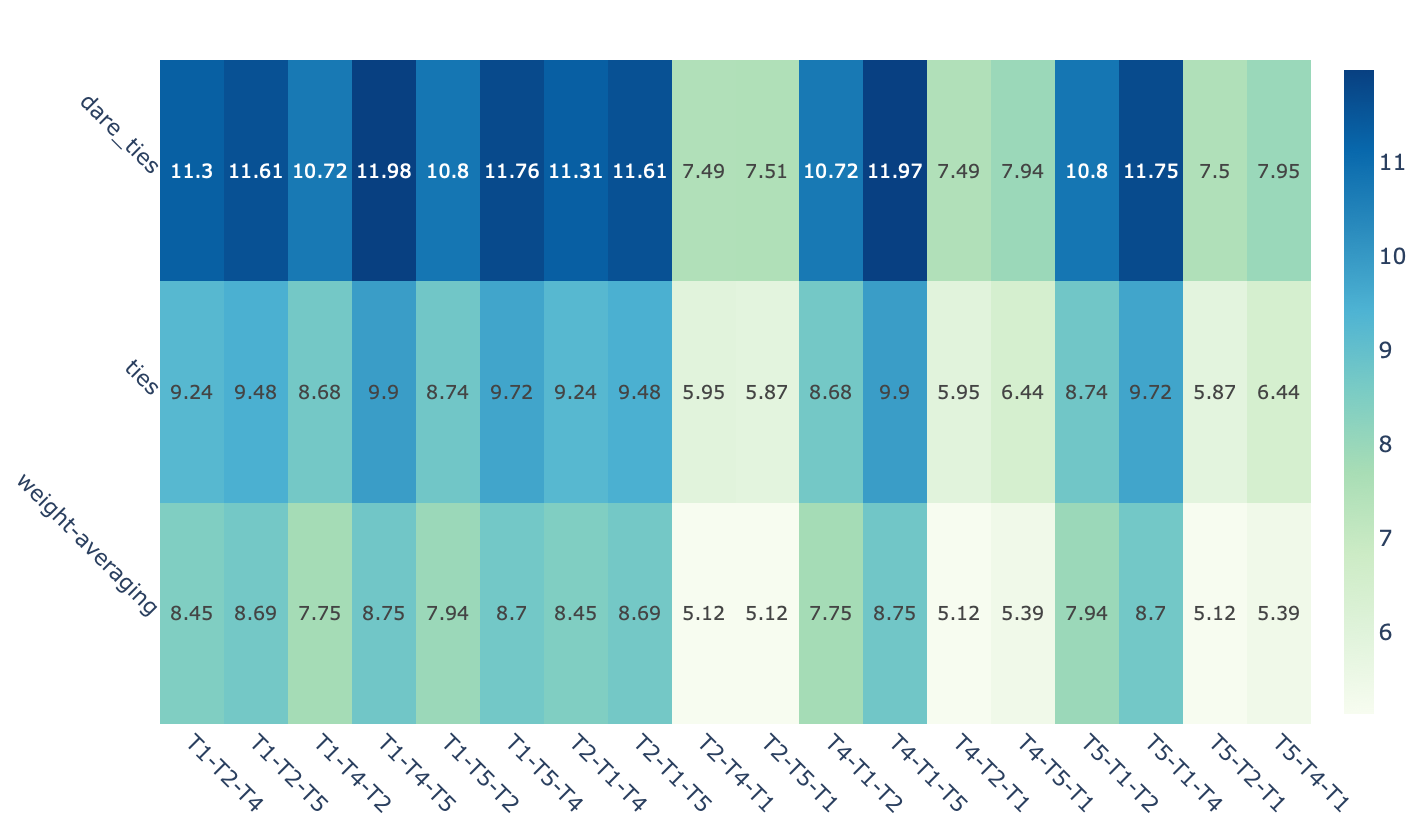}
\caption{The Fraction of Sign Difference (FSD) between the merged adapters of RQ3 having length of 3 vs. APR adapter trained on \starcoder model.}
\label{fig:fsd-rq3-1}
\end{figure}

\begin{figure}
\includegraphics[width=\textwidth]{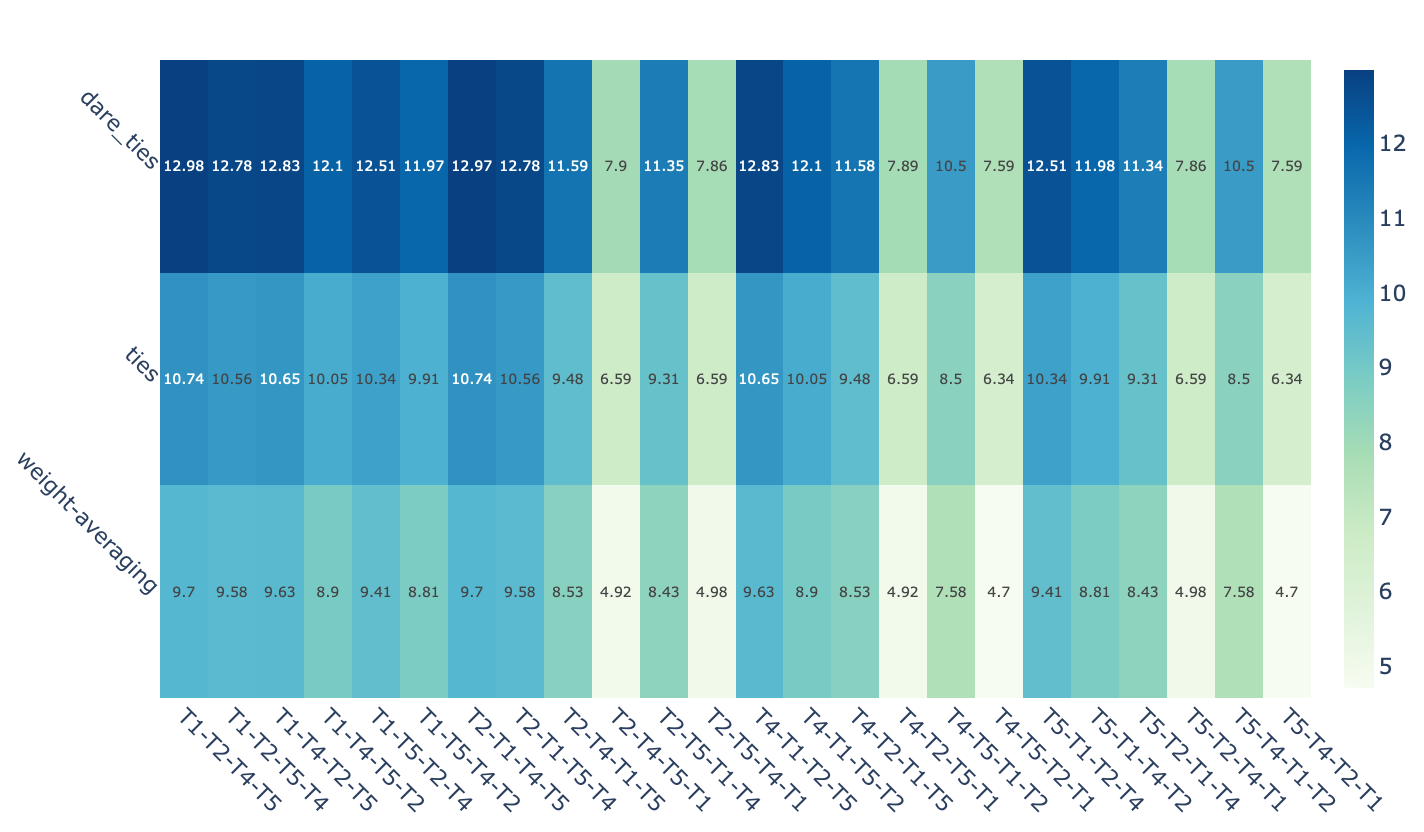}
\caption{The Fraction of Sign Difference (FSD) between the merged adapters of RQ3 having length of 4 vs. APR adapter trained on \starcoder model.}
\label{fig:fsd-rq3-2}
\end{figure}

Figures \ref{fig:fsd-rq3-1} and \ref{fig:fsd-rq3-2} show the FSD results of merged adapters having 3 and 4 adapters, respectively. 
The first difference of the FSD results of RQ3 is that in continual merging approach, the FSD values of the ties and weight-averaging methods are no longer the same. This can be due to different weights considered in the continual merging, so the sign differences can vary. 

Recall that in the third RQ, the position of the adapters in the merged adapters played a significant role. By having a closer look at these two figures, we notice that as we involve the APR adapter in the last step of the continual merging, the FSD values of the merged adapters vs. the APR adapter becomes lower. This is in agreement with the fact that the lastly added adapters in the merged adapters of RQ3 have the highest impact in the merged adapter. Similarly, the merged adapters that contain the APR adapter in their first steps of merging have a higher FSD value which further justifies the above mentioned fact. 

The \passk results of the third RQ showed that the performance of the merged adapters are highly dependent on the performance of the last adapters in the merged adapter. By analyzing these results and the FSD values for this RQ, the impact of the last adapters on the merged adapters can be further observed. This observation is however different from our previous discussions in this section. For the other two RQs, and single task adapters, would could not observe a direct relation among the \passk scores and the FSD values. But we can have a stronger conclusion for this relation in case of continual merging. 

\begin{figure}[h]

\begin{subfigure}{0.4\textwidth}
    \includegraphics[width=\linewidth]{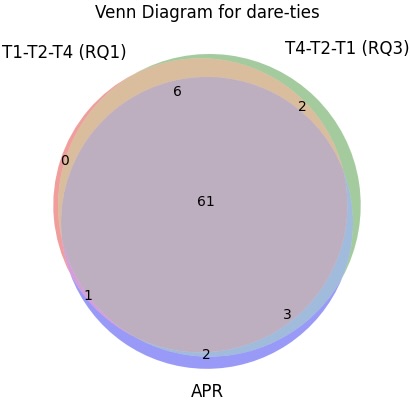} 
    \label{fig:venn1}
\end{subfigure}
\begin{subfigure}{0.4\textwidth}
    \includegraphics[width=\linewidth]{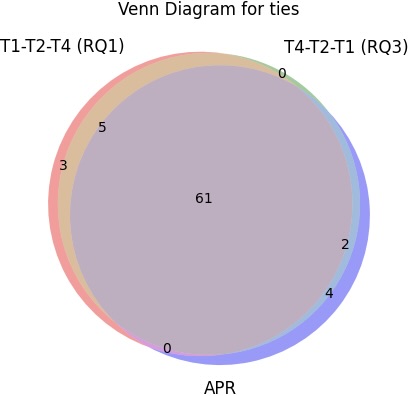}
    \label{fig:venn2}
\end{subfigure}
\centering
\begin{subfigure}{0.4\textwidth}
    \includegraphics[width=\linewidth]{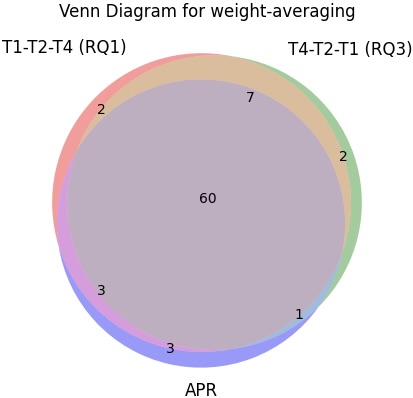}
    \label{fig:venn3}
\end{subfigure}

\caption{Venn diagrams of the number of problems solved by two merged adapters obtained by equal-weight merging (RQ1) and continual merging (RQ3) approaches along with the APR adapter across three merging methods. }
\label{fig:venns}

\end{figure}


\subsection{The State of the Solved Problems}

To analyze the ratio of the problems that were solved by different merged adapters, we illustrate the number of unique problems that were solved by each adapter for T4-T2-T1 adapter as shown in Venn diagrams of Figure \ref{fig:venns}. As it is observed, many problems are solved by any of the adapters, i.e., T4-T2-T1 from RQ3, T1-T2-T4 from RQ1 and APR adapter itself. 
However, there are some problems that can only be solved using merged adapters (either equal-weight averaging or continual merging); and the number of problems that are uniquely solved by merged adapters are more than the number of problems that are uniquely solved by APR adapter. 
Different merging methods, i.e., dare-ties, ties, and weigh-averaging affect these numbers.
For example, with dare-ties method, no problem is solved only with T1-T2-T4, but with the ties and weight-averaging methods, there are 3 and 2 problems that are solved only with these merged adapter (see plots related to the RQ1 of Figure~\ref{fig:venns}). Similarly, when considering T4-T2-T1 adapter, we can observe that 2 problems are solved using each of dare-ties and weight-averaging methods and no problem is with ties.
But overall, there are 8, 8, and 11 problems that can only be solved with the merged methods using dare-ties, ties, and weight-averaging, respectively, compared to 2, 0, and 3 problems that can only be solved by APR adapter. 
This examination emphasizes our previous discussions about the benefits of using merged adapters for APR task. 

\subsection{Incorrect and Correct Problems When Changing Merging Technique}

We also calculated the number of problems that were correct using APR adapter and became incorrect using the merged adapter, and vice versa. These results are calculated both in RQ1 and RQ3, using the merged adapters T1-T2-T4 and T4-T2-T1, respectively. Note that the merged adapters are the same, and only their order matters in RQ3. 
Table~\ref{tab:ratio1-pass10} shows the results for \passten scores. The \textit{A} $\rightarrow$ \textit{B} in the table represents the number of problems that were correct using \textit{A}, but were incorrect in \textit{B}. For example, \textit{APR} $\rightarrow$ \textit{weight-averaging} value for equal-weight averaging (RQ1) being 5 means that there were five problems that were correctly solved using \textit{APR} adapter, but are incorrect using \textit{weight-averaging}. 
Two patterns can be seen from the values of this table. 
First, number of incorrect problems increases when we have a \textit{merged-model} $\rightarrow$ \textit{APR}, compared to \textit{APR} $\rightarrow$ \textit{merged-model}. 
This shows the benefits of merged models, even though being small.
Second, when using continual merging on the weight-averaging merging technique, the number of incorrect problems decreases, compared to equal-weight averaging. This number from continual merging is on-par for other two merging techniques, ties and dare-ties compared to equal-weight averaging. 
It is worth mentioning that, the numbers are small, representing 2--5\% of the 165 benchmark problems that are incorrectly solved when we change the models. 
Using \passone scores, we noticed a slightly different results, where the number of incorrect problems increases when we have \textit{APR} $\rightarrow$ \textit{merged-model}, compared to \textit{merged-model} $\rightarrow$ \textit{APR}. A similar trend is seen for continual merging vs. equal-weight averaging.


\begin{table}[]
\centering
\caption{Number of problems that are correctly solved by one model and become incorrect when solving with the other model, using \passten scores from \starcoder model for T1-T2-T4 and T4-T2-T1 merged adapters. The A $\rightarrow$ B in the table represents the number of problems that were correct using A, but were incorrect in B. }
\label{tab:ratio1-pass10}
\begin{tabular}{@{}lcc@{}}
\toprule
Models                             & \# Incorrect problems equal-weight averaging (RQ1) & \# Incorrect problems continual merging (RQ3) \\ \midrule
APR $\rightarrow$ weight-averaging & 5                                                  & 3                                             \\
weight-averaging $\rightarrow$ APR & 6                                                  & 8                                             \\
APR $\rightarrow$ ties             & 6                                                  & 6                                             \\
ties $\rightarrow$ APR             & 8                                                  & 9                                             \\
APR $\rightarrow$ dare-ties        & 4                                                  & 4                                             \\
dare-ties $\rightarrow$ APR        & 9                                                  & 5                                             \\ \bottomrule
\end{tabular}
\end{table}

Table~\ref{tab:correct-pass10} represents the number of correctly solved problems using APR and three merging methods, with equal-weight averaging compared to continual merging. The results are based on \passten scores for T1-T2-T4 and T4-T2-T1 merged adapters, RQ1 and RQ3, respectively. 
Using continual merging, the number of correctly solved problems increases slightly using weight-averaging and ties, and decreases for dare-ties. 
This result emphasize the benefit of continual merging, which also can depend on the merging approach.

\begin{table}[]
\centering
\caption{Number of problems that are correctly solved by each model, using \passten scores of \starcoder model for T1-T2-T4 and T4-T2-T1 merged adapters.}
\label{tab:correct-pass10}
\begin{tabular}{@{}lcc@{}}
\toprule
Models           & \# Correct problems equal-weight averaging (RQ1) & \# Correct problems continual merging (RQ3) \\ \midrule
APR              & 67                                               & 67                                          \\
weight-averaging & 68                                               & 72                                          \\
ties             & 69                                               & 70                                          \\
dare-ties        & 72                                               & 68                                          \\ \bottomrule
\end{tabular}
\end{table}

\section{Implications and Future Research Directions}
\label{sec:implications}

\subsection{Implications for Researchers}
Our study is the first to explore merging the learned parameters of models without additional training. We investigated the capabilities of merged models within the context of parameter-efficient fine-tuning approaches for automated program repair.
Our results demonstrate that merging adapters from related software engineering tasks can benefit complex tasks like APR. Interestingly, the number of merged adapters or even the degree to which their parameter signs differ from those of the APR adapter, does not strongly correlate with performance. Instead, the effectiveness of the merged adapter depends more on the nature of the tasks being merged.
Moreover, not all task-specific adapters are beneficial to a given target task. In our experiments, for instance, merging the T5 adapter negatively impacted the performance of the merged model. This highlights the importance of carefully selecting which task adapters to include in the merge.
A promising direction for future work is to investigate which tasks contribute the most useful knowledge when merging adapters for a specific target task.

In this study, we empirically explored the number and order of tasks for merging. However, exhaustively evaluating all possible task combinations and orders is nontrivial. A promising direction for future research is to develop automated techniques that can identify which tasks and in which order they should be merged to best benefit a given target task.

Additionally, our continual merging results suggest that the order in which adapters are merged affects the final performance, either positively or negatively. In this study, we explored different merging orders empirically. Based on the initial performance of individual adapters on the APR task, we were able to anticipate which tasks might be more beneficial and where they should be positioned in the merged model.
Although we currently lack an automated method to identify the most beneficial tasks for a given target task, our findings indicate that placing the most effective adapter last in the continual merging sequence can improve performance. This is because the last adapter tends to have the greatest influence in the merged model, effectively assigning it a higher weight. For APR, this strategy can help ensure that the most relevant knowledge is prioritized, leading to better overall results.

In this study, we used an existing dataset comprising five tasks, including APR. This choice was particularly important, as we required instruction-tuning data compatible with the \starcoder and \granite models.
After evaluating the results, we observed that the individual adapters performed similarly to APR on the APR benchmark (i.e., the HumanEvalFix dataset). Upon manual inspection, we found that while the task-specific datasets were well-categorized for each task, the APR benchmark contained a more diverse set of samples, including instances from tasks beyond APR.
This may explain why the APR-specific adapter did not outperform other task-specific adapters on the APR benchmark.
Although this observation does not impact the main findings or conclusions of our study, we recommend that future research efforts focus on curating more clearly defined task-specific datasets and dedicated benchmark datasets to ensure accurate evaluation.

Additionally, researchers can investigate the impact of merging a broader variety of tasks for APR or other target tasks. For example, tasks such as comment generation or clone detection could be incorporated into the merged adapters to enhance performance on APR. More generally, researchers can curate new datasets and explore merging techniques across a wider range of software engineering tasks to better understand their effectiveness and transferability.


\subsection{Implications for Practitioners}
Merging models without any additional training, as explored in this work, offers several practical benefits. Given the high cost of dataset curation and labeling, model merging enables practitioners to reuse existing knowledge from related tasks for new tasks. While we do not claim that merged models generalize across all software engineering tasks, our findings suggest that merged models exhibit generalization capabilities for automated program repair. Similar trends have been reported in the NLP domain, where merged models demonstrate promising performance across diverse tasks~\cite{akiba2025evolutionary}.

This indicates that merging existing models may be a viable strategy worth exploring before investing in dataset creation for new SE tasks. The technique is particularly valuable when working with private or confidential codebases, as it allows knowledge transfer without requiring access to sensitive data for training, by leveraging the learned knowledge of other models or tasks. 
Beyond reusing pre-trained models, merging also contributes to energy savings, aligning with the goals of sustainable and environmentally responsible AI~\cite{shi2024efficient}.

Another notable finding from our results is that while different merging methods yield varying scores for the merged adapters, the performance differences across these techniques were not substantial. This suggests that in similar settings, practitioners could opt for simpler and more computationally efficient methods, such as straightforward weight averaging, instead of more complex merging approaches.

In the continual merging setting, where the objective is to maintain a single merged adapter at a time, we observed that the most recently added adapters have a significant impact on the final performance. Therefore, to optimize the merged adapter's effectiveness, higher-performing adapters should be added later in the merging sequence.
Our continual merging approach has two important implications for practitioners. 
First, it enables adapters to be merged sequentially as they become available over time, removing the need to train adapters for all tasks simultaneously.
Second, it allows to weight the adapters differently, giving practitioners the ability to control their influence on the merged model based on the target task.

\section{Threats to Validity}
\label{sec:threats}

Threats to \textbf{internal validity} refer to internal factors that might affect the reliability of the results. 
These factors in our study are related to model checkpoints and the dataset used.
The model checkpoints that we selected are pre-trained without any further fine-tuning process such as instruction-tuning. The instruction-tuning process will align both models to generate more proper output code snippets. 
In all cases, we used the best hyper-parameters and configurations, as recommended by the model developers.  
As we used the open source code and APIs, we anticipate low threat related to this factor. 
The other potential threat concerns the dataset used in our experiments. 
Although the task type of the records in the original dataset has been obtained by prompting GPT-4 model, and might include incorrect labels, the same dataset is used for all experiments and models. Therefore, the results are affected in the same way by the used dataset. 
Please note that such noisy labels might exist in other datasets as well, and their effect needs to be evaluated in a separate study, which is out of the scope of our current study.
Similarly, for testing the trained models, we used existing benchmark datasets to alleviate any threats related to the choice of datasets. 
Another threat can be related to the effect of model architecture and the model size. We did not conduct experiments on the effect of model architecture on the results, and we chose same size models, to reduce the related threats. It is worthy to note that our results are concerned with the merging capability of the trained adapters. To reduce threats related to this point, we chose two models in our study.


Threats to \textbf{external validity} relate to the generalizability of our findings. In this work, we conducted experiments only on automatic program repair, along with the other four tasks for Python language. Although the techniques used in the merged models/adapters can be adopted for other tasks and languages, the obtained results are limited to the used tasks and programming language and might not be applicable to other areas or programming languages.

Threats to \textbf{construct validity} refers to misalignment among the test and what needs to be measured.
In our work, we intend to compare the results for APR task, which is evaluated using \passk metric. 
We found this metric to be more reliable than others, as in the benchmark, the fixed codes generated by the models will be evaluated by test cases, rather than pure consideration of textual similarity.
We also tested the differences of the obtained results using statistical tests and also reported effect size. 

Lastly, we expect low threats to the \textbf{conclusion validity}. As the models are frozen and we kept all the models' configurations the same, the results are expected to be related to the merging techniques. We used different evaluation metrics, supported by statistical tests and additional analysis in the discussion section. 
It is worth noting the difference in the input format of the benchmark dataset when computing \passk and \robustpass results. The bigcode-evaluation-harness library used for \passk eliminates the docstrings from the benchmark, while the \robustpass library \cite{recode} requires docstrings, as some of the perturbations are also applied on them. Therefore, the direct comparison among the scores might not be the best approach. That being said, the conclusion that perturbations affect the performance in a negative way are still valid. 

\section{Conclusion}
\label{sec:con}
In this study, we explored the effect of merging different task-specific adapters in Code LLMs on automatic program repair. The merged models and adapters have shown promising results in other fields including computer vision and natural language processing, and have improved the performance of the model for in-domain data. However, their effect on code-related tasks is not explored. Our study is the first to address this gap by conducting experiments on two merging paradigms, as well as generalizability of the merged adapters to the APR task. Our results suggest that merging task-specific adapters can improve the performance of the models on APR benchmark. Additionally, our proposed continual merging approach can be effective if the order of adapters that are merged together is determined efficiently. Such results and insights could shed light on the use of other code-related tasks to improve the performance of a new task. This work provides insights on how to learn from other tasks and develop adapters that are obtained from multiple adapters by different merging techniques. 
Our code and datasets are shared publicly. 
Future directions for this research could consider other tasks to combine for APR or study the effect of merging techniques for other SE tasks, as well as experimenting with other LLMs and industrial evaluations.

\section*{Data Availability Statement}\label{sec:data-ack}
We include all scripts used to obtain the results our \href{https://github.com/mqddd/mergerepair}{GitHub repository}.

\section*{Conflict of Interest}
The authors declare that they have no conflict of interest. \\
This research is supported by a grant from the Natural Sciences
and Engineering Research Council of Canada RGPIN-2019-05175, as well as support from computational resources and services provided by Advanced Research Computing at the University of British Columbia.

\bibliographystyle{spbasic}      
\bibliography{bibliography}   


\end{document}